\documentclass[12pt]{article}
\usepackage{mathrsfs}
\makeatletter \@addtoreset{equation}{section} \makeatother
\renewcommand{\theequation}{\thesection.\arabic{equation}}
\newcommand{\ba}{\begin{array}}
\newcommand{\ea}{\end{array}}
\newcommand{\beq}{\begin{equation}}
\newcommand{\eeq}{\end{equation}}
\newcommand{\bea}{\begin{eqnarray}}
\newcommand{\eea}{\end{eqnarray}}

\def\bce{\begin{center}}
\def\ece{\end{center}}

\def\nonu{\nonumber}

\def\pa{\partial}
\def\al{\alpha}
\def\be{\beta}
\def\ga{\gamma}

\def\de{\delta}

\def\la{\lambda}

\def\si{\sigma}

\def\eps6{{\displaystyle \mathop{\epsilon}^{6}}{}}
\def\g6{{\displaystyle \mathop{g}^{6}}{}}

\def\nab6{{\displaystyle \mathop{\nabla}^{6}}{}}

\def\0{{\sst{(0)}}}
\def\1{{\sst{(1)}}}
\def\2{{\sst{(2)}}}
\def\3{{\sst{(3)}}}
\def\4{{\sst{(4)}}}
\def\5{{\sst{(5)}}}
\def\6{{\sst{(6)}}}
\def\7{{\sst{(7)}}}
\def\8{{\sst{(8)}}}

\def\ba{\begin{array}}
\def\ea{\end{array}}
\def\beq{\begin{equation}}
\def\eeq{\end{equation}}
\def\be{\begin{equation}}
\def\ee{\end{equation}}

\def\la{\lambda}
\def\eps{\epsilon}

\def\ba{\begin{array}}
\def\ea{\end{array}}
\def\beq{\begin{equation}}
\def\eeq{\end{equation}}
\def\be{\begin{equation}}
\def\ee{\end{equation}}

\def\la{\lambda}
\def\eps{\epsilon}

\def\eps6{{\displaystyle \mathop{\epsilon}^{6}}{}}

\def\nab6{{\displaystyle \mathop{\nabla}^{6}}{}}

\newcommand{\bean}{\begin{eqnarray*}}
\newcommand{\eean}{\end{eqnarray*}}

\begin{document}
\thispagestyle{empty} \addtocounter{page}{-1}
   \begin{flushright}
\end{flushright}

\vspace*{1.3cm}
  
\centerline{ \large \bf
  The Grassmannian-like  Coset Model
  and  
  the Higher Spin Currents
}
\vspace*{1.5cm}
\centerline{ {\bf  Changhyun Ahn}
} 
\vspace*{1.0cm} 
\centerline{\it 
 Department of Physics, Kyungpook National University, Taegu
41566, Korea} 
\vspace*{0.5cm}
\vskip2cm

\centerline{\bf Abstract}
\vspace*{0.5cm}

In the Grassmannian-like coset model,
$\frac{SU(N+M)_k}{SU(N)_k \times U(1)_{k N M (N+M)}}$,
Creutzig and Hikida
have found the charged spin-$2,3$ currents and the neutral
spin-$2,3$ currents previously. In this paper,
as an extension of  Gaberdiel-Gopakumar conjecture 
found ten years ago,
we calculate the operator product expansion (OPE)
between the charged
spin-$2$ current
and itself, the OPE between the charged 
spin-$2$ current and  the charged 
spin-$3$ current
and the OPE
 between the neutral
spin-$3$ current
and itself
for generic $N, M$ and $k$.
From the second OPE, we obtain
the new charged quasi primary spin-$4$
current while from  the last one,
the new neutral primary spin-$4$ 
current is found implicitly.
The infinity limit of $k$ in the structure constants
of the OPEs is described in the context of asymptotic symmetry
of $M \times M$ matrix generalization of $AdS_3$ higher spin theory.
Moreover, 
the OPE between the charged
spin-$3$ current
and itself is determined for fixed $(N,M)=(5,4)$
with arbitrary $k$ up to the third order pole.
We also obtain the OPEs between
charged spin-$1,2,3$ currents and neutral spin-$3$ current.
From the last OPE, we realize that there exists
the presence of the above charged quasi primary spin-$4$
current in the second order pole for fixed $(N,M)=(5,4)$.
We comment on the complex free fermion realization.

\vspace*{4cm}
 \begin{flushright}
{\it On the occasion of my sixtieth birthday}
\end{flushright}

\baselineskip=18pt
\newpage
\renewcommand{\theequation}
{\arabic{section}\mbox{.}\arabic{equation}}

\tableofcontents


\section{ Introduction}

The Grassmannian-like coset model is described by \cite{CHR1306}
\bea
\frac{SU(N+M)_k}{SU(N)_k \times U(1)_{k N M (N+M)}}.
\label{coset}
\eea
By introducing the 't Hooft-like coupling constant $\la \equiv
\frac{k}{(k+N)}$ and taking the infinity limit of ${N}$
with fixed $\la$ and $M$, it has been proposed in
\cite{CH1812} that
the above coset model is dual to $M \times M$ matrix generalization
of $AdS_3$ Vasiliev higher spin theory \cite{PV1,PV2}.
For $M=1$, their proposal leads to the Gaberdiel-Gopakumar
conjecture \cite{GG1011} via level-rank duality.
See also \cite{GG1205,GG1207,AGKP} for review of \cite{GG1011}.
The central charge of the coset model with infinity limit of
level $k$ with fixed $\la$ and $M$ coincides with the one
in the asymptotic symmetry of above $AdS_3$ higher spin theory.
The charged spin-$2,3$ currents and the neutral
(higher) spin-$2, 3$ current in terms of the coset realization
characterized by five spin-$1$ currents
have been found explicitly.
At $\la=2$ (or $k=-2N$), the operator product expansion (OPE)
between the charged spin-$2$ current and itself for general
$(N,M)$, by decoupling
the charged spin-$3$ current, leads to
the one of the ``rectangular'' $W$-algebra with $SU(M)$ symmetry
of $AdS_3$ higher spin theory.

In this paper, we will compute the OPE between the charged
spin-$2$ current and itself by hand,
for generic $k$ as well as generic
$N$ and $M$. It turns out that the above
charged spin-$3$ current, for generic $\lambda$,  should appear
in the right hand side of the OPE.
The structure constants appearing in the
right hand side of this OPE in terms of these
three parameters will be determined completely.

\begin{itemize}
  \item[]
At each singular term, we should rearrange the coset
composite operators in terms of the known currents,
i) the stress energy tensor of spin-$2$,
ii) the spin-$1$ current of $SU(M)$,
iii) the charged spin-$2$
current by allowing all the possible nonlinear terms.
\end{itemize}

It is known that after subtracting the descendant terms,
we are left with the sum of quasi primary operators
\cite{Blumenhagenetal,Nahm1,Nahm2}.
We should determine the structure constants appearing
in these quasi primary operators of the right hand side of the OPE.
Because there are free adjoint indices $a$ and $b$ of $SU(M)$
in the left hand side of the OPE,
it is rather nontrivial to exhaust all the possible
quasi primary operators which will be contracted with some
$SU(M)$ invariant
tensors.
For example, in general, the first order pole of this OPE 
can contain the cubic terms in the spin-$1$ current which possesses
a single adjoint index. Then those invariant tensors will contain
fifth order invariant ones maximally.
That is, two of them will be the above free indices
while three of them will be contracted with each index of cubic
terms. This is the reason why the OPE between the nonsinglet
charged operators even their spins are low is more complicated
to analyze,
compared to the OPE between the singlet operators.
Note that in the examples of \cite{AK1509,AKK1703,AGK2004},
there exist some OPEs having
nonsinglet indices associated with the $SO(4)$ but for these cases
it is not so difficult to figure out its structures in the right hand
sides of the OPEs because we can determine
the vector and adjoint indices and  the invariant tensors in $SO(4)$
for fixed rank. 

Furthermore,
we will obtain the OPE
between the charged spin-$2$ current and
the charged spin-$3$ current which occurs at the first order pole
of the previous
OPE between the charged spin-$2$ current and itself.
Now we
should include both the charged spin-$3$
current and the neutral spin-$3$ current
as the candidates for the quasi primary operators
in the list of known currents we described in
previous paragraph.
The presence of the neutral spin-$3$ current
is due to the fact that the left hand side of this OPE
has two different operators, contrary to the previous OPE
between the charged spin-$2$ current and itself.
The point is how we can write down the singular terms
described by the coset realization
in terms of the known currents.
We expect that up to the second order pole
of this OPE, we should express them
by using the known currents with various $SU(M)$ invariant tensors.

\begin{itemize}
  \item[]
By analyzing the first order pole of this OPE,
we will determine the new quasi primary charged spin-$4$
current in terms of coset realization.
By construction, all the relative coefficients appearing in the
coset composite operators are determined automatically although
the careful analysis
should be performed.
\end{itemize}


From the explicit result for the OPE
between the neutral spin-$3$ current and itself
for fixed $(N,M)$ values, we will extract this OPE
for generic $(N,M)$ case and at the second order pole of
this OPE we will observe that there should be new
primary neutral spin-$4$ current in terms of coset realization.

\begin{itemize}
  \item[]
In obtaining this result, we realize that the $k$-dependent
structure constant can be rewritten as the modified central charge
which is equal to the coset central charge subtracted by
the central term due to the stress energy tensor for the
quadratic Sugawara term in the spin-$1$
current of $SU(M)$.
\end{itemize}

Then i) the modified stress energy tensor of spin-$2$,
ii) the neutral spin-$3$
current and iii) the neutral spin-$4$ current will consist of the
generators of
the standard $W$ algebra and their OPEs with the spin-$1$ current
do not have any singular terms. That is, the spin-$1$ current is
decoupled in the OPEs between these singlet currents.

\begin{itemize}

\item[]
  In section $2$, we review the results of \cite{CH1812}
by emphasizing that the spin-$2$ current and the spin-$3$ currents
can be obtained by hands without trying to perform for several $(N,M)$
values. Those currents were determined previously. The derivations
for obtaining these are new.
%
 In section $3$, the simplest nontrivial OPE between the charged
 spin-$2$ current and itself can be obtained.
The structure constants are new.
 We will
observe the charged spin-$3$ current at the first order pole.
%
  In section $4$,
the next nontrivial  OPE between the charged
spin-$2$ current and the charged spin-$3$ current can be obtained.
The new charged quasi primary
spin-$4$ current at the first order pole is determined.

  \item[]
In section $5$, the
new OPE between the charged
 spin-$3$ current and itself can be determined for
 specific $N$ and $M$ values
\footnote{The integer $M=4$ is the lowest value
  in order to have an independent $SU(M)$ invariant tensors \cite{CH1812}.
  We take $N$ which is different from $M$ as five.}.
%
In section $6$,
the
 new OPE between the uncharged
 spin-$3$ current and itself can be determined and
 the new uncharged spin-$4$ current appears
 at the second order pole.
%
 In section $7$, the new OPEs between the charged spin-$1,2,3$ currents
 and the uncharged spin-$3$ current are described.
 %
 In section $8$, we present the future directions with a summary
 of this paper. 
%
 In Appendices, we will describe some detailed calculations
 based on the previous sections.
  The free field realization of \cite{OS} is reviewed and
 we explain how their results can be related to the previous
 results by taking the appropriate limits for the parameters we are
 considering.
\end{itemize}
 
 The Thielemans package \cite{Thielemans}
 is used together with the mathematica package \cite{mathematica}.
 The similar coset in the work of
 \cite{EP} where the possibility of
 four parameters in the specific coset is described
 is studied \footnote{There is a similar construction,
   a matrix extended $W_{1+\infty}$
   algebra \cite{EP1}, defined in terms of matrix extended Miura
   transformation (See also \cite{AM} for some mathematics for
   the ``rectangular'' $W$-algebra). The truncation of
 this  matrix extended $W_{1+\infty}$
 algebra can be realized the one in (\ref{coset}) without
 $U(1)$ factor in the denominator.
 The three parameters of the algebra are given by
 $N, M$ and $k$ in the subsection $3.5$ of \cite{EP1}.
  We thank Lorenz
 Eberhardt for
 pointing this out.}.
 
 The charged spin-$1,2,3,4$ currents and uncharged
 spin-$2,3,4$ currents we are considering in this paper are given by
 \bea
 && \mbox{spin-1}: J^a(z), \qquad \mbox{spin-2}: K^a(z), \qquad
 \mbox{spin-3}: P^a(z), \qquad
 \mbox{spin-4}: \hat{R}^a(z),
 \nonu \\
 && \mbox{spin-2}: T(z), \qquad
 \mbox{spin-3}: W^{(3)}(z), \qquad
 \mbox{spin-4}: W^{(4)}(z).
 \label{fields}
 \eea
 Here $T(z)$ is the stress energy tensor
 and the index $a$ in (\ref{fields})
 is an adjoint index of $SU(M)$ and $a =1,2, \cdots,
 (M^2-1)$. Except of $T(z)$ and $\hat{R}^a(z)$ which are quasi
 primary currents, the remaining currents are
 primary ones under the stress energy tensor.
 In the context of \cite{GG1011}, the OPEs between the
 neutral higher spin currents are relevant to this conjecture
 and
 the algebra between them is closed under the neutral higher spin
 currents. In addition to that, there are
 also the OPEs between the charged higher spin currents and
 the neutral ones and the OPEs between the charged higher
 spin currents. The right hand sides of these OPEs
 will contain the composite charged or neutral higher spin
 operators.

 \begin{itemize}
   \item[]
     The main work of this paper is to start with the charged and
     neutral
 higher
 spin currents \cite{CH1812} and construct their algebra
 explicitly as an extension
 of \cite{GG1011} in the above coset model (\ref{coset}).
 \end{itemize}

 What we have found newly in this paper
 is the higher spin spin-$4$ currents
 in (\ref{fields}). The remaining ones were found
 in \cite{CH1812} previously.
 
\section{ Review with some new derivations}

The normalization of the generators $(t^{\al}, t^a, t^{u(1)},
t^{(\rho \bar{i})}, t^{(\bar{\si} j)})$ in $SU(N+M)$ of the coset (\ref{coset})
can be fixed by taking the following simple metric 
\cite{CH1812}
\bea
\mbox{Tr} (t^{\al} t^{\beta}) =\de^{\al \beta},
\qquad
\mbox{Tr} (t^{a} t^{b}) =\de^{a b},
\qquad
\mbox{Tr} (t^{u(1)} t^{u(1)}) =1,
\qquad
\mbox{Tr} (t^{(\rho \bar{i})} t^{(\bar{\si} j)}) =\de^{\rho \bar{\si}} \, \de^{j \bar{i}}.
\label{metric}
\eea
Under the decomposition of $SU(N+M)$ into the
$SU(N) \times SU(M)$,
the adjoint representation of $SU(N+M)$ breaks into
\bea
({\bf N+M})^2-{\bf 1} \longrightarrow ({\bf N}^2-{\bf 1},{\bf 1})
\oplus ({\bf 1},{\bf M}^2-{\bf 1}) \oplus ({\bf 1}, {\bf 1})
\oplus ({\bf N},
\overline{{\bf M}}) \oplus (\overline{{\bf N}},{\bf M}).
\label{branching}
\eea
The fundamental indices $\rho$ and $j$ among (\ref{branching})
run over $\rho =1,2,\cdots, N$ and $j =1,2,\cdots, M$,
while  the antifundamental indices $\bar{\si}$ and $\bar{i}$
run over $\bar{\si} =1,2,\cdots, N$ and $\bar{i} =1,2,\cdots, M$. 
Note that the barred index in (\ref{metric})
becomes the unbarred one  when we raise or lower
it 
and vice versa.
For the $\al, a$ and $u(1)$ indices where
the adjoint indices are given by $\al =1,2, \cdots, (N^2-1)$
and $a=1,2,\cdots, (M^2-1)$ respectively, we can raise or lower them without
any change \footnote{Sometimes we use the $SU(M)$ indices $a,b,c, \cdots$
as superscripts.}. We will use the metric in (\ref{metric}) all the time.

For the above given generators, the
totally antisymmetric $f$ and totally symmetric $d$
symbols can be expressed as follows:
\bea
\mbox{Tr} ([t^{\al}, t^{\beta}] t^{\ga})  & = & i f^{\al \beta \ga},
\qquad
\mbox{Tr} ([t^{a}, t^{b}] t^{c}) = i f^{a b c},
\qquad
\mbox{Tr} (\{t^{\al}, t^{\beta}\} t^{\ga}) =  d^{\al \beta \ga},
\nonu \\
\mbox{Tr} (\{t^{a}, t^{b}\} t^{c}) & = & d^{a b c},
\qquad
\cdots,
\label{fdtrace}
\eea
where the abbreviated parts
can be written similarly.
We use the following nontrivial $f$ symbols \cite{CH1812} which are
totally antisymmetric
\bea
i f^{(\rho \bar{i})(\bar{\si} j) u(1)} & = & \sqrt{\frac{M+N}{M N}} \, \de^{j \bar{i}}\,
\de^{\rho \bar{\si}}, \qquad
i f^{(\rho \bar{i})(\bar{\si} j) \al} = \de^{\rho \bar{\rho_1}}\, \de^{\si_1 \bar{\si}} \,
\de^{j \bar{i}} \, t^{\al}_{\si_1 \bar{\rho_1}}, \nonu \\
i f^{(\rho \bar{i})(\bar{\si} j) a} & = & -\de^{\rho \bar{\si}}\, \de^{i_1 \bar{i}} \,
\de^{j \bar{j_1}} \, t^{a}_{i_1 \bar{j_1}}.
\label{fdexp}
\eea
Due to the traceless property of the generators,
when the indices $\rho$ and $\bar{\si}$ are equal to each other
in the second relation of (\ref{fdexp}), the corresponding
$f$ symbols are zero.
Similarly, for the equal $\bar{i}$ and $j$ in the third relation,
the $f$ symbols vanish.

The nontrivial
$SU(N+M)$ currents satisfy the following OPEs \cite{CH1812}
\bea
J^{\al}(z) \, J^{\beta}(w) & = & \frac{1}{(z-w)^2} \, k \, \de^{\al \beta}+
\frac{1}{(z-w)} \, i \, f^{\al\beta}_{\,\,\,\,\,\,\,\,\ga} \, J^{\ga}(w) + \cdots,
\nonu \\
J^{\al}(z) \, J^{(\rho \bar{i})}(w) & = & 
\frac{1}{(z-w)} \, i \,
f^{\al (\rho \bar{i})}_{\,\,\,\,\,\,\,\,\,\,\,\,(\si \bar{j})} \,
J^{(\si \bar{j})}(w) + \cdots,
\nonu \\
J^{\al}(z) \, J^{(\bar{\rho} i)}(w) & = & 
\frac{1}{(z-w)} \, i \,
f^{\al (\bar{\rho} i)}_{\,\,\,\,\,\,\,\,\,\,\,\,(\bar{\si} j)} \,
J^{(\bar{\si} j)}(w) + \cdots,
\nonu \\
J^{a}(z) \, J^{b}(w) & = & \frac{1}{(z-w)^2} \, k \, \de^{a b}+
\frac{1}{(z-w)} \, i \, f^{a b}_{\,\,\,\,\,\,\,\,c} \, J^{c}(w) + \cdots,
\nonu \\
J^{a}(z) \, J^{(\rho \bar{i})}(w) & = & 
\frac{1}{(z-w)} \, i \,
f^{a (\rho \bar{i})}_{\,\,\,\,\,\,\,\,\,\,\,\,(\si \bar{j})} \,
J^{(\si \bar{j})}(w) + \cdots,
\nonu \\
J^{a}(z) \, J^{(\bar{\rho} i)}(w) & = & 
\frac{1}{(z-w)} \, i \,
f^{a (\bar{\rho} i)}_{\,\,\,\,\,\,\,\,\,\,\,\,(\bar{\si} j)} \,
J^{(\bar{\si} j)}(w) + \cdots,
\nonu \\
J^{u(1)}(z) \, J^{u(1)}(w) & = & \frac{1}{(z-w)^2} \, k + \cdots,
\nonu \\
J^{u(1)}(z) \, J^{(\rho \bar{i})}(w) & = & 
\frac{1}{(z-w)} \, i \,
f^{u(1) (\rho \bar{i})}_{\,\,\,\,\,\,\,\,\,\,\,\,\,\,\,\,\,\,\,\,(\si \bar{j})} \,
J^{(\si \bar{j})}(w) + \cdots,
\nonu \\
J^{u(1)}(z) \, J^{(\bar{\rho} i)}(w) & = & 
\frac{1}{(z-w)} \, i \,
f^{u(1) (\bar{\rho} i)}_{\,\,\,\,\,\,\,\,\,\,\,\,\,\,\,\,\,\,\,\,(\bar{\si} j)} \,
J^{(\bar{\si} j)}(w) + \cdots,
\nonu \\
J^{(\rho \bar{i})}(z) \, J^{(\bar{\si} j)}(w) & = & \frac{1}{(z-w)^2} \, k \,
\de^{\rho \bar{\si}} \, \de^{j \bar{i}}
\label{OPEspin1spin1}
 \\
& + & 
\frac{1}{(z-w)} \, \Bigg[
  i \,
  f^{(\rho  \bar{i}) (\bar{\si} j)}_{\,\,\,\,\,\,\,\,\,\,\,\,\,\,\,\,\,\,\,\,\,\,u(1)}
  \, J^{u(1)}+
  i \,
  f^{(\rho  \bar{i}) (\bar{\si} j)}_{\,\,\,\,\,\,\,\,\,\,\,\,\,\,\,\,\,\,\,\,\,\,\al}
  \, J^{\al}+
  i \,
  f^{(\rho  \bar{i}) (\bar{\si} j)}_{\,\,\,\,\,\,\,\,\,\,\,\,\,\,\,\,\,\,\,\,\,\,a} \, J^{a}
  \Bigg](w) + \cdots.
\nonu
  \eea
  The second order pole in (\ref{OPEspin1spin1}) has
  the explicit $k$ dependence with weight $1$.
  From the nonzero $f$ symbols in (\ref{fdexp}),
  the spin-$1$ currents transforming as
  $({\bf N},\overline{{\bf M}})$ or $(\overline{{\bf N}},{\bf M})$
  appear in many places of (\ref{OPEspin1spin1}).
Due to the last OPE in (\ref{OPEspin1spin1}),
the contraction between the spin-$1$ current and its conjugated one
in the OPEs later
will provide the remaining three kinds of spin-$1$ currents in the right hand
side.

Note that there are also the five regular OPEs besides the
above ten OPEs
  \bea
  J^{\al}(z) \, J^a(w) & = & 0 + \cdots, \qquad
  J^{\al}(z) \, J^{u(1)}(w) =0 + \cdots, \qquad
   J^{a}(z) \, J^{u(1)}(w) =0 + \cdots,
   \nonu \\
    J^{(\rho \bar{i})}(z) \, J^{(\si \bar{j})}(w) & = & 0 +\cdots, \qquad
    J^{(\bar{\rho} i)}(z) \, J^{(\bar{\si} j)}(w) =0 + \cdots.
    \label{regular}
  \eea
  These come from the trivial results from both
  metric (\ref{metric}) and $f$ symbols in
  (\ref{fdtrace}) and (\ref{fdexp}).
  In particular, the first and the third relations in (\ref{regular})
  can be generalized to
  the spin-$2,3,4$ currents with the adjoint index $a$ according to the coset
  (\ref{coset}) we are considering.
  
  We can express the stress energy tensor \cite{CH1812}, by Sugawara
  construction,
  \bea
  T(z)  & =  &\frac{1}{2(k+N+M)} \Bigg[ J^{\al}  J^{\al} + J^a  J^a +
    \de_{\rho \bar{\si}} \de_{j \bar{i}} \,
    J^{(\rho \bar{i})}  J^{(\bar{\si} j)} + \de_{\rho \bar{\si}} \de_{j \bar{i}} 
   J^{(\bar{\si} j)}  J^{(\rho \bar{i})} + J^{u(1)}  J^{u(1)} 
    \Bigg](z)
  \nonu \\
  &-& \frac{1}{2(k+N)} \,  J^{\al} \, J^{\al}(z) - \frac{1}{2k} \,
  J^{u(1)} \, J^{u(1)}(z).
  \label{T}
  \eea
  The first five terms of (\ref{T}) come from the $SU(N+M)$ of the coset
  (\ref{coset}) while the remaining ones come from the $SU(N) \times U(1)$
  of the coset.
  Note that we can move the $J^{(\rho \bar{i})}$ in the fourth term of
  (\ref{T})
  to the left
  and combine it with the third term together with a derivative term
  according to  the relation
  $\de_{\rho \bar{\si}}\, \de_{j \bar{i}} \,
  [ J^{(\bar{\si} j)},   J^{(\rho \bar{i})}]= - M N \sqrt{\frac{M+N}{M N}} \,
  \pa \, J^{u(1)}$ which will be used several times in this paper.
Then we have the following OPE  
  \bea
  T(z) \, T(w) = \frac{1}{(z-w)^4} \, \frac{c}{2} + \frac{1}{(z-w)^2} \,
  2 \, T(w) + \frac{1}{(z-w)} \, \pa \, T(w) + \cdots.
  \label{TT}
  \eea
  It is rather nontrivial to check this OPE (\ref{TT}) explicitly
  by using the (\ref{OPEspin1spin1}).
Here the central charge in (\ref{TT}) is given by  \cite{CH1812}
\bea
c  & = &  \frac{k((N+M)^2-1)}{(k+M+N)}-\frac{k(N^2-1)}{(k+N)}-1
\nonu \\
&= &
\frac{(-k^2+k^2 M^2-2k N-M N+2k^2 M N + k M^2 N-N^2 + k M N^2)}{(k+N)(k+M+N)}.
\label{charge}
\eea
Furthermore, the spin-$1$ current is primary operator
under the stress energy tensor (\ref{T})
\bea
T(z) \, J^a(w) =\frac{1}{(z-w)^2} \, J^a(w) + \frac{1}{(z-w)} \, \pa \,
J^a(w) + \cdots.
\label{OPETJ}
\eea
Note that
$T(z)$ is a singlet under the horizontal subalgebra
$SU(M)$ \cite{BS}.
The OPEs between $T(z)$ and $J^{\alpha}(w)$ (and $J^{u(1)}(w)$)
are regular. When we further divide
the $SU(M)$ piece in the coset (\ref{coset}) and subtract
the corresponding stress energy tensor, $\frac{1}{2(k+M)}\, J^a\, J^a(w)$,
from (\ref{T}),
then this modified stress energy tensor is no longer singular OPE
with spin-$1$ current $J^a(w)$.


\subsection{A charged spin $2$ current}

The next question is whether
the spin-$2$ current transforming as adjoint
representation of $SU(M)$ exists or not.
If there exists, then how do we construct explicitly?
It is natural to require that
it should transform as a primary operator under the stress energy tensor
(\ref{T}). The nontrivial requirement is the relation between the
previous spin-$1$ current and this spin-$2$ current.
In general, the second order pole of this OPE
contains the spin-$1$ current with two free adjoint indices
while the first order pole contains the composite spin-$2$ operators
contracted with the appropriate indices.
In the specific basis, the spin-$2$ current can transform as
the ``primary'' operator under the spin-$1$ current \cite{BCG}.
Furthermore, the spin-$2$ current should transform
under the adjoint representation of the horizontal finite dimensional
Lie algebra $SU(M)$ \cite{BS}.

Among five spin-$1$ currents, we can make the quadratic terms between
them with derivative terms in order to have spin-$2$ operator.
The nontrivial term is given by
the $SU(M)$ generator multiplied by
the spin-$1$ current
 transforming
 as   $({\bf N},\overline{{\bf M}})$
 and its conjugated one.
 Moreover, the fundamental and antifundamental indices
 of $SU(N)$ should be contracted each other.
 We expect that there should be the spin-$2$ operator
 contracted by $d$ symbol \cite{BBSS,Ahn1111} from the adjoint
 spin-$1$ current $J^a(z)$.
It turns out that
a charged spin-$2$ current
\cite{CH1812}
is given by \footnote{\label{derinK} We have the
relation $\de_{\rho \bar{\si}} \,
t^a_{j\bar{i}} \,
J^{(\bar{\si} j)} \,  J^{(\rho \bar{i})}(z) =  \de_{\rho \bar{\si}} \,
t^a_{j\bar{i}} \, J^{(\rho \bar{i})} \,  J^{(\bar{\si} j)}(z) + N \, \pa J^a(z)$
from the last OPE of (\ref{OPEspin1spin1}) with the help of
\cite{BBSS}.}
\bea
K^a(z) & = & \de_{\rho \bar{\si}} \,
t^a_{j\bar{i}} \, (J^{(\rho \bar{i})} \,  J^{(\bar{\si} j)} +
J^{(\bar{\si} j)} \,  J^{(\rho \bar{i})})  (z)
 -\frac{N}{(M+2k)} \, d^{abc} \, J^b\, J^c(z)
 \nonu \\
 & + & \frac{2N}{k} \sqrt{\frac{M+N}{M N}} \, J^a  \,  J^{u(1)}(z).
 \label{spin2expression}
 \eea
 Note that the third term of (\ref{spin2expression})
occurs in \cite{BBSS,Ahn1111}. 
 Instead of introducing the
 arbitrary coefficients, we will check whether the above
 result is consistent with other conditions.

 Now we can compute the OPE between $J^{u(1)}(z)$ and $K^a(w)$
 and it turns out that the second order pole of this OPE coming from
 the first two and last terms of (\ref{spin2expression}) has
 $J^a(w)$ term whose coefficient vanishes, similar to the third one of
 (\ref{regular}).
 Moreover, the OPE between $J^{\al}(z)$ and
 $K^a(w)$ can be obtained from the first two terms of (\ref{spin2expression})
 and this leads to the vanishing of this OPE, along the line of
 the first relation of (\ref{regular}),
 where the traceless conditions
 for the generators $t^{\al}_{\rho \bar{\si}}$ and
 $t^a_{i \bar{j}}$ are used.
 From the OPE between $J^{a}(z)$ and $K^b(w)$, the second order pole vanishes
 by using the identity that the triple product $d f f$ is proportional to
 $d$ symbol \cite{BBSS,NPB97,Ahn1111}.
 We also consider $\pa \, J^a(z)$ term in (\ref{spin2expression})
   but the vanishing of third order pole of
   the OPE between $J^{a}(z)$ and $K^b(w)$ does not allow us to
   add this term.
 Finally, the first order pole of this OPE
 can be written in terms of $i f^{ab}_{\,\,\,\,\,\,c} \,
 K^c(w)$.

 Therefore, we summarize that the charged spin-$2$ current
 has the following OPE
 \bea
 J^a(z) \, K^b(w) = \frac{1}{(z-w)} \, i \, f^{ab}_{\,\,\,\,\,\,c} \, K^c(w) +
 \cdots.
 \label{JK}
 \eea
 We can compute the commutator
 $[J_0^a, K^b(w)]$ and this leads to
 $i \, f^{a b c} \, K^c(w)$ from the result of (\ref{JK}).
 In other words, the spin-$2$ current transforms under the
 adjoint representation of the horizontal finite dimensional
 Lie algebra $SU(M)$ as mentioned before.
 Here $J_0^a$ is the Laurent zero mode of spin-$1$ current $J^a(z)$
 \cite{BS}.
 Because the complete expression of this charged
 spin-$2$ current is given by (\ref{spin2expression}), we can calculate
 the OPE with the stress energy tensor (\ref{T})
 and it is given by 
 \bea
 T(z) \, K^a(w) = \frac{1}{(z-w)^2} \, 2 \, K^a(w) + \frac{1}{(z-w)} \, \pa \,
 K^a(w) + \cdots,
\label{TK}
 \eea
 where the relation (\ref{OPETJ})
 and other ones  are used.
So far,
the currents are given by
the stress energy tensor (\ref{T}),
the spin-$1$ current and the spin-$2$ current (\ref{spin2expression}).
Their OPEs are given by (\ref{TT}), (\ref{OPETJ}), (\ref{TK}),
the fourth relation of (\ref{OPEspin1spin1}), and (\ref{JK}).


\subsection{A charged spin $3$ current}

We would like to construct the charged spin-$3$ current as we did in
previous subsection.
This charged spin-$3$ current should be a primary operator
under the stress energy tensor (\ref{T}).
We expect that the cubic term of $SU(M)$ adjoint
spin-$1$ current with the fourth order $d$ symbols
\cite{Schoutens,Ahn1111}
as a nonderivative term can arise.
For the OPE with the spin-$1$ current,
we require the previous ``primary'' condition under the spin-$1$ current.

It turns out that
the charged spin-$3$ current which was obtained by using
the works of \cite{GKO1,GKO2,BBSS1} has the following terms
\footnote{The coefficients $a_6$,$a_{10},a_{14}$ and $a_{15}$
  are vanishing where the corresponding terms are given by
  $a_6 \, i \, f^{a b c} \, \de_{\rho \bar{\rho}} \,
  t^b_{j\bar{i}} \, J^c   (J^{(\rho \bar{i})}   J^{(\bar{\rho} j)}+
  +
J^{(\bar{\rho} j)}   J^{(\rho \bar{i})})  (z)+
  a_{10}  \, d^{a b c} \, \pa \,   J^b  \,  J^c(z)+
  a_{14} \, J^a \, \pa \, J^{u(1)}(z) + a_{15} \,
  \pa \, J^a  \, J^{u(1)}(z) $. In order to check (\ref{spin3exp})
  we keep these terms also.
\label{zerofour}}
 \bea
 P^a(z) & = & a_1 \,
 t^{\al}_{\rho \bar{\si}} \, t^a_{j\bar{i}} \, J^{\al} 
 J^{(\rho \bar{i})}  \, J^{(\bar{\si} j)}(z)
 + a_2 \, J^{\al}  \, J^{\al}  \, J^a(z)+a_3 \, J^b  \, J^b \, J^a(z)
 +a_4 \, J^a  \, J^{u(1)}  \, J^{u(1)}(z)
 \nonu \\
 &+& a_5 \, d^{abc} \, \de_{\rho \bar{\rho}} \,
t^b_{j\bar{i}} \, J^c   (J^{(\rho \bar{i})}   J^{(\bar{\rho} j)} +
J^{(\bar{\rho} j)}   J^{(\rho \bar{i})})  (z)
+ a_7 \, \de_{\rho \bar{\si}} \,
t^a_{j\bar{i}}\, J^{u(1)}   (J^{(\rho \bar{i})}  J^{(\bar{\si} j)} +
J^{(\bar{\si} j)}   J^{(\rho \bar{i})})  (z)
\nonu \\
&+& a_8 \, \de_{\rho \bar{\si}} \, \de_{j\bar{i}}\,
J^a   \, (J^{(\rho \bar{i})} \, J^{(\bar{\si} j)} +
J^{(\bar{\si} j)}  \,  J^{(\rho \bar{i})})  (z)
+ a_9 \,  d^{abc} \, J^b  \, J^c  \,  J^{u(1)}(z)
+ a_{11} \, i \, f^{abc} \, \pa \,   J^b  \,  J^c(z)
\nonu \\
&+& a_{12} \, \de_{\rho \bar{\si}} \,
t^a_{j\bar{i}} \, \pa \, J^{(\rho \bar{i})} \, J^{(\bar{\si} j)}(z)
+ a_{13} \, \de_{\rho \bar{\si}} \,
t^a_{j\bar{i}}  \, \pa  \, J^{(\bar{\si} j)} \,  J^{(\rho \bar{i})}(z)
+a_{16} \, \pa^2  \, J^a(z)
\nonu \\
&+& a_{17} \, 6 \, \mbox{Tr} \, (t^{a} \,
t^{\left(b \right.} \, t^c \, t^{\left. d \right)}) \, 
J^b  \, J^c  \, J^d(z).
\label{spin3exp}
 \eea
 The $a_3, a_{11}, a_{16}$ and $a_{17}$ terms
 contain the spin-$1$ current only
 \footnote{The $a_{17}$ term can be written as
   $\frac{6}{M} \, J^a \, J^b \, J^b +\frac{3}{2} \, (i f +d)
   ^{a b e} \, d^{e c d}\, J^b \, J^c \, J^d+
   \frac{3}{2}\, (i f +d)^{a d e}\, d^{e b c} \, i\, f^{b d f}\,
   \pa \, J^f \, J^c -\frac{1}{4} \, (i f + d)^{a d e} \, f^{c d f}
   \, f^{b f g} \, \pa^2 \, J^g+
   \frac{6}{M} \, i\, f^{b a c}  \, \pa \, J^c \, J^b-
   \frac{1}{M}\, f^{b a c} \, f^{b c d}\, \pa^2 \, J^d$.}.
 The $a_{17}$ term is related to the above cubic term with
 the fourth order $d$ symbols mentioned before.
 We understand the $a_2, a_{4}$ and $a_9$ terms
 because the indices except the free adjoint index $a$
 are contracted properly. The nontrivial parts are given by the
 remaining six terms.
 The free index $a$ arises in the generator $t^a_{j \bar{i}}$,
 the spin-$1$ current $J^a$ and the $d^{a b c} $ symbols.
 They contain the spin-$1$ currents
transforming
as   $({\bf N},\overline{{\bf M}})$ or $(\overline{{\bf N}},{\bf M})$.
For $a_5, a_7, a_{12}$ and $a_{13}$ terms, the Kronecker
delta symbols are multiplied in order to contract with the
fundamental and antifundamental indices of $SU(N)$ each other.
For $a_8$ term, there exists further Kronecker
delta symbols associated with the fundamental and antifundamental
indices of $SU(M)$. Note that in the $a_1$ term, there is an additional
generator $t^{\alpha}_{\rho \bar{\si}}$ contracted with three other
indices. 
We do not get the $a_1$  term from the each term of
charged spin-$2$ current and other operators.
Therefore, the $a_1$ term is crucial for the construction of
an independent charged spin-$3$ current.

Now we would like to construct the spin-$3$ current step by step
explicitly by assuming the operator contents of \cite{CH1812}.
We calculate the OPEs by hand without using the method given in
\cite{CH1812} where they have obtained this charged spin-$3$ current
for several fixed low $(N,M)$ values and extracted
the $(N,M)$ dependence of relative coefficients
as well as $k$ dependence.
 By requiring that we should have
 the condition
 $J^{\al}(z) \, P^a(w)=0$, along the line of
 the first relation of (\ref{regular}),
 where the corresponding terms in
 (\ref{spin3exp}) are given by $a_1,a_2,a_8,a_{12}$ and $a_{13}$ terms,
 the second order pole provides the following equations
 \bea
&& \Bigg[ (2N+k) \, a_1 +(a_{12}-a_{13}) \Bigg] \, t^{\al}_{\rho \bar{\si}}
 \, t^a_{j \bar{i}} \, J^{(\rho \bar{i})} \, J^{(\bar{\si} j)}(w)  =  0,
 \nonu \\
&& \Bigg[ a_1 -2(N+k)\, a_2-2 M \,a_8 \Bigg] \, J^{\al} \, J^a(w)  =  0.
 \label{tworel}
 \eea
 Here the relation $[t^{\al}, t^{\beta}] = i
 f^{\al \beta}_{\,\,\,\,\,\,\ga} \, t^{\ga}$ is used  in (\ref{tworel}).
 Moreover, from (\ref{OPEspin1spin1}),
 we have the following identity (See also \cite{BBSS})
 \bea
[J^{(\rho \bar{i})}, \, J^{(\bar{\si} j)}]=
 i \,
  f^{(\rho  \bar{i}) (\bar{\si} j)}_{\,\,\,\,\,\,\,\,\,\,\,\,\,\,\,\,\,\,\,\,\,\,u(1)}
  \, \pa \, J^{u(1)}+
  i \,
  f^{(\rho  \bar{i}) (\bar{\si} j)}_{\,\,\,\,\,\,\,\,\,\,\,\,\,\,\,\,\,\,\,\,\,\,\al}
  \, \pa \, J^{\al}+
  i \,
  f^{(\rho  \bar{i}) (\bar{\si} j)}_{\,\,\,\,\,\,\,\,\,\,\,\,\,\,\,\,\,\,\,\,\,\,a} \,
  \pa \, J^{a}.
 \label{comm}
 \eea
 Then the contribution from the second term of $a_8$ in (\ref{spin3exp})
 is the same as the one from the first term because the additional two
 delta symbols in $a_8$ term can act on (\ref{comm}) which leads to
 zero value.
 
 Similarly, the regularity condition
 $J^{u(1)}(z) \, P^a(w)=0$, similar to the third relation of
 (\ref{regular}),
 gives the following
 equations we should have
 \bea
&& \Bigg[-N \, \sqrt{\frac{M+N}{M N}}  \, a_{12} + 2 \, k \,
   a_{14} - N \, \sqrt{\frac{M+N}{M N}} 
 \, a_{13} \Bigg] \, J^a(w) =0,
 \nonu \\
&& \Bigg[ 2 k \, a_4 -2  N \, \sqrt{\frac{M+N}{M N}} \, a_7 + 2(M+N) \,a_8 \Bigg]
\, J^a \, J^{u(1)}(w)
 =0,
 \nonu \\
&& \Bigg[ -2  N \, \sqrt{\frac{M+N}{M N}} \,a_5 + k \, a_9  \Bigg] \, d^{abc}
 J^b \, J^c(w) =0,
 \nonu \\
&& \Bigg[ 2k \, a_7 +   \sqrt{\frac{M+N}{M N}} \,
   a_{12} -  \sqrt{\frac{M+N}{M N}}
   \, a_{13} \Bigg] \, \de_{\rho \bar{\si}} \, t^a_{j \bar{i}} \,
 J^{(\rho \bar{i})} \, J^{(\bar{\si} j)}(w)
 =0,
 \nonu \\
&& \Bigg[ - 2  M N \, \sqrt{\frac{M+N}{M N}} \, a_6 + k \, N \, a_7 -
   N \, \sqrt{\frac{M+N}{M N}} \, a_{13} + k \,
   a_{15} \Bigg] \, \pa \, J^a(w) =0.
 \label{Ju1P}
 \eea
 Each term of the last four terms can be seen from the
 charged spin-$2$ current in (\ref{spin2expression}).
 In the last relation of (\ref{Ju1P}), the identity
 $f^{a b}_{\,\,\,\,\,\,c} \, J^b \, J^c(w)= i M \, \pa \, J^a(w)$
 \cite{BBSS,Ahn1111}
 is used.
 In the computation of $a_7$ term,
there exists  the relation
 $\delta_{\rho \bar{\si}} \, t^a_{j \bar{i}} \, J^{(\bar{\si} j)} \, J^{(\rho \bar{i})}(w)=
 \delta_{\rho \bar{\si}} \, t^a_{j \bar{i}} \,  J^{(\rho \bar{i})}\, J^{(\bar{\si} j)}(w)+
 N \, \pa \, J^a(w)$ which can be obtained from the
 relation (\ref{comm}).

Let us consider the OPE between $J^a(z)$ and $P^b(w)$.
The fourth and third order poles of this OPE  
give us
\bea
&& \Bigg[ 2 M k \, a_3 - 2 M \, k \, a_{11} -k \, N \,
  a_{12} + k \, N \, a_{13} + 6k \, a_{16}
    \Bigg] \, \de^{ab}  =  0,
  \nonu \\
&&  \Bigg[ 2(M-k) \, a_3 + (M^2-4) \, \frac{N}{M} \,
    a_5 + 2N \, a_8 -(2k+M) \, a_{11}
    -\frac{N}{2} \, a_{12}  \nonu \\
&&    +\frac{N}{2} \, a_{13} + 2 a_{16} +
    (M^2+6) \, a_{17} \Bigg] \, i \, f^{a b}_{\,\,\,\,\,\,c} \, J^c(w)  =  0,
  \nonu \\
&&  \Bigg[4 N M \,
    \sqrt{\frac{M+N}{M N}} \, a_6 -N \, \sqrt{\frac{M+N}{M N}} \,
    a_{12} -
    N \, \sqrt{\frac{M+N}{M N}} \, a_{13} + 2 k \,
    a_{15} \Bigg] \, \de^{ab} \, J^{u(1)}(w)= 0,
  \nonu \\
&&  \Bigg[-N M \, a_6 +(2k+M) \, a_{10} +\frac{N}{2} \, a_{12} +
    \frac{N}{2} \, a_{13}
    \Bigg]  \, d^{a b c} \, J^c(w)  =  0.
  \label{JP43}
  \eea
  In the calculation of the second relation of (\ref{JP43}),
  we use the following relation 
  \bea
  \mbox{Tr} (t^i\, t^b\, t^c\, t^d) = \frac{1}{M} \, \de^{ib}\, \de^{cd}
  +\frac{1}{4} \, ( i f + d)^{i b e}\, (i f +d )^{e c d}.
  \label{fourt}
  \eea
  This can be obtained
  by recalling the fact that
  the product of two generators
  can be written in terms of Kronecker delta symbol
  with identity matrix, $f$ and
  $d$ symbols with generator and we can multiply further generators
  successively.
  By multiplying three $f$ symbols into (\ref{fourt}),
  we obtain the intermediate result
  \bea
  &&
  \mbox{Tr} (t^i\, t^b\, t^c\, t^d) \, f^{abf} \, f^{f c g} \, f^{gdh}
  = \nonu \\
  &&
  -2 f^{a i h} + \frac{1}{4}  \Bigg[M^2  f^{i a h} - i  M^2  d^{h a i}-
  i  M^2  d^{i a h} -(M^2-4)  f^{i h a} \Bigg],
  \label{fourt1}
  \eea
  where the identities for the triple products $ f f f$ and $d f f$
  \cite{BBSS,Ahn1111}
  are used in (\ref{fourt1}).
  Then the remaining five similar terms can be obtained and by adding these
  we arrive at the final contribution $(M^2+6) \,a_{17}
  \, i f^{a b c} \, J^c(w)$ in (\ref{JP43}).

  Let us describe the second order pole which will be more
  complicated.
  We have the following result
  \bea
&&  \Bigg[ -a_1 + k \, a_2 \Bigg] \, \de^{ab} \, J^{\al} \, J^{\al}(w) = 0,
  \qquad
  \Bigg[ k \, a_4 - 2 N \, \sqrt{\frac{M+N}{M N}} \, a_7 \Bigg] \,
  \de^{ab} \,
J^{u(1)} \, J^{u(1)}(w)  =  0,
\nonu \\
&& \Bigg[ -2 N \, \sqrt{\frac{M+N}{M N}} \, a_5 + N \, a_7 +(2k+M) \, a_9
  \Bigg]
\, d^{a b c} \, J^c \, J^{u(1)}(w)
 =  0,
\nonu \\
 && \Bigg[ -2 N \, \sqrt{\frac{M+N}{M N}} \, a_6 + a_{15}
  \Bigg] \, i \, f^{a b c} \, J^c \, J^{u(1)}(w)  =  0,
\nonu \\
&&\Bigg[ k a_{14} - k N M \sqrt{\frac{M+N}{M N}} a_8 -N \sqrt{\frac{M+N}{M N}}
  a_{13} \Bigg] \de^{a b} \, \pa \, J^{u(1)}(w)  =  0,
\nonu \\
&& \Bigg[ 2(k+M) \, a_5 -\frac{1}{2} \, a_{12} + \frac{1}{2} \, a_{13} \Bigg]
\, d^{a b c} \, \de_{\rho \bar{\si}} \, t^c_{j \bar{i}} \, J^{(\rho \bar{i})}
\, J^{(\bar{\si} j)}(w)  =  0,
\nonu \\
&& \Bigg[ -2(k+M) \, a_6 +\frac{1}{2} \, a_{12} + \frac{1}{2} \, a_{13} \Bigg]
 \, i\,
 f^{a b c} \, \de_{\rho \bar{\si}} \, t^c_{j \bar{i}} \, J^{(\rho \bar{i})}
\, J^{(\bar{\si} j)}(w)  =  0,
\nonu \\
&& \Bigg[ 2 k \, a_8 -\frac{1}{M} \, a_{12} +\frac{1}{M} \, a_{13} \Bigg]
\, \de^{a b}  \, \de_{\rho \bar{\si}} \, \de_{j \bar{i}} \, J^{(\rho \bar{i})}
\, J^{(\bar{\si} j)}(w)  =  0,
\nonu \\
&& \Bigg[ -N \, a_6 + a_{10} \Bigg] \, i\, f^{a c e} \, d^{b c d} \, J^e \, J^d(w)=0,
\nonu \\
 &&\Bigg[ k \, a_{10} + N \, (k+M) \, a_5 +\frac{N}{2} \, a_{13} + M N \, a_6
  \Bigg] \, d^{a b c} \, \pa \, J^c(w)  =  0,
\nonu \\
&& \Bigg[ -2 \, a_3 - a_{11} + \frac{3}{2} \, k \, a_{17}
  \Bigg] \, d^{a b e} \, d^{e c d} \, J^c \, J^d(w)  =  0,
\nonu \\
&& \Bigg[ 2(k+M) \, a_3 + 2 N \, a_8 + \frac{4}{M} \, (2a_3 +a_{11}) +
  (\frac{12 k}{M}+18)\, a_{17} \Bigg] \, J^a \, J^b(w)  =  0,
\nonu \\
&& \Bigg[ k \, a_3 - \frac{4}{M} \, (2a_3 +a_{11})+
  (\frac{6k}{M}-6) \, a_{17} \Bigg] \, \de^{a b} \, J^c \, J^c(w)  =  0,
\nonu \\
&& \Bigg[ N \, a_5 +(2a_3 +a_{11}) + 3(k+M) \, a_{17}
  \Bigg] \, d^{a c d} \, d^{b c e} \, J^d \, J^e(w)  =  0,
\nonu \\
&& \Bigg[ k \, a_{11} + 2 a_{16} + \frac{N}{2} \, a_{13} -N \, (k+M) \, a_6-
  \frac{N}{M} \, (M^2-4) \, a_5-\frac{1}{M} \, (M^2-4) \,
  (2a_{3}+a_{11})
  \nonu \\
&&   + (-\frac{3}{2} \, k \, M +(-2M^2-6)) \, a_{17} - 2 N \, a_8 -
  \frac{4}{M} \, (2a_{3}+a_{11}) \Bigg] \, i \, f^{a b c} \, \pa \, J^c(w)
 =  0.
\label{JPpole2}
\eea
We rewrite the term $f^{a c d} \, f^{d b e}\, J^c \, J^e(w)$
in terms of Kronecker delta $\de$ and $d$ symbols by using the
corresponding identity \cite{BS,Ahn1111}.
For the calculation of last five relations associated with
$a_{17}$ term in (\ref{JPpole2}),
the identities containing the
quartic products of $f f f f$, $f f f d$ and $f f d d$
\cite{NPB97,Ahn1111} are used.
Note that although there are also $f^{a e c} \, d^{b e d} \, J^c \, J^d(w)$
and $d^{a b c} \, \pa \, J^c(w)$ in general, those contributions
from the coefficient $a_{17}$ become zero.

By solving the above equations (\ref{tworel}), (\ref{Ju1P}),
(\ref{JP43}) and
(\ref{JPpole2}),
we obtain the coefficients appearing in
 the spin-$3$ current as follows:
 \bea
 a_2  & = &
 \frac{a_1}{k}, \qquad a_3= \frac{ N (k+2 N)}{k (k+M) (3 k+2 M)} \, a_1,
 \qquad
 a_4 = \frac{ (k+2 N) (M+N)}{k^2 M} \, a_1,
 \nonu \\
 a_5  & = & -\frac{ (k+2 N)}{4 (k+M)} \, a_1,
 \qquad
 a_7 =   \frac{(k+2 N)}{2 k} \, \sqrt{\frac{M+N}{M N}} \, a_1,
 \qquad
 a_8 = -\frac{ (k+2 N)}{2 k M} \, a_1,
 \nonu \\
 a_9 &=& -  \frac{(k+2 N)N}{2 k (k+M)} \,
 \sqrt{\frac{ (M+N)}{M N}} \, a_1,
\qquad
a_{11} =\frac{ (k^2-8) N (k+2 N)}{4 k (k+M) (3 k+2 M)} \, a_1,
\nonu \\
a_{12} & = & -\frac{1}{2} \, (k+2 N) \, a_1,
\qquad
a_{13}  =\frac{1}{2}  \, (k+2 N) \, a_1,
\label{avalues}
 \\
a_{16} & = & -\frac{ N (6 k^3+9 k^2 M+4 k M^2+12 M)
  (k+2 N)}{12 k (k+M) (3 k+2 M)} \, a_1,
\qquad
a_{17}  =  \frac{ N (k+2 N)}{6 (k+M) (3 k+2 M)} a_1.
\nonu
 \eea
 Except the coefficient $a_2$, all the coefficients contain
 the factor $(k+2N)$. These are the same as the ones in \cite{CH1812}.
 As described in the footnote \ref{zerofour},
 the four coefficients, $a_6$, $a_{10}$, $a_{14}$ and $a_{15}$
 are vanishing.
 
 Also we have  the primary condition under the stress energy tensor
 mentioned before
 \bea
 T(z) \, P^a(w) = \frac{1}{(z-w)^2} \, 3 \, P^a(w) + \frac{1}{(z-w)} \,
 \pa \, P^a(w) + \cdots.
 \label{TP}
 \eea
 In order to check this condition (\ref{TP}), the relations
 (\ref{OPETJ})
 can be used.
 
 After using the vanishing of
 the fourth, third and second order poles  we are left with
 the first order pole and can be written as 
 \bea
 J^a(z) \, P^b(w) =  \frac{1}{(z-w)} i \, f^{a b c}
  \, P^c(w) + \cdots,
 \label{JP}
 \eea 
 where the fundamental relations (\ref{OPEspin1spin1})
 can be used in (\ref{JP}).
 As explained in (\ref{JK}), the spin-$3$ current
 transforms under the adjoint representation of $SU(M)$.
 Once again, the charged spin-$3$ current is primary
 operator via (\ref{TP}) and (\ref{JP}).

 
\subsection{An uncharged spin $3$ current}

How do we construct the higher spin-$3$ current which is neutral
under the spin-$1$ current?
We should write down the possible composite spin-$3$ operators
and determine the relative coefficients by
imposing the basic conditions coming from the coset (\ref{coset}).
As explained before, we should require
that this spin-$3$ current transforms as the primary operator
under the stress energy tensor (\ref{T}).

It turns out that
the uncharged spin-$3$ current \cite{CH1812}
has the following independent terms
\footnote{The coefficients $b_9$, $b_{10}$ and $b_{11}$ are vanishing
  and the corresponding terms are given by
  $b_9 \, \pa\, J^{\alpha} \, J^{\alpha}(z)+
  b_{10} \,\pa\, J^{a} \, J^{a}(z)+ b_{11}\,
  \pa\, J^{u(1)} \, J^{u(1)}(z)$.}
 \bea
 W^{(3)}(z) & = &
b_1 \, d^{\alpha \beta \gamma} \, J^{\alpha} J^{\beta} J^{\gamma}(z)
+ b_2 \, d^{a b c} \, J^a J^b J^c(z)+
b_3 \, J^{u(1)}  J^{u(1)}  J^{u(1)}(z) + b_4 \, J^{\al} J^{\al} J^{u(1)}(z)
\nonu \\
&+& b_5 \, J^a \, J^a \, J^{u(1)}(z) + b_6 \, 
t^{\al}_{\rho \bar{\si}} \, \de_{j\bar{i}}\,
J^{\al}   \, (J^{(\rho \bar{i})}  \, J^{(\bar{\si} j)} +
J^{(\bar{\si} j)} \,   J^{(\rho \bar{i})})  (z)
\nonu \\
&+& b_7  \, \de_{\rho \bar{\si}} \,
t^a_{j\bar{i}} \, J^a  (J^{(\rho \bar{i})}  J^{(\bar{\si} j)} +
J^{(\bar{\si} j)}  J^{(\rho \bar{i})})  (z)
+b_8 \, \de_{\rho \bar{\si}} \,
\de_{j\bar{i}} \, J^{u(1)}  (J^{(\rho \bar{i})}  J^{(\bar{\si} j)} +
J^{(\bar{\si} j)}  J^{(\rho \bar{i})})  (z) 
\nonu \\
&+& b_{12} \, 
\de_{\rho \bar{\si}} \,
\de_{j\bar{i}} \, \pa  \, J^{(\rho \bar{i})}  \,J^{(\bar{\si} j)}(z)
+ b_{13} \,
\, \de_{\rho \bar{\si}} \,
\de_{j\bar{i}}  \, \pa  \, J^{(\bar{\si} j)} \,  J^{(\rho \bar{i})}(z)
+ b_{14} \, \pa^2 \, J^{u(1)}(z).
\label{W}
\eea
The second term can be seen from the work of \cite{BBSS}.
The $b_2$, $b_5$ and $b_7$ terms 
can be seen from the terms of spin-$2$ current in
(\ref{spin2expression}).
When we differentiate the stress energy tensor (\ref{T}), then
we observe the $b_{12}$ and $b_{13}$ terms.
For the $b_6$ term, we have seen similar $a_1$
term in the charged
spin-$3$ current.

The regularity condition $J^{\al}(z) \, W(w) =0$ implies the following
relations coming from the third and second order poles 
\bea
&& \Bigg[ 2(k+N) \, b_9 + M \, b_{12} + M \, b_{13} \Bigg] \, J^{\al}(w)  =  0,
\nonu \\
&& \Bigg[ 3(k+N) \, b_1 + M \, b_6 \Bigg] \, d^{\al \beta \ga} \,
J^{\beta} \, J^{\ga}(w)  =
 0,
\nonu \\
&& \Bigg[ 2(k+N) \, b_4 + 2 M \, b_8 + 2 M \, \sqrt{\frac{M+N}{M N}} \,
  b_6\Bigg]
\, J^{\al} \, J^{u(1)}(w)  =  0,
\nonu \\
&&
\Bigg[ (k+N) \, b_9 + M \, b_{13} -(k M+ 2 M N) \, b_6 \Bigg] \, \pa \, J^{\al}(w)
 =  0,
\nonu \\
&& \Bigg[ 2(k+2N) \, b_6 + b_{12} - b_{13} \Bigg] \, \de_{j \bar{i}} \,
t^{\al}_{\rho \bar{\si}} \, J^{(\rho \bar{i})} \, J^{(\bar{\si} j)}(w)  = 0.
\label{JalW}
\eea
In this calculation, we have the identities $f^{\al \beta \ga} \,
J^{\beta} \, J^{\ga}(w)= i N \pa \, J^{\al}(w)$ and $\mbox{Tr} (t^{\al} \,
t^{ \beta} \, t^{ \ga} ) =\frac{1}{2}( i f+d )^{\al \beta \ga}$
as described before.

Similarly, from the OPE between $J^{u(1)}(z)$ and $W(w)$,
we have the following relations from the fourth, the third and the second order
poles
\bea
&& \Bigg[ k M N \, \sqrt{\frac{M+N}{M N}} \,
  b_{12} -  k M N \, \sqrt{\frac{M+N}{M N}}
  \, b_{13} + 6 k \, b_{14} \Bigg] = 0,
\nonu \\
&& \Bigg[ 2 k \, b_{11} + (M+N) \, b_{12} + (M+N) \,
  b_{13}\Bigg] J^{u(1)}(w) =  0,
\nonu \\
&& \Bigg[ k \, b_4 + 2 M   \,
  \sqrt{\frac{M+N}{M N}} \, b_6 \Bigg] \, J^{\al} \, J^{\al}(w)= 0,
\qquad
\Bigg[ k \, b_5 -2 \,N\, \sqrt{\frac{M+N}{M N}} \, b_7 \Bigg] \,
J^a \, J^a(w)= 0,
\nonu \\
&& \Bigg[ 2(M+N) \, b_8 + 3 k \, b_3 \Bigg] \, J^{u(1)} \, J^{u(1)}(w) =  0,
\nonu \\
&&
\Bigg[ 2 k \, b_8 + \sqrt{\frac{M+N}{M N}} \, b_{12} - \sqrt{\frac{M+N}{M N}}
  \, b_{13} \Bigg] \, \de_{\rho \bar{\si}} \, \de_{j \bar{i}} \,
J^{(\rho \bar{i})} \, J^{(\bar{\si} j)}(w) = 0,
\nonu \\
&&
\Bigg[ k \, b_{11} + k M N \, \sqrt{\frac{M+N}{M N}} \, b_8 -(M+N) \, b_{13}
  \Bigg] \, \pa \, J^{u(1)}(w) = 0.
\label{Ju1W}
\eea
The identity $\de_{\rho \bar{\si}} \, \de_{j \bar{i}} \,
[J^{(\rho \bar{i})} , J^{(\bar{\si} j)}](w)=
\sqrt{\frac{M+N}{M N}}\, M \, N\, \pa \,
J^{u(1)}(w)$ coming from (\ref{comm}) is used
in the calculation of last two equations of (\ref{Ju1W}).
If we use the relations (\ref{JalW}) and (\ref{Ju1W}) only,  
then the coefficients are not determined completely.

In order to  calculate the OPE between $T(z)$
and $W(w)$, we should obtain the following nontrivial OPEs
\bea
T(z) \, J^{(\rho \bar{i})}(w) & = &
\frac{1}{(z-w)^2} \,
\Bigg[\frac{(-k-M+2k^2 M-N + k M N)}{2 k M (k+N)}\Bigg]
\,  J^{(\rho \bar{i})}(w)
\nonu \\
& + & 
\frac{1}{(z-w)} \, \Bigg[ \frac{1}{(k+N)} \, i
  f^{(\rho \bar{i}) \al}_{\,\,\,\,\,\,\,\,\,\,\,\,\,\,(\si \bar{k})}
  \, J^{\al}  J^{(\si \bar{k})}
+  \frac{1}{k} \, i
  \, f^{(\rho \bar{i}) u(1)}_{\,\,\,\,\,\,\,\,\,\,\,\,\,\,\,\,\,\,(\si \bar{j})}
  \, J^{u(1)}  J^{(\si \bar{j})}
  \nonu \\
  & + &  \pa   J^{(\rho \bar{i})} \Bigg](w) +  \cdots,
\nonu \\
T(z) \, J^{(\bar{\rho} j)}(w) & = &
\frac{1}{(z-w)^2} \,
\Bigg[\frac{(-k-M+2k^2 M-N + k M N)}{2 k M (k+N)} \Bigg]
\,  J^{(\bar{\rho} j)}(w)
\nonu \\
& + & 
\frac{1}{(z-w)} \, \Bigg[ \frac{1}{(k+N)} \, i
  \, f^{(\bar{\rho} j) \al}_{\,\,\,\,\,\,\,\,\,\,\,\,\,\,(\bar{\si} k)}
  \, J^{\al}  J^{(\bar{\si} k)}
+  \frac{1}{k} \, i
  \, f^{(\bar{\rho} j) u(1)}_{\,\,\,\,\,\,\,\,\,\,\,\,\,\,\,\,\,\,(\bar{\si} k)}
  \, J^{u(1)}  J^{(\bar{\si} k)}
  \nonu \\
  & + &  \pa   J^{(\bar{\rho} j)} \Bigg](w)
+  \cdots.
\label{TJJ}
\eea
In the first order term of (\ref{TJJ}),
there exist nontrivial nonlinear terms.
Even the second order term has nontrivial coefficients
which depend on $N$, $M$ and $k$ explicitly.
In this calculation we use the following
identity
\bea
t^{\al}_{\rho_1 \bar{\si}_1} \, t^{\al}_{\rho_2 \bar{\si}_2}
=\de_{\rho_1 \bar{\si}_2} \, \de_{\rho_2 \bar{\si}_1} -\frac{1}{N} \,
\de_{\rho_1 \bar{\si}_1} \, \de_{\rho_2 \bar{\si}_2},
\qquad
t^{a}_{i \bar{j}} \, t^{a}_{k \bar{l}}
=\de_{i \bar{l}} \, \de_{k \bar{j}} -\frac{1}{M} \,
\de_{i \bar{j}} \, \de_{k \bar{l}}.
\label{ttrelations}
\eea
In (\ref{ttrelations}), they satisfy for any four indices and
similar relations for contracted indices can be obtained from these
identities.

We summarize the fifth, fourth and third order poles
in the OPE between $T(z)$ and $W^{(3)}(w)$ as follows:
\bea
&& \Bigg[ 2k \, (M^2-1) \, b_{10} + \frac{N(-k-M+2k^2 M-N + k M N)}{(k+N)} \,
  (b_{12}+b_{13}) \Bigg] = 0,
\nonu \\
&& \Bigg[k \, (M^2-1) \, b_5 -2N (M^2-1) \, \sqrt{\frac{M+N}{M N}}\,
  b_7+  \frac{N(-k-M+2k^2 M-N + k M N)}{(k+N)}
  \, b_8  \nonu \\
&&  +  \frac{(-k-M+2k^2 M-N + k M N)}{2 k (k+N)} \, N
  \, \sqrt{\frac{M+N}{M N}}
  \, (b_{12}-b_{13}) \, \Bigg] \, J^{u(1)}(w)  =  0,
\nonu \\
&& 2 b_{10} \, J^a \, J^a(w) + (b_{12}+b_{13}) \, \Bigg[ 2
\de_{\rho \bar{\si}} \, \de_{j \bar{i}} \,
J^{(\rho \bar{i})} \, J^{(\bar{\si} j)}-
\frac{(M+N)}{k} \, J^{u(1)} \, J^{u(1)}  \nonu \\
&& -\frac{M}{(k+N)} \, J^{\al}
\, J^{\al} - M N \, \sqrt{\frac{M+N}{M N}} \, \pa \,
J^{u(1)} \Bigg](w)  =  0.
\label{TWcon}
\eea
It can be checked that the contribution from the coefficient $b_6$ term
vanishes by using the various further contractions
between the operators
appearing in the contributions from the $b_{12}$ or $b_{13}$
term.
We have the following primary condition
under the stress energy tensor 
\bea
 T(z) \, W^{(3)}(w) = \frac{1}{(z-w)^2} \, 3 \, W^{(3)}(w) +
 \frac{1}{(z-w)^2} \, \pa \, W^{(3)}(w) + \cdots.
 \label{TW}
 \eea
 It will be rather complicated to check this
 by hand explicitly.
 If we identify some of the factors in the spin-$3$ current with
 the previous known currents, then the corresponding computations
 will be easier.
 
 By solving (\ref{JalW}), (\ref{Ju1W}) and (\ref{TWcon}), we arrive at the
 following intermediate result for the coefficients
 \bea
 b_3 &=&  \frac{2(k+N)(M+N)(k+2N)}{k^2 M} \, \sqrt{\frac{M+N}{M N}} \, b_1,
 \qquad
 b_4 = \frac{6(k+N)}{k} \, \sqrt{\frac{M+N}{M N}} \, b_1,
 \nonu \\
 b_5 &=& \frac{2 N}{k} \, \sqrt{\frac{M+N}{M N}} \, b_7,
 \quad
 b_6 = -\frac{3(k+N)}{M} \, b_1, \nonu \\
 b_8  & = & -\frac{3(k+N)(k+2N)}{k M} \, \sqrt{\frac{M+N}{M N}} \, b_1,
 \qquad
b_{12}  =  \frac{3(k+N)(k+2N)}{M} \, b_1, 
\nonu \\
b_{13} &=& -\frac{3(k+N)(k+2N)}{M}\, b_1,
\qquad
b_{14} = -N (k+N)(k+2N) \, \sqrt{\frac{M+N}{M N}} \, b_1.
\label{interbvalue}
 \eea
 The coefficients are written in terms of $b_1$ and $b_7$ and
 moreover the coefficient $b_2$
 is not determined yet.
 Except the coefficients of $b_4$ and $b_6$, all the
 coefficients contain the factor $(k+2N)$.
 We will analyze further in section $7$ and determine
 the remaining coefficients completely. Therefore, we
 have checked that the expressions for the spin-$3$ current
 is correct for any $(N,M)$ and $k$.

%
%
 
\section{ The OPE between the charged  higher spin-$2$
  current and itself}

In this section, we would like to construct the
OPE $K^a(z) \, K^b(w)$ which did not appear in \cite{CH1812}
by using the explicit realization in (\ref{spin2expression})
with the help of (\ref{OPEspin1spin1}).
What they have observed in \cite{CH1812} is that
the above OPE is found
by assuming that there exist the spin-$1,2$
currents as well as the stress energy tensor (\ref{T}).
Of course, they have constructed the uncharged spin-$3$ current
which does not appear in the above OPE.
Moreover, they have used the Jacobi identities
between these currents and the relative coefficients
appearing in this OPE depend on $(N,M)$ and $k$ explicitly
by collecting some of the results for fixed $(N,M)$ values.
Furthermore, their construction
does not tell us any information on the coset model.

On the other hand, in our construction we use the explicit
realization of coset and the currents are given by
(\ref{T}), (\ref{spin2expression}), (\ref{spin3exp}) and
(\ref{W}).
We will observe that there exists a charged spin-$3$
current described in (\ref{spin3exp}) in the first order pole
of the OPE.

It is useful to calculate the OPEs between $K^a(z)$ and other
spin-$1$ operators. We have (\ref{JK})
and the OPE between $K^a(z)$ and $J^{\alpha}(w)$
and the  OPE between $K^a(z)$ and $J^{u(1)}(w)$
have trivial results from the analysis of the subsection $2.1$.
Then the remaining nontrivial OPEs are 
given by
\bea
K^a(z) \, J^{(\rho \bar{i})}(w) &=&
\frac{1}{(z-w)^2} \,
\Bigg[\frac{2(k^2-1)(2k+M+N)}{k(2k+M)} \Bigg] \,
\de^{k \bar{i}} \, t^a_{k \bar{j}} \, J^{(\rho \bar{j})}(w)
\nonu \\
& + &
\frac{1}{(z-w)} \Bigg[ 2(k+N) \de_{k \bar{j}} \, (t^{a})^{\bar{i} k} \,
  \pa \, J^{(\rho \bar{j})} -\frac{2(k+N)}{k}  \sqrt{\frac{M+N}{M N}}
   \de^{k \bar{i}}  t^a_{k \bar{j}}    J^{u(1)}    J^{(\rho \bar{j})}
  \nonu \\
  &+& \frac{2}{k M}(k+M+N) \, J^a \, J^{(\rho \bar{i})}-(i \, f -\frac{(2k+
    M+2N)}{
    (2k+M)} \, d)^{a b c} \, \de^{k \bar{i}} \, t^c_{k \bar{j}} \, J^b \,
  J^{(\rho \bar{j})} \nonu \\
  &-& 2 \de^{k \bar{i}} \, \de_{\si \bar{\si_1}} \, (t^{\al})^{\bar{\si_1} \rho}
  \, t^a_{k \bar{j}} \, J^{\al} \, J^{(\si \bar{j})} \Bigg](w)
+ \cdots,
\nonu \\
K^a(z) \, J^{(\bar{\rho} j)}(w) &=&
\frac{1}{(z-w)^2} \,
\Bigg[
\frac{2(k^2-1)(2k+M+N)}{k(2k+M)} \Bigg]
\, \de^{j \bar{l}} \, t^a_{k \bar{l}} \, J^{(\bar{\rho} k)}(w)
\nonu \\
& + &
\frac{1}{(z-w)} \Bigg[ 2(k+N) \, \de_{k \bar{k_1}} \, (t^{a})^{\bar{k_1} j} \,
  \pa \, J^{(\bar{\rho} k)} +\frac{2(k+N)}{k}  \sqrt{\frac{M+N}{M N}}
   \de^{j \bar{l}}  t^a_{k \bar{l}}  J^{u(1)}  J^{(\bar{\rho} k)}
  \nonu \\
  &-& \frac{2}{k M}(k+M+N) \, J^a \, J^{(\bar{\rho} j)}-(i \, f +
  \frac{(2k+M+2N)}{
    (2k+M)} \, d)^{a b c} \, \de^{j \bar{j_1}} \, t^c_{k \bar{j_1}} \, J^b \,
  J^{(\bar{\rho} k)} \nonu \\
  &+& 2 \de^{j \bar{l}} \, \de_{\si \bar{\si_1}} \, (t^{\al})^{\bar{\rho} \si}
  \, t^a_{k \bar{l}} \, J^{\al} \, J^{(\bar{\si_1} k)} \Bigg](w)
+ \cdots.
\label{KJJ}
\eea
These two OPEs look similar but they are different from each other.
Based on these OPEs, we can calculate the OPEs
between the charged spin-$2$ current and the derivative of spin-$1$
currents by simply taking the derivative with respect to the argument
$w$.
We use the identity of two and triple products of generators
\bea
t^a \, t^b & = &
\frac{1}{M} \, \de^{a b} \,{\bf 1}_{M} + \frac{1}{2} (i \, f +d)^{a b c}
\, t^c,
\nonu \\
t^a \, t^b \, t^c & = &
\frac{1}{M} \, \de^{b c} \,t^a + \frac{1}{2M} \, \de^{a d} \, (i \, f +
d)^{b c d}\,
     {\bf 1}_M 
+ \frac{1}{4} (i \, f +d)^{b c d} \, (i \, f + d)^{a d f} t^f,
\label{ttandttt}
\eea
where the first relation can be obtained from the $f$ and $d$ symbols
in (\ref{fdtrace}) 
together with the metric in (\ref{metric}) and the second relation
can be determined by acting other generator on the first relation.


\subsection{The fourth, third and second order poles}

Then the fourth order pole can be determined
by the OPE between the spin-$2$ current and the first two terms
of spin-$2$ current.
If we use the property of the footnote \ref{derinK},
then the contribution from the second term of the spin-$2$ current
can be expressed as the contribution from the first term and the
contribution from the OPE between the spin-$2$ current
and the derivative of spin-$1$ current which can be easily
obtained from the defining relation in (\ref{JK}).

It turns out that the fourth order pole of this OPE
is given by
\bea
K^a(z) \, K^b(w) \Bigg|_{\frac{1}{(z-w)^4}} =
\frac{4(k^2-1)N(2k+M+N)}{(2k+M)} \, \de^{ab},
\label{pole4}
\eea
which is equal to $\frac{c_1}{2} \, \de^{ab}$ in the notation of
\cite{CH1812}.
Then we can determine the coefficient
\bea
c_1 =  \frac{8(k^2-1)N(2k+M+N)}{(2k+M)}. 
\label{c1}
\eea
The free indices $a$ and $b$ arise in the form of invariant
Kronecker delta symbols.

How do we obtain the third order pole?
By using the
trace of triple product of generators appearing in (\ref{ttandttt})
leading to the second contribution because the first and last
contributions provide zero due to the tracelessness of
the generator,
the final result
can be expressed as a $f$ symbols with spin-$1$ current.
It turns out that
the third order pole of this OPE
is given by
\bea
K^a(z) \, K^b(w) \Bigg|_{\frac{1}{(z-w)^3}} =
\frac{4(k^2-1)N(2k+M+N)}{k(2k+M)} \, i f^{a b c} \,J^c(w),
\label{pole3}
\eea
which is given by $c_2  \, i \,f^{a b c} \,J^c(w)$
in the notation of \cite{CH1812}.
Therefore, we have the coefficient 
\bea
c_2 = \frac{4(k^2-1)N(2k+M+N)}{k(2k+M)}.
\label{c2value}
\eea

Let us present the final result first.
The second order pole can be written as
\bea
&& K^a(z) \, K^b(w) \Bigg|_{\frac{1}{(z-w)^2}} =
\nonu \\
&& -\frac{4(k+N)(M+N)}{k M} \, \de^{a b} \, J^{u(1)} \, J^{u(1)}(w)
-4 \de^{a b} \, J^{\al} \, J^{\al}(w)
-\frac{4 N}{k M} (k+M+N) \, J^a \, J^b(w)
\nonu \\
&& + 2 \sqrt{\frac{M+N}{M N}} \, ( -\frac{N^2}{k} \,i \, f +
\frac{N(4k^2 + 2k M + 4 k N +M N)}{k (2k+M)} \, d )^{a b c}
\, J^c \, J^{u(1)}(w) 
\nonu \\
&& + N ( i \, f - \frac{(k+M+2N)}{(2k+M)} \, d )^{a e c} ( i \, f
+ d )^{d b c}
\, J^e\, J^d(w) 
+ \frac{8(k+N)}{M} \, \de^{a b} \, \de^{\rho\bar{\si}} \, \de_{j \bar{l}} \,
J^{(\rho \bar{l})} \, J^{(\bar{\si} j)}(w)
\nonu \\
&& - 2 N \, i \, f^{a b c} \, \de^{\rho \bar{\si}} \, t^c_{j \bar{l}} \,
J^{(\rho \bar{l})} \, J^{(\bar{\si} j)}(w)
+ \frac{2(4k^2 + 2 k M+ 4 k N + M N)}{(2k+M)} \, d^{a b c} \,
 \de^{\rho \bar{\si}} \, t^c_{j \bar{l}} \,
J^{(\rho \bar{l})} \, J^{(\bar{\si} j)}(w)
\nonu \\
&& - 4 N (k+N)\, \sqrt{\frac{M+N}{M N}} \, \de^{a b} \, \pa \, J^{u(1)}(w)
+ 2 k N \, i \, f^{a b c} \, \pa J^c(w)
\nonu \\
&& +
\frac{2 k N(2k +M +2N)}{(2k+M)} \, d^{a b c} \, \pa \, J^c(w) 
 -\frac{M N}{(2k+M)} \, d^{a b c} \, K^c(w) + N \, i \, f^{a b c} \, K^c(w).
\label{Rel}
\eea
The contribution from the third term of (\ref{spin2expression})
is given by
the second term of the last line of (\ref{Rel}).
The last term of (\ref{Rel}) comes from the
expression of the second term having a derivative term of $J^b(w)$
in the footnote \ref{derinK}.
Then the remaining expressions come from the first two terms in
(\ref{spin2expression}).
Then
the operator contents of (\ref{Rel}) is the same as
the ones in (\ref{JPpole2}) as expected.

The next question is how we can write down
the above expression (\ref{Rel}) in terms of
previous known currents, spin-$1,2$ currents as well as the
stress energy tensor?
Of course, there should be a descendant term originating from the
third order pole. This is a simple derivative term of spin-$1$ current
with fixed known coefficient.
Moreover, it is obvious that there are stress energy tensor and
spin-$2$ current of spin-$2$.
Now it is clear to simplify
(\ref{Rel}) by comparing it with (\ref{T}) and (\ref{spin2expression}). 

It is easier to look at the terms of singlet operator without having
any group indices first. 
By identifying $J^{u(1)}\, J^{u(1)}(w)$ term in both (\ref{Rel}) and
(\ref{T}),
we observe that
the coefficient of $T(w)$ in the second order pole should be equal to
\bea
\frac{8}{M}(k+N)(k+M+N),
\label{three}
\eea
by focusing on the first term of (\ref{Rel}).
This is equivalent to $2 \frac{c_1 \, a_{1,CH}}{c}$
of \cite{CH1812} with (\ref{charge}) and (\ref{c1}).
Then we can extract the coefficient of $a_{1,CH}$ from (\ref{three})
as follows:
\bea
a_{1,CH} =
\frac{(2k+M)(-k^2+k^2 M^2-2k N-M N+2k^2 M N + k M^2 N-N^2 + k M N^2)}{
  2(k^2-1)M N(2k+M+N)}.
\label{a1}
\eea
Then the structure constant (\ref{three})
appearing in the stress energy tensor
of the second order pole is determined.
Of course, other terms of the stress energy tensor
in the second order pole can be checked.

Let us move to the other structure constant and  
the coefficient of $d^{a b c} \, K^c(w)$ is given by
\bea
\frac{2k(2k+M+2N)}{(2k+M)},
\label{2c6rel}
\eea
which is equal to $2 c_6$ in \cite{CH1812}.
Note that the contribution (\ref{2c6rel}) comes from
the $d$ term of the second line of (\ref{Rel})
and the second term in the last line of (\ref{Rel})
by focusing on the singlet term of (\ref{spin2expression}).
Then
the coefficient of $c_6$ of \cite{CH1812} from (\ref{2c6rel})
is given by
\bea
c_6 = \frac{k(2k+M+2N)}{(2k+M)}.
\label{c6}
\eea 
Then the structure constant (\ref{2c6rel})
appearing in the spin-$2$ current
of the second order pole is determined.

After subtracting the descendant term, the stress energy
tensor term and
spin-$2$ current term from the second order pole, there exists the sum
of 
some nonzero composite operators which corresponds to
a quasi primary operator.
We can collect the following nonderivative quadratic $J^a$
dependent terms in (\ref{Rel})
\bea
&& \frac{4N}{M} \,
\de^{a b} \, J^c \, J^c + \frac{N(4 k^2+4 k M+M^2+M N)}{(2k+M)^2}
\, d^{a b e} \, d^{e c d} \, J^c \, J^d \nonu \\
&& -\frac{4N(2k+M+N)}{k M}
\, J^a \, J^b -\frac{2N(2k+M+N)}{(2k+M)} \, d^{a c e} \,
d^{e b d} J^c \, J^d.
\label{fourquantity}
\eea
From the expression of (\ref{Rel}), it is easy to
see that the above terms (\ref{fourquantity})
come from the last term of the first line (entering into the third term
of (\ref{fourquantity})),
the first term of the third line,
and the second  term of the last line (contributing to the
second term
of (\ref{fourquantity})) of (\ref{Rel}).
Because we are looking at the particular composite operators,
the other terms in (\ref{Rel}) including the derivative terms
should be checked explicitly.

On the other hand, the two invariant fourth order $d$ symbols
are studied in \cite{CH1812} as well as
the two product of Kronecker delta symbols.
Then we can express the above quantities
by writing down their invariant tensors in terms of $f$ and $d$
symbols via the first two relations in Appendix (\ref{tensor}).
In other words, 
we have
\bea
&& \Bigg[c_{31} + \frac{4}{M} \, c_{32}-\frac{4}{M} \, c_{33} +
  \frac{2 c_1 a_1}{c}
  \frac{1}{2(k+M+N)} \Bigg] \, \de^{a b} \, J^c \, J^c
\label{samefourquantity} \\
&& + \Bigg[(c_{32}-c_{33}) - 2 c_6 \,
\frac{N}{(2k+M)} \Bigg]\, d^{a b e} \, d^{e c d} \, J^c \, J^d
+\Bigg[\frac{8}{M} \, c_{33} + c_{34}\Bigg] 
J^a \, J^b + 2c_{33} \, d^{a c e} \, d^{e b d} \, J^c \, J^d.
\nonu
\eea
Note that these four independent operators
appear in (\ref{JPpole2}).
For the $c_{33}$ term, as we can see in
the second relation of Appendix (\ref{tensor}),
the various identities can be used. After using the
symmetric property of the free indices, then
half of them can be rewritten as the other half.
It turns out that
$f d$ term and the derivative term with $d$ symbols are
vanishing.

Then we obtain the following expressions, by using the two equations
(\ref{fourquantity}) and (\ref{samefourquantity}),
\bea
c_{31} & = & -\frac{4(4k^3+4k^2 M+k M^2+ 8 k^2 N+ 6 k M N +M^2 N+ 4k N^2+
  M N^2)}{M(2k+M)^2},
\label{c3134} \\
c_{32}  & = & \frac{2 k N(2k+M+N)}{(2k+M)^2}, \qquad
c_{33} = -\frac{N(2k+M+N)}{(2k+M)}, \qquad
c_{34} = -\frac{4N(2k+M+N)}{k(2k+M)}. 
\nonu
\eea
Therefore, we have determined the second order pole
with (\ref{a1}), (\ref{c6}) and (\ref{c3134})
completely.
As we emphasized before, the structure constants we have found here
are different from the their $(3.27)$  in \cite{CH1812}.

\subsection{The first order pole and charged spin-$3$ current}

Now we can collect all the contributions
entering into the first order pole and
we arrive at the final results as follows:
\bea
&& K^a(z) \, K^b(w) \Bigg|_{\frac{1}{(z-w)}} =
i \, f^{a b c} \, N \pa \, K^c(w)
\nonu \\
&& -\frac{N}{(2k+M)} \, \Big( i \, f^{a c e} \, d^{b c d} \, K^e \, J^d +
i\, f^{a d e} \, d^{b c d} \, J^c \, K^e \Big)(w)
+ \frac{2 N}{k} \, \sqrt{\frac{M+N}{M N}} \, i \, f^{a b c} \,
K^c \, J^{u(1)}(w)
\nonu \\
&&+ \de_{\rho \bar{\si}} \,
t^b_{j \bar{i}} \, \Bigg[ 4(k+N) \de_{k \bar{l}} \, (t^a)^{\bar{i} k}  \,
\pa \, J^{(\rho \bar{l})}\,
J^{(\bar{\si} j)}  - \frac{4}{k} \, (k+N) \, \sqrt{\frac{M+N}{M N}} \,
\de^{k \bar{i}} \, t^a_{k \bar{l}} \,
((J^{u(1)} \, J^{(\rho \bar{l})}) J^{(\bar{\si} j)})
\nonu \\
&& + \frac{4}{k M}\, (k+M+N) \,
((J^a\, J^{(\rho\bar{i})})\, J^{(\bar{\si} j)})
\nonu \\
&&-2 ( i \, f -\frac{(2k+M+2N)}{(2k+M)} \, d )^{a c d}\,
\de^{k \bar{i}} \, t^d_{k \bar{l}} \,
((J^c \, J^{(\rho \bar{l})}) \, J^{(\bar{\si} j)})
\nonu \\
&&- 4 \de^{k \bar{i}} \, \de_{\si_1 \bar{\si_1}}\, (t^{\al})^{\bar{\si_1} \rho} \,
t^a_{k \bar{l}} \, ((J^{\al} \, J^{(\si_1 \bar{l})}) \, J^{(\bar{\si} j)}) 
+ 4(k+N) \de_{k \bar{l}} \, (t^a)^{\bar{l} j} \, 
 J^{(\rho \bar{i})}\,
\pa \, J^{(\bar{\si} k)} \nonu \\
&& + \frac{4}{k} \, (k+N) \, \sqrt{\frac{M+N}{M N}} \,
\de^{j \bar{l}} \, t^a_{k \bar{l}} \,
J^{(\rho \bar{i})}\, J^{u(1)} \, J^{(\bar{\si} k)}
 - \frac{4}{k M}\, (k+M+N) \,
 J^{(\rho\bar{i})} \, J^a \, J^{(\bar{\si} j)}
\nonu \\
&&-2 ( i \, f + \frac{(2k+M+2N)}{(2k+M)} \, d )^{a c d}\,
\de^{j \bar{j_1}}\, t^d_{k \bar{j_1}} \,
J^{(\rho \bar{i})}\, J^c \, J^{(\bar{\si} k)}
\nonu \\
&&+ 4 \, \de^{j \bar{l}} \, \de_{\si_1 \bar{\si_1}}\, (t^{\al})^{\bar{\si} \si_1} \,
t^a_{k \bar{l}} \, J^{(\rho \bar{i})}\, J^{\al} \, J^{(\bar{\si_1} k)}\Bigg](w).
\label{KKpole1}
\eea
Compared to the previous second order pole, it is rather easy to
obtain this first order pole because we do not have to consider
the additional contractions between the operators.
The first two terms in the second line of (\ref{KKpole1})
are determined from the OPE between the spin-$2$ current and the
third term of (\ref{spin2expression}) while
the last term  in the second line of (\ref{KKpole1})
comes from the OPE between the spin-$2$ current and the
last term of (\ref{spin2expression}).

According to the observation of \cite{CH1812},
there exist five quasi primary operators including the
spin-$3$ current after subtracting the various descendant operators
properly.
Let us look at the $J^{\alpha}$ term in (\ref{KKpole1}).
It appears in the sixth line and the last line.
We can easily see that they have the product of two generators
and this contains
the $f$ symbols with numerical value $\frac{1}{2}$.
Then the overall numerical factor will be $4$ by adding the above
two contributions.
Because the operator contents are the same as the
one of the first term of spin-$3$ current
(\ref{spin3exp}), by extracting the first term of $P^c(w)$ in
the above first order term (\ref{KKpole1}),
we determine the structure constant,
the coefficient of $ P^c(w)$ in the right hand side
of the OPE
\bea
C_{K^a \,K^b}^{P^c} = \frac{4}{a_1} \, i \, f^{a b c}.
\label{Struct}
\eea
Of course, this is one of the terms among thirteen terms in
(\ref{spin3exp}). 
Further analysis on this direction can be done without any difficulty.

Note that the second term of spin-$3$ current
contains only $J^{\alpha}$ and $J^c$ term.
We can check that this term cannot be seen from (\ref{KKpole1}).
However, among the list of the five
quasi primary operators we mentioned,
we can find that term. This implies that
we should have exact coefficient in the two places,
in the quasi primary operator and the spin-$3$ current
with opposite signs.
Then we can determine the coefficient $a_3^{CH}$ in \cite{CH1812}
by focusing on the second term of $P^c(w)$
\bea
i\, \Bigg[\frac{1}{2(k+M+N)}-\frac{1}{2(k+N)} \Bigg] \,
c_2 \, a_{3, CH} + i \, \frac{4}{a_1} \, a_2 =0,
\label{a3rel}
\eea
where (\ref{Struct}) is used.
Note that the two terms inside the bracket in (\ref{a3rel})
are coming from the explicit stress energy tensor in (\ref{T}).
From this (\ref{a3rel})
together with (\ref{c2value}) and (\ref{avalues}),
we have determined the coefficient
\bea
a_{3,CH} = \frac{2(2k+M)(k+N)(k+M+N)}{(k^2-1) M N (2k+M+N)}. 
\label{a3}
\eea
Then the structure constant appearing in
this quasi primary operator is given by the first term of
(\ref{a3rel}) with (\ref{c2value}) and (\ref{a3}).

Now we move to the other quasi primary operator.
Let us determine the coefficient of $c_{73}$ appearing in
the first order pole in \cite{CH1812}
by looking at $f^{ab c} \, d^{c d e} \, J^d \, J^e \, J^{u(1)}(w)$.
Then we have the following relation
\bea
-i \frac{4 N^2}{k(2k+M)} \, \sqrt{\frac{M+N}{M N}}
-i \frac{4}{a_1} \, a_9- \frac{2N}{k} \, \sqrt{\frac{M+N}{M N}} \,
c_{73}=0,
\label{a9rel}
\eea
where the first term originates from the
second, third
and fourth terms of (\ref{KKpole1}).
In the $c_{73}$ term of (\ref{a9rel}),
the relation of third line in Appendix (\ref{tensor})
is used.
In the $a_9$ term, the relation (\ref{Struct}) is used.
By substituting the value of $a_9$ in (\ref{avalues})
into (\ref{a9rel}),
we obtain 
\bea
c_{73} = i \frac{k(2k+M+2N)}{(k+M)(2k+M)}.
\label{c73}
\eea

For the $c_{72}$ term having $ f^{a c e} \, d^{b e d}$  in \cite{CH1812},
we should focus on the $a_5$ term of
the spin-$3$ current $P^c(w)$. See also the relations in
Appendix (\ref{tensor}).
Then we have
\bea
- 2 i + i \frac{4}{a_1} \, 2 \, a_5 + 2 \, c_{73} - 2 \, i \, c_{72} =0.
\label{c72rel}
\eea
There are two contributions from (\ref{KKpole1}) for the first
term in (\ref{c72rel}). The corresponding terms are $f$ terms in the
fifth and eighth line of (\ref{KKpole1}).
In the $a_5$ term here, the second term of $a_5$
appearing in (\ref{spin3exp}) can be
written in terms of the first term and derivative term.
Then the number $2$ exists in (\ref{c72rel}).
We determine the coefficient $c_{72}$ from
(\ref{c72rel}) by using  
(\ref{avalues}) and (\ref{c73})
as follows:
\bea
c_{72} = -\frac{(2k+M+2N)}{(2k+M)}.
\label{c72}
\eea
Therefore, the structure constant associated with
$c_{72}$ and $c_{73}$ terms is completely determined.

Now we consider the quasi primary operator
which is cubic terms in the spin-$1$ currents. 
For the coefficient $c_{53}$, we consider
$d^{a c f} \, f^{f b g} \, d^{g d e}\, J^c \, J^d \, J^e(w)$ term.
In this case, we
have
\bea
i \, \frac{2 N^2}{(2k+M)^2} - 3 \, i \, c_{53} + 2 \, \frac{N}{(2k+M)} \,
i \, c_{72}=0.
\label{c53c72}
\eea
It is rather nontrivial to extract the exact contribution from the
$c_{53}$ term with corresponding $d_{52}^{a b c d e}$ tensor.
The other contribution from $c_{72}$ can occur here.
Therefore,
from (\ref{c72}) and (\ref{c53c72}),
we determine the coefficient $c_{53}$
\bea
c_{53} =-\frac{2N(2k+M+N)}{3(2k+M)^2}.
\label{c53}
\eea

By considering the 
$f^{a b f} \, d^{f c g} \, d^{g d e}\, J^c \, J^d \, J^e(w)$ term,
we have
\bea
i \, \frac{2N^2}{(2k+M)^2} - i \, \frac{3}{2} \, c_{52} -
i \, \frac{3}{2} \, c_{53} + i \, \frac{N}{(2k+M)} \, c_{72}
-i \, \frac{4}{a_1} \, \frac{3}{2} \, a_{17} +
\frac{N}{(2k+M)} \, c_{73} =0.
\label{c52rel}
\eea
Again the
the first term
can be obtained 
from the first two terms in the second line of
(\ref{KKpole1}) with Jacobi identity.
In this case also, the corresponding invariant tensors
associated with $c_{52}$ and $c_{53}$ terms look complicated
in Appendix (\ref{tensor}) but if we use the symmetric property
of the indices between $c$, $d$ and $e$ we will obtain
simpler expression and we can extract the exact coefficients
we presented above.
For the $c_{72}$ term, the Jacobi identity is used.
It is easy to obtain the coefficient $c_{52}$
by substituting (\ref{avalues}), (\ref{c53}), (\ref{c72}) and (\ref{c73})
into the above (\ref{c52rel})
\bea
c_{52} =\frac{2 k N(2k+M+N)}{3(2k+M)^2(3k+2M)}.
\label{c52}
\eea

For the $f^{a b c} \, J^c \, J^d \, J^d(w)$ term,
we have
\bea
i \, \frac{4}{a_1} \, a_3 + i \, \frac{4}{a_1} \, \frac{6}{M} \, a_{17}
+\frac{1}{2(k+M+N)} \, c_2 \, a_{3,CH} + c_{51} + i\, \frac{6}{M} \,
c_{52} -i \, \frac{6}{M} \, c_{53} =0.
\label{c51rel}
\eea
We can observe the first term with previous structure constant
(\ref{Struct})
in the spin-$3$ current.
It is obvious to see the $a_{51}$ term
and we obtain the $c_{52}$ and $c_{53}$ terms with above coefficients.
Again, from (\ref{avalues}), (\ref{c2value}), (\ref{a3}), (\ref{c52})
and (\ref{c53}), we determine the coefficient $c_{51}$
from (\ref{c51rel})
\bea
c_{51} = -i \, \frac{4 (6k^3+7k^2 M+2 k M^2 +12 k^2 N + 10 k M N + 2 M^2 N
  +6 k N^2 + 2 M N^2)}{k M (2k+M)(3k+2M)}.
\label{c51}
\eea
We also realize that
$a_{2,CH}$ can be obtained from $a_{3,CH}$ in (\ref{a3})
\bea
a_{2,CH} & = & \frac{1}{6} (1- 3 a_{3,CH})
\nonu \\
& = & 
\frac{1}{6(k^2-1)M N (2k+M+N)}
\, (-12 k^3 -18 k^2 M - 6 k M^2 -24 k^2 N -26 k M N
\nonu \\
& + & 2 k^3 M N
-  7 M^2 N
  + k^2 M^2 N-12 k N^2 - 7 M N^2 + k^2 M N^2).
\label{a2}
\eea
Note that this (\ref{a2}) is not an independent structure constant
because this can be obtained from $a_{3,CH}$.
Therefore, we have determined the structure constants
with $c_{51}$, $c_{52}$ and $c_{53}$ terms appearing in the
the cubic spin-$1$ current terms.

We are left with one final quasi primary operator
of spin-$3$ which contains the derivative terms.
This is the most nontrivial parts to extract the correct
structure constants because the derivative terms
appear all over the places.
Let us determine the remaining two coefficients, $c_{41}$ and $c_{43}$.
For the former, by looking at the $f^{a b c} \, f^{c d e} \, \pa \, J^d \, J^e(w)$,
we eventually have
\bea
&& \frac{4}{a_1} \, \Bigg(
-2 a_3 -a_{11} -\frac{3M}{2} \, a_{17} \Bigg)
+ 2 c_{41} + i \, c_{51} -\frac{3M}{2} \, (c_{52}+c_{53}) \nonu \\
&& + \Bigg(\frac{12}{M}-3
\, \frac{8-M^2}{2M}\Bigg) \, c_{53} + \frac{N^2}{(2k+M)} =0.
\label{nontri}
\eea
It is not difficult to check the coefficient for the
$c_{51}$ term because it contains already one of the $f$ symbols.
For the $c_{52}$ term, we should move the spin-$1$ currents
to the left in order to obtain the above derivative term with
some identity including the $f$ or $d$ symbols.
For the $c_{53}$ term, the identity for $f f d d$
\cite{NPB97,Ahn1111} is used.
The last term of (\ref{nontri}) comes from the
fifth line of (\ref{KKpole1}) which should be simplified further.
Then this will give us the final expression as above.
The above (\ref{nontri}) leads to
\bea
c_{41} & = & \frac{1}{k M(2k+M)^2(3k+2M)} (-24k^4 -40k^3 M-22 k^2 M^2 -4 k M^3
-48 k^3 N-64 k^2 M N \nonu \\
&+ & 2 k^4 M N -28 k M^2 N+ 3k^3 M^2 N- 4 M^3 N+ k^2 M^3 N-24 k^2 N^2
-20 k M N^2 + k^3 M N^2 \nonu \\
& - & 4 M^2 N^2 +k^2 M^2 N^2),
\label{c41}
\eea
where
the previous results (\ref{avalues}), (\ref{c51}),
(\ref{c52}) and (\ref{c53}) are used in (\ref{nontri}).

Now we would like to determine the final undetermined coefficient.
For the $c_{43}$ coefficient,
we consider the expression of $f^{a b i} K^a(z) \, K^b(w)\Bigg|_{\frac{1}{(z-w)}}$.
Then we have the relation
\bea
&& -N \, \frac{4}{k M}(k+M+N) + N M -N \,
\frac{(2k+M+2N)}{(2k+M)} \, \frac{(M^2-4)}{M}
\nonu \\
&& -\frac{4}{a_1} \, i \, 2 M \Bigg[ 2 \, i \, a_3 + i \, a_{11} +i\,
  \Big( \frac{4}{M}+\frac{2}{M}+
  \frac{(M^2-4)}{M} +\frac{(M^2-4)}{2M} \Big) \, a_{17} \Bigg]
\nonu \\
&&-2(2 M \, c_{41} +c_{43}) - 2 M \, i \, c_{51}-
\Bigg(-2M \, \frac{6}{M}-\frac{3}{2} \, 2 M \, \frac{(
  M^2-4)}{M} \Bigg)\, (c_{52}+c_{53})
\nonu \\
&&-\Bigg(2M \, \frac{12}{M} + 3 M \, \frac{(M^2-4)}{M}
\Bigg) \, c_{53} =0.
\label{this}
\eea
The fourth line of (\ref{KKpole1}) contributes
the first term of (\ref{this}) if we further simplify nonstandard
normal ordering product in the composite operator.
Again the fifth line of (\ref{KKpole1})
can be simplified and we can check the contribution from
this will be the remaining two terms in the first line of
(\ref{this}). Now we can move to the next line.
For the $a_3$ term, we obtain the above factor by moving
the spin-$1$ current to the left. For the $a_{11}$ term, we will
have $f f$ term which is proportional to $2M$.
We collect all the contributions from the $a_{17}$ term.
From the above (\ref{this}) by substituting (\ref{avalues}),
(\ref{c41}), (\ref{c51}), (\ref{c52}) and (\ref{c53}),
we arrive at 
\bea
c_{43} =-\frac{2 M N(2k+M+N)}{k(2k+M)^2}. 
\label{c43}
\eea
Therefore, we have determined all the structure constants
associated with $c_{41}$ and $c_{43}$ appearing in
the first order pole
\footnote{We have checked that all the structure constants
  are consistent with each other
  when we consider the $(N,M)=(6,5)$ case
  and the $(N,M)=(7,6)$ case.}.

\subsection{The final OPE}

After collecting the previous results (\ref{pole4}), (\ref{pole3}) ,
(\ref{Rel}) and (\ref{KKpole1}),
we summarize the OPE, in the notation of \cite{CH1812},
between the charged spin-$2$
current and itself as follows:
\bea
K^a(z) \, K^b(w) & = &
\frac{1}{(z-w)^4} \, \frac{c_1}{2} 
+\frac{1}{(z-w)^3} \, i\, c_2 \, f^{a b c} \, J^c(w)
\nonu \\
& + & \frac{1}{(z-w)^2} \, \Bigg[ \frac{1}{2} \, i\, c_2 \, f^{a b c} \, \pa \, J^c + \frac{2 a_{1,CH} \, c_{1}}{c}\, \de^{a b}
  \, T + 2 \, c_6 \, d^{a b c}\, K^c
  \nonu \\
  & + & \Big( 
  c_{31} \,  \de^{a b} \, \de^{c d} + c_{32} \, d^{a b c d}_{4SS1}
 +c_{33} \, 
  d^{a b c d}_{4SS2}
   +  c_{34}
  \,  \de^{a c} \, \de^{b d}
  \Big) \, \frac{1}{2} \, (J^c \, J^d + J^d \, J^c)
  \Bigg](w)
\nonu \\
& + & \frac{1}{(z-w)} \, \Bigg[  \frac{1}{6} \, i\, c_2 \, f^{a b c}
  \, \pa^2 \, J^c+ \frac{1}{2} \, \pa \, \Big( \frac{2 a_{1,CH} \, c_{1}}{c}\, \de^{a b}
  \, T + 2 \, c_6 \, d^{a b c}\, K^c
  \nonu \\
  & + & \Big( 
  c_{31} \,  \de^{a b} \, \de^{c d} + c_{32} \, d^{a b c d}_{4SS1}
 +c_{33} \, 
  d^{a b c d}_{4SS2}
   +  c_{34}
  \,  \de^{a c} \, \de^{b d}
  \Big) \, \frac{1}{2} \, (J^c \, J^d + J^d \, J^c) \Big)
  \nonu \\
  & + & i \, c_2 \, a_{3,CH}  \,
  f^{a b c} \,\Big( T \, J^c -\frac{1}{2} \, \pa^2 \, J^c\Big)
  \nonu \\
  &+&\Big( 
  c_{41} \, d^{a b c d}_{4AA1}
   +  c_{43}
  \,  \de^{a c} \, \de^{b d}
  \Big)\, (\pa J^c \, J^d -\pa J^d \, J^c -\frac{1}{3} \, i\,
  f^{c d e} \pa^2 \, J^e)
  \nonu \\
  &+& \Big( c_{51} \, f^{a b c}\, \de^{d e} +
    c_{52} \, d^{a b c d e}_{51}+ c_{53}\, d^{a b c d e}_{52} \Big)
    \nonu \\
    & \times &  \frac{1}{6} (J^c \, J^d \, J^e +J^c \, J^e \, J^d+
    J^e \, J^c \, J^d
    + J^d \, J^c \, J^e +J^d \, J^e \, J^c+J^e \, J^d \, J^c)\nonu \\
    & + & \Big( c_{72} \,d^{a b c d}_{4AA2}+ c_{73}\, d^{c d a b}_{4SA}
    \Big) \, J^c \, K^d 
  + i \, \frac{4}{a_1} \, f^{a b c} \, P^c \Bigg](w) + \cdots,
\label{fullKK}
\eea
where the structure constants are given by
(\ref{c1}),(\ref{c2value}),(\ref{a1}),(\ref{charge}),(\ref{c6}),
(\ref{c3134}),(\ref{a3}),(\ref{c41}),
(\ref{c43}),(\ref{c51}),(\ref{c52}),(\ref{c53}),(\ref{c72}),
and (\ref{c73}) and we present them here
\bea
c_1 & = &  \frac{8(k^2-1)N(2k+M+N)}{(2k+M)}, \qquad
c_2 = \frac{4(k^2-1)N(2k+M+N)}{k(2k+M)},
\nonu\\
a_{1,CH} & = &
\frac{(2k+M)(-k^2+k^2 M^2-2k N-M N+2k^2 M N + k M^2 N-N^2 + k M N^2)}{
  2(k^2-1)M N(2k+M+N)},
\nonu \\
c & = &
\frac{(-k^2+k^2 M^2-2k N-M N+2k^2 M N + k M^2 N-N^2 + k M N^2)}{(k+N)(k+M+N)},
\nonu \\
c_6 & = &  \frac{k(2k+M+2N)}{(2k+M)}, 
\nonu \\
c_{31} & = & -\frac{4(4k^3+4k^2 M+k M^2+ 8 k^2 N+ 6 k M N +M^2 N+ 4k N^2+
  M N^2)}{M(2k+M)^2},
\nonu \\
c_{32}  & = & \frac{2 k N(2k+M+N)}{(2k+M)^2}, \qquad
c_{33} = -\frac{N(2k+M+N)}{(2k+M)},
\nonu \\
c_{34} & = &  -\frac{4N(2k+M+N)}{k(2k+M)}, \qquad
a_{3,CH} = \frac{2(2k+M)(k+N)(k+M+N)}{(k^2-1) M N (2k+M+N)},
\nonu \\
c_{41} & = & \frac{1}{k M(2k+M)^2(3k+2M)} (-24k^4 -40k^3 M-22 k^2 M^2 -4 k M^3
-48 k^3 N\nonu \\
& - & 64 k^2 M N 
+  2 k^4 M N -28 k M^2 N+ 3k^3 M^2 N- 4 M^3 N+ k^2 M^3 N-24 k^2 N^2
\nonu \\
& - & 20 k M N^2 + k^3 M N^2  -  4 M^2 N^2 +k^2 M^2 N^2),
\nonu \\
c_{43} & = & -\frac{2 M N(2k+M+N)}{k(2k+M)^2}, 
\nonu \\
c_{51} & = &
-i \, \frac{4 (6k^3+7k^2 M+2 k M^2 +12 k^2 N + 10 k M N + 2 M^2 N
  +6 k N^2 + 2 M N^2)}{k M (2k+M)(3k+2M)},
\nonu \\
c_{52} & = & \frac{2 k N(2k+M+N)}{3(2k+M)^2(3k+2M)},
\qquad
c_{53} =-\frac{2N(2k+M+N)}{3(2k+M)^2},
\nonu \\
c_{72} & = & -\frac{(2k+M+2N)}{(2k+M)}, \qquad
c_{73} = i \frac{k(2k+M+2N)}{(k+M)(2k+M)}.
\label{KKcoeff}
\eea
In the last line of the second order pole in (\ref{fullKK}),
there exists a quasi primary
spin-$2$ operator.
In the first two lines of the first order pole
there are descendants for the spin-$1$ and spin-$2$ operators. 
In the next five lines, there are quasi primary spin-$3$ operators.
More precisely, the last one is a primary spin-$3$ current
where the coefficient $a_1$ is the overall factor in (\ref{spin3exp}).
In general, the quasi primary spin-$3$ operator in the last line
is given by $(J^c \, K^d -\frac{1}{4}\, i \, f^{c d e} \, \pa K^e)(w)$.
However, the derivative term vanishes when we multiply the tensors
of $c_{72}$ and $c_{73}$ terms
\footnote{Due to the symmetric or antisymmetric
  properties of the right hand side of this OPE, we can obtain
  the quantities by multiplying the antisymmetric $f$ symbols,
  the symmetric $d$ symbols, or symmetric Kronecker
  delta symbols. The $c_{31}$-$c_{34}$ terms are symmetric,
  the $c_{41}$-$c_{43}$ terms are antisymmetric,
  the $c_{51}$-$c_{53}$ terms are antisymmetric and
  the $c_{72}$-$c_{73}$ terms are antisymmetric under the exchange of
  the indices $a$ and $b$.}.

Let us emphasize here that although the operator contents
appearing in the right hand side of (\ref{fullKK}) except the spin-$3$
current 
are the same as
the ones in \cite{CH1812}, the structure constants are completely different
from theirs.
We can check that the difference between our results and theirs
will provide the factor $(k+2N)$.

When we take the infinity limit of $k$ after substituting
$N = \frac{(1-\la)}{\la} \, k$ into the various structure
constants (\ref{KKcoeff}) we have determined, we obtain the corresponding
values in terms of $\la$, $k$ and  $M$. We present them in Appendix
$C$. Although we do not compare here the exact  values
for the structure constants
with the ones in \cite{JKKR}, we can check the $k$ dependence
as well as $M$ dependence. We observe that their $(4.40)-(4.42)$
are consistent with our results with $\la =2$ in
Appendix $C$ by focusing on
the $k$ dependence. Moreover, our coefficients
$c_{31}$ and $c_{51}$ do depend on the factor $\frac{1}{M}$
which can be seen from \cite{JKKR} also
\footnote{We regard $d_{4SS2}^{a b c d}$ as $\frac{4}{M} \, \de^{a d}
  \, \de^{b c}-f^{a c e}\, f^{e b d} +
  i\, f^{a c e}\, d^{e b d}+ i\, d^{ a c e}\, f^{e b d}+
  d^{a c e} \, d^{e b d}$ by using the symmetric property
  in the indices of $c$ and $d$ in (\ref{fullKK}) from the general
  definition in Appendix $A$.
  Similarly, $d_{51}^{a b c d e}$ is given by $i \, f^{a b f} \,
  (\frac{6}{M} \, \de^{f c} \, \de^{d e}+\frac{3}{2}\, (i f +d)^{
    f c g} \, d^{g d e})$ by imposing the symmetric property
  between the indices $c,d$ and $e$. We also have
  $d_{52}^{a b c d e} = d_{51}^{a b c d e}+ (\frac{12}{M} \, i \,
  f^{c b a} \, \de^{d e}+\frac{3}{2} \, i \, f^{c b f}\,
  d^{a f g} \, d^{ g d e} -\frac{3}{2}\, i\, f^{c a f} \,
  d^{b f g} \, d^{g d e})$. Finally,
  we have
  $d_{4 A A 2 }^{a b c d} = i\, (d^{e a c } \, f^{b d e}+ f^{a c e}\,
  d^{e b d})$.}.


\section{ The OPE between the charged  higher spin-$2$
  current and the charged  higher spin-$3$ current}


\subsection{The fifth, fourth and third order poles}

First of all, we can calculate
the fifth order pole
of the OPE $K^a(z) \, P^b(w)$ for the fixed $(N,M)=(5,4)$.
It turns out that the nonzero contribution appears
when the indices $a$ and $b$ are the same.
The coefficients contain $a_6, a_{12}$ and $a_{13}$
from $P^b(w)$ and moreover
the common factor appears in the sum of $a_{12}$ and $a_{13}$.
Then this contribution becomes zero from the footnote \ref{zerofour}
and (\ref{avalues}).

For the
fourth order pole
of the OPE $K^a(z) \, P^b(w)$ for the fixed $(N,M)=(5,4)$,
the contribution appears in the coefficients, $a_5, a_{6}, a_7, a_{12}$
and $a_{13}$ of $P^b(w)$
and the relevant fields are given by $J^{u(1)}$ and $J^c$.
Again by substituting the values of (\ref{avalues}),
all these terms are vanishing.

Now we move on the third order pole of
$K^a(z) \, P^b(w)$ where the nonzero results appear explicitly.
The relevant coefficients are given by $a_3$, $a_5$, $a_7$,
$a_8$, $a_{11}$,
$a_{12}$, $a_{13}$, $a_{16}$, and $a_{17}$.
For the calculation of $a_5$ terms in (\ref{spin3exp}),
it is better to rewrite them by using
the charged spin-$2$ current in (\ref{spin2expression})
because the first two terms of (\ref{spin2expression}), which are
equal to the factor of $a_5$ terms, can be written in terms of
the remaining three quantities.
That is,
\bea
 \de_{\rho \bar{\si}} \,
t^c_{j\bar{i}} \, (J^{(\rho \bar{i})} \,  J^{(\bar{\si} j)} +
J^{(\bar{\si} j)} \,  J^{(\rho \bar{i})})  (w)
 & = & K^c(w)
 +\frac{N}{(M+2k)} \, d^{c d e} \, J^d \, J^e(w)
 \nonu \\
 & - & \frac{2N}{k} \sqrt{\frac{M+N}{M N}} \,
 J^c  \,  J^{u(1)}(w).
\label{a5relation}
\eea
Then the $a_5$ term
contains $d^{b c d} \, J^d$ multiplied by the above expression
to the right.
The nontrivial calculation comes from the OPE between
$K^a(z)$ and $d^{b c d} \, J^d \, K^c(w)$.
Due to the fact that there is a relation in (\ref{JK}),
the contribution of the third order pole in the above OPE
can be obtained from the second  order pole of the OPE
$K^e(z) \, K^c(w)$ and
 the third  order pole of the OPE
$K^a(z) \, K^c(w)$
we have determined in previous section.

It is also nontrivial to calculate the $a_8$
term of (\ref{spin3exp}).
Then we 
should calculate the second order pole of the OPE
between $K^c(z)$ and $ \de_{\rho \bar{\si}} \,
\de_{j\bar{i}} \,   (J^{(\rho \bar{i})}  J^{(\bar{\si} j)} +
J^{(\bar{\si} j)}  J^{(\rho \bar{i})})(w)$
and the third order pole of similar OPE
with different index we have
obtained in previous
section.

Because the $a_{12}$ and $a_{13}$ terms of (\ref{spin3exp}) cannot be
written in terms of other known quantities, it is rather complicated to
extract the corresponding third order poles.
Let us consider the OPE between the current $K^a(z)$
and the composite operator $\de_{\rho \bar{\si}} \,
t^b_{j\bar{i}} \, J^{(\rho \bar{i})} \,  \pa \, J^{(\bar{\si} j)}(w)$
which is not exactly the $a_{13}$ term because there exists
$-\frac{1}{2}\, N \, \pa^2 \, J^b(w)$ from the normal ordering in the
above composite operator. That is, the commutator
$ \de_{\rho \bar{\si}} \,
t^b_{j\bar{i}} \,[ \pa \, J^{(\bar{\si} j)},  J^{(\rho \bar{i})}]$
provides the above second derivative term although there are other
two terms and the OPEs with $K^a(z)$
do not contribute to the final result.

For the coefficient $a_{12}$ term,
we have the following relation
\bea
&& \de_{\rho \bar{\si}} \,
t^b_{j\bar{i}} \, \pa \, J^{(\rho \bar{i})}  \, J^{(\bar{\si} j)}(w)
 =  \frac{1}{2} \, \pa \, K^b(w) -
\de_{\rho \bar{\si}} \,
t^b_{j\bar{i}} \, J^{(\rho \bar{i})} \,  \pa \, J^{(\bar{\si} j)}(w)
-\frac{N}{2}\, \pa^2 \, J^b(w)
\nonu \\
& & + \frac{N}{2(2k+M)}\, d^{b c d} \, \pa \, (J^c \, J^d)(w)
-\frac{N}{k} \, \sqrt{\frac{M+N}{M N}}\,
\pa \, ( J^b \, J^{u(1)})(w).
\label{1213}
\eea
For the second term of (\ref{1213}),
we have analyzed them in the context of $a_{13}$ term in previous
paragraph.
It is easy to observe that the third order pole from the OPE
between $K^a(z)$ and $\pa \, K^b(w)$ is given by
$(\pa \, (K^a \, K^b)_{pole-3} + 2 (K^a \, K^b)_{pole-2})(w)$
from the previous section.

For the $a_{17}$ term of (\ref{spin3exp}),
in general, there are quintic products in the $f$ and $d$ symbols.
After collecting the three products here correctly
we are left with $f$ or $d$ symbols and we can further use
the identities between the triple products by combining
these single $f$ or $d$ symbols with the remaining quadratic
products between them.

It turns out that the third order pole, by collecting the above
results, is summarized by
\bea
&& K^a(z) \, P^b(w) \Bigg|_{\frac{1}{(z-w)^3}}=
2 M \, a_3\, i \, f^{a b c} \, K^c(w) 
\nonu \\
&& + a_5 \, \Bigg[ -\frac{1}{k M (2k+M)}
  2N(-8k^2+ 4 k M-4 k^3 M + 4 M^2- 2k^2 M^2 -8 k N + 4 M N
  \nonu \\
 && 
    - 2k^2 M N + k M^2 N) \, i\, d^{a c e} \, f^{e b d} \, J^c \, J^d
    -\frac{1}{k (2k +M)}  N (8k^2 -4 k M+ 4k^3 M-4M^2 \nonu \\
    && +  2k^2 M^2 
    + 8 k N - 4 M N+ 2k^2 M N - k M^2 N) \, d^{a b c} \, \pa \, J^c
    \nonu \\
    && + \frac{(M^2-4)  (4k^2 + 2k M + 4 k N +M N)}{M(2k+M)} \,
    ( N \, i \, f^{a b c} \, \pa \, J^c +
    \frac{2 N}{k}\, \sqrt{\frac{M+N}{M N}}
    \, i\, f^{a b c} \, J^c \, J^{u(1)} \nonu \\
    &&+ 2 \, i\, f^{a b c} \, \de_{\rho \bar{\si}} \, t^c_{j \bar{i}} \,
    J^{(\rho \bar{i})} \, J^{(\bar{\si} j)}) \Bigg](w)  +
\frac{4(k^2-1)N(2k+M+N)}{k(2k+M)}\, a_7
\, i\, f^{a b c} \, J^c \, J^{u(1)}(w)
\nonu \\
&& + a_8 \,
\Bigg[  \frac{4N}{k} \, \sqrt{\frac{M+N}{M N}}\, (2k+M+2N) \,
  i\, f^{a b c} \, J^{u(1)} \, J^c  -2N \, \frac{(2k+M+2N)}{(2k+M)} \, i\,
  f^{a b c} \, d^{c d e} J^d \, J^e
  \nonu \\
  && + 4(2k +M+2N)
  \, i \, f^{a b c} \, \de_{\rho \bar{\si}} \, t^c_{j \bar{i}}\,
  J^{(\rho \bar{i})} \, J^{(\bar{\si} j)}
 + 2 N (2k +M+2N) \,
\,
i \, f^{a b c} \, \pa \, J^c
\Bigg](w) 
\nonu \\
&& +  \Bigg[ - M \, a_{11}
+ 2 \, a_{16} + (M^2+6) \, a_{17} \Bigg]
i \, f^{a b c} \, K^c(w). 
\nonu \\
&& + (a_{13}-a_{12}) \, \Bigg[
  -2 \, \sqrt{\frac{M+N}{M N}}\, (k N+3 N^2-50) \,
  \de^{a b} \, 
  \pa \, J^{u(1)} \nonu \\
  && +
\frac{N (6 k^3+3 k^2 M+2 k^2 N-4 k-2 M-2 N)}{k (2 k+M)}\,
i \, f^{a b c} \, \pa \, J^c
+  \frac{k N (2 k+M+2 N)}{(2 k+M)} \, d^{a b c} \, \pa \, J^c
\nonu \\
&& +\frac{(4 k^3+2 k^2 M-k M N-4 k-2 M-2 N)}{k (2 k+M)}
 \, i \, f^{a b c} \, \de_{\rho \bar{\si}} \, t^c_{j \bar{i}}\,
 J^{(\rho \bar{i})} \, J^{(\bar{\si} j)}
 \nonu \\
 &&+ \frac{(4 k^2+2 k M+4 k N+M N)}{2 k+M} \, d^{a b c} \, \de_{\rho \bar{\si}} \, t^c_{j \bar{i}}\,
 J^{(\rho \bar{i})} \, J^{(\bar{\si} j)} +
 \frac{4 (k+N)}{M}\, \de^{a b } \, \de_{\rho \bar{\si}} \, \de_{j \bar{i}}\,
 J^{(\rho \bar{i})} \, J^{(\bar{\si} j)}
 \nonu \\
 && - \frac{2 (k+N) (M+N)}{k M}\,
 \de^{a b } \, J^{u(1)} \, J^{u(1)} -
\frac{N^2 }{k}\, \sqrt{\frac{M+N}{M N}}\,
 i\, f^{a b c} \, J^{u(1)} \, J^c
 \nonu \\
 &&+\sqrt{\frac{M+N}{M N}}\, \frac{N 
   (4 k^2+2 k M+4 k N+M N)}{k (2 k+M)} \,
 d^{a b c} \, J^{u(1)} \, J^c
 - \frac{2 N (k+M+N)}{k M} \, J^a \, J^b
 \nonu \\
 &&+\frac{N}{2}\,
 ( i\, f -\frac{2 k+M+2 N}{2 k+M} \,
 d )^{a e c} \,  ( i \, f + d )^{d b c}\,
 J^{e} \, J^{d} - 2 \, \de^{a b} \, J^{\alpha} \, J^{\alpha}
 + N \, i\,  f^{a b c} \, K^c
 \Bigg](w)
\nonu \\
&& + a_{12} \, \Bigg[\frac{1}{2} \, \pa \, (K^a \, K^b)_{pole-3}
  + (K^a \, K^b)_{pole-2} +\frac{M N}{(2 k+M)} \,
  d^{a b c} \, K^c
  \Bigg](w).
\label{KP3}
\eea

We expect that the spin of third order pole
is given by $2$ and it is natural to consider $K^c(w)$ term.
Let us focus on the
term $ i\, f^{a b c} \, J^c \, J^{u(1)}(w)$ in (\ref{KP3})
by remembering the
explicit form in (\ref{spin2expression}).
We obtain the following result
\bea
            && \frac{4 M N }{k} \, \sqrt{\frac{ M+N}{M N}}
\, a_3 
+\frac{2 (M^2-4)
  N (4 k^2+2 k M+4 k N+M N)}{M k (2 k+M)}\,
 \sqrt{\frac{M+N}{M N}}
\, a_5
\nonu \\
&& + \frac{4 (k^2-1) N (2 k+M+N)}{k (2 k+M)} \,
a_7 + \frac{ 4 N (2 k+M+2 N)}{k}\,
                        \sqrt{\frac{M+N}{M N}}
\, a_8
\nonu \\
&&-\frac{2  M N}{k} \,\sqrt{\frac{ (M+N)}{M N}}\,
a_{11} -\frac{ N^2 }{k} \, \sqrt{\frac{M+N}{M N}}\,
a_{12} +
\frac{ N^2 }{k} \, \sqrt{\frac{M+N}{M N}}\,
a_{13}
\nonu \\
&&+\frac{4 N }{k} \,\sqrt{\frac{ M+N}{M N}}\, a_{16}
+ \frac{2 (M^2+6) N }{k} \, \sqrt{\frac{ (M+N)}{M N}}\, a_{17}.
\label{combi}
\eea
Note that the $K^c(w)$ term in (\ref{KP3}) can participate in
the expression of (\ref{combi}).
By substituting the coefficients in (\ref{avalues}) into the
above (\ref{combi}), we obtain the final coefficient of
$K^c(w)$ in the third order pole.
 
Therefore, finally we determine the third order pole
of the OPE $K^a(z) \, P^b(w)$ as follows:
\bea
(K^a \, P^b)_{pole-3} =
\frac{(k^2-4)(2k+M)(k+2N)(3k+2M+2N)}{2k(k+M)(3k+2M)}
\, a_1 \, i\, f^{a b c} \, K^c(w).
\label{KP3-1}
\eea
Because the factor $(k+2N)$ appears in all the coefficients
except $a_1$ and $a_2$ in the spin-$3$ current,
it is obvious to see that this factor
appears in (\ref{KP3-1}).

\subsection{The second order pole}


\subsubsection{Complete second order pole in the coset realization}

For $a_1$ term in the spin-$3$ current,
we should calculate the OPEs between the
first order poles of the first OPE in (\ref{KJJ})
and $J^{(\bar{\si} j)}$. Compared to the OPE $K^a(z) \, W^{(3)}(w)$
associated with $b_7$ term,
the $a_1$ term of (\ref{spin3exp}) contains
the generator $t^{\alpha}_{\rho \bar{\si}}$ rather than
$\de_{\rho \bar{\si}}$. In this case, we have similar relations to
(\ref{ttandttt}) where the indices $a, b,c, \cdots$
are replaced by $\alpha, \beta, \gamma, \cdots$ and
$M$ is replaced by $N$. The identities involving $f$ or $d$
symbols for $SU(N)$ are used.

For $a_5$ term, from the previous relation in (\ref{a5relation}),
we need to calculate the first and second order poles
of the OPE between the charged spin-$2$ current.
For the former, due to the additional quadratic product
of $f$ and $d$ symbols, the identities involving
$f f f d$, $f d f d$, $f f d d$ and $ f d d d$ can be used
\cite{NPB97,Ahn1111}.

For $a_7$ term, by using the previous relation in (\ref{a5relation})
where the index $c$ is replaced by $b$, we can calculate the OPEs
between $K^a$ and the right hand sides of (\ref{a5relation}).
Then as before, the second order pole of the OPE between $K^a(z)$
and $K^b(w)$ can be used.

For $a_8$ term, as an alternative method,
we can use the stress energy tensor and
the second and third terms of (\ref{T})
can be written as
\bea
&& \frac{1}{2(k+N+M)} (
    \de_{\rho \bar{\si}} \,\de_{j \bar{i}} 
    \, J^{(\rho \bar{i})} \, J^{(\bar{\si} j)} +
    \de_{\rho \bar{\si}} \,\de_{j \bar{i}} 
    \,J^{(\bar{\si} j)} \,  J^{(\rho \bar{i})})    =  
\label{otherrel}
    \\
 &&    T -
    \frac{1}{2(k+N+M)} \,\Bigg[ J^{\al} \, J^{\al} + J^a \, J^a  +
      J^{u(1)} \, J^{u(1)} 
    \Bigg]
  - \frac{1}{2(k+N)} \,  J^{\al} \, J^{\al} - \frac{1}{2k}\, 
  J^{u(1)}\,  J^{u(1)}.
\nonu
\eea
We can regard the $a_8$ term as the product of
$J^b$ with the right hand side of (\ref{otherrel}).
Then the nonzero contributions of the OPE with
$K^a$ can be calculated from the $T$ term 
and $J^c \, J^c$ terms in (\ref{otherrel})
by using (\ref{JK}) and (\ref{TK}) because the OPEs between $K^a$ and
both $J^{\alpha}$ and $J^{u(1)}$
do not have any singular terms. 

For $a_{12}$ term,
due to the relation in (\ref{1213}),
 the second order pole from the OPE
between $K^a(z)$ and $\pa \, K^b(w)$ is given by
$\frac{1}{2} \,
(\pa \, (K^a \, K^b)_{pole-2} +  (K^a \, K^b)_{pole-1})(w)$
from the previous section.

For $a_{13}$ term,
the second order pole of the first OPE in (\ref{KJJ}) can combine
with $\pa \, J^{(\bar{\si} j)}$
and similarly the operator $ J^{(\rho \bar{i})}$
can be multiplied by the second order pole of the OPE
$K^a(z)$ and  $\pa \, J^{(\bar{\si} j)}(w)$.
Moreover, 
there are also contributions from the second order pole
between the first order pole
of the first OPE in (\ref{KJJ}) and $\pa \, J^{(\bar{\si} j)}$
and contributions from the second order pole
between $K^a$ and $\pa \, J^{(\bar{\si} j)}$.

For $a_{17}$ term,
the identities for the quartic products in the $f$ and
$d$ symbols are used \cite{NPB97,Ahn1111}.

We present the complete second order pole in Appendix $D$.

\subsubsection{How to rearrange the second order pole}

At first sight,
because the spin is given by $3$ in this particular pole, 
we do not expect that there should be other independent
spin-$3$ current. It is natural to consider the possibility of
spin-$3$ currents, $P^c(w)$ and $W^{(3)}(w)$ with an appropriate
additional $SU(M)$ invariant  tensors because the right hand side
of the OPE $K^a(z) \, P^b(w)$
should contain the free indices $a$ and $b$.
Of course, the descendant of (\ref{KP3-1}) with fixed known
coefficient should also appear in the right hand side
\bea
\frac{1}{4} \,
   \frac{(k^2-4)(2k+M)(k+2N)(3k+2M+2N)}{2k(k+M)(3k+2M)}
   \, a_1 \, i\, f^{a b c} \, \pa \, K^c.
   \label{descendant}
   \eea
The nontrivial things to check explicitly is to write down
the remaining composite operators in terms of the known currents
for generic $N, M$ and $k$.

The simplest term we can consider is
the $b_3$ term of $W^{(3)}(w)$ in (\ref{W}).
From the $a_7$ term in the second order pole in the OPE
$K^a(z) \, P^b(w)$, the corresponding cubic term in $J^{u(1)}$,
$\de^{a b} \, J^{u(1)} \, J^{u(1)} \, J^{u(1)}$,  
is given by 
$ J^{u(1)} \, (K^a \, K^b)_{pole-2}(w)$ and the coefficient is
\bea
-\frac{4  (k+N) (M+N)}{k M} \, a_7.
\label{3ju1}
\eea
By substituting the $a_7$ in (\ref{avalues}) into (\ref{3ju1}),
then this leads to $-\frac{b_1}{a_1} \, b_3$ with (\ref{interbvalue}).
This implies that there should be
\bea
 - \de^{a b} \, \frac{a_1}{b_1} \, W^{(3)}(w)
\label{pole2W}
\eea
in the second order pole of the OPE we are considering.

We can check also other simple term.
For example,
the
$a_2$ term of (\ref{spin3exp}), $
 d^{a b c} \,
J^{\alpha} \, J^{\alpha} \, J^c(w)$, can be seen from
both $a_1$ and $a_5$ terms in the second order pole.
They are given by
\bea
\frac{2 (2 k+M+N)}{(2 k+M)} \, a_1 - 4\, a_5. 
\label{alphaalpha}
\eea
By substituting the $a_5$ value in (\ref{avalues}) into
(\ref{alphaalpha}),
this can be written as
\bea
\frac{k (3 k+2 M) (2 k+M+2 N)}{(k+M) (2 k+M)}\, a_2,
\label{coeffa2}
\eea
where the relation (\ref{avalues}) is used.
Then the second order pole should contain, from (\ref{coeffa2}),
\bea
\frac{k(3k+2M)(2k+M+2N)}{(k+M)(2k+M)} \, d^{a b c} \, P^c(w).
\label{pole2P}
\eea

After subtracting (\ref{descendant}), (\ref{pole2W}), and
(\ref{pole2P})
from the second order pole, we have checked that 
we are left with the following seven terms for fixed $(N,M)=(5,4)$
\bea
&& d_{30} \, J^a \, K^b(w)  +  d_{32} \,
J^b \, K^a(w) +   d_{33}  \,
f^{a c e} \, f^{b d e} \,
J^c \, K^d(w)
\label{seven}
\\
&& +    d_{38} \,
d^{a c e} \, d^{b d e}  \, J^d \, K^c(w) 
+  d_{39} \,
d^{a c e} \, d^{b d e}  \, J^c \, K^d(w)
+ d_{41} \,
\de^{a b} \,
J^c \, K^c(w)
 + d_{42} \,
i \, f^{a b c} \, \pa K^c(w),
\nonu
\eea
where the ordering in the coefficients is not important.
Of course, these coefficients are known for the above
fixed values of $(N,M)$.
We have obtained (\ref{seven})
by assuming the possible terms with arbitrary coefficients
in the right hand side of
the second order in the OPE. 
Note that the above terms (\ref{seven}) also
arise in the coefficient of $a_{17}$ term of the second order pole.
This implies that the second order pole can be written
in terms of the known currents we mentioned before.

Then the next thing we should consider
is to determine the above seven undetermined coefficients
in terms of $N$, $M$ and $k$.
Let us consider the
$d_{33}$ term in (\ref{seven}).
Recall that there exists a relation
we mentioned several times before
\bea
f^{a b c} \, f^{c d e} =-\frac{4}{M} \, (\de^{a e} \, \de^{b d}-
\de^{a d} \, \de^{b e}) - (d^{b d c} \, d^{c a e} -
d^{a d c} \, d^{c b e}).
\label{ffrel}
\eea
When we meet the $f f$ terms, we should always
use this identity in order to collect the independent terms.
Then by remembering the spin-$2$ current, the $d_{33}$ term has 
$d^{a b e} \, d^{e c d} \,
J^c \, K^d(w)$, where we can see
$\frac{2 N }{k} \, \sqrt{\frac{M+N}{M N}} \, d^{a b e} \, d^{e c d} \,
J^c \, J^d \, J^{u(1)}(w)$. We collect
the corresponding terms in the second order pole
as follows:
\bea
 -\frac{4 N }{k}\, \sqrt{\frac{M+N}{M N}}\, a_3
-\frac{M N}{(2 k+M)} \, a_9
-\frac{2 N }{k} \, \sqrt{\frac{M+N}{M N}}\, a_{11}
+ N \, a_7. 
\label{LHS}
\eea
Note that there are also contributions from
$a_{17}$ term we do not write down here
but they are cancelled each other.
It turns out that the $f f $ term
with above cubic operators in $J^{u(1)} \, (K^a \, K^b)_{pole-2}$
provides the final contribution
with the help of (\ref{ffrel}).
This should be equal to
\bea
\frac{k (3 k+2 M) (2 k+M+2 N)}{(k+M) (2 k+M)} \, a_9
+ \frac{2 N }{k} \, \sqrt{\frac{M+N}{M N}}\, d_{33},
\label{RHS}
\eea
where the first term comes from (\ref{pole2P}).
Therefore, we determine the coefficient $d_{33}$,
by using (\ref{LHS}) and (\ref{RHS}) together with (\ref{avalues}),
as follows:
\bea
d_{33} =
\frac{ (2 k+M)^2 (k+2 N) (3 k+2 M+2 N)}{4 (k+M)^2 (3 k+2 M)}
\, a_1,
\label{d33}
\eea
which can be substituted into (\ref{seven}).

We can move on the term $\de^{a b} \, J^c K^c$
where there exists $\de^{a b} \, J^c \, J^c \, J^{u(1)}$
with an appropriate coefficient concerning on the
coefficient $d_{41}$.
From the second order pole, we have
\bea
-\frac{16 N }{k M} \, \sqrt{\frac{M+N}{M N}}\, a_3
+ \frac{4 N}{M} \, a_7- \frac{8 N }{k M}\,
\sqrt{\frac{M+N}{M N}}\, a_{11}
-\frac{12  N }{k} \, \sqrt{\frac{M+N}{M N}}\, a_{17},
\label{lhs}
\eea
which (there are two contributions from
the $a_{17}$ term with (\ref{ffrel})
and the final result by summing over
them is given as above) is equal to
\bea
-\frac{a_1}{b_1} \, b_5 + \frac{8 N }{k M} \,
\sqrt{\frac{M+N}{M N}} \, d_{33} +
\frac{2 N }{k}\, \sqrt{\frac{M+N}{M N}}
\, d_{41}.
\label{rhs}
\eea
The first term is obtained from (\ref{pole2W}).
By equating these two (\ref{lhs}) and (\ref{rhs}) together with
(\ref{d33}),
we have determined the corresponding coefficient as follows:
\bea
d_{41} =
-\frac{ k (2 k+M) (k+2 N) (3 k+2 M+2 N)}{(k+M)^2 (k+2 M) (3 k+2 M)}
\, a_1.
\label{d41}
\eea
Now this can be substituted into the (\ref{seven}) again.

We consider the $d_{30}$ term where there exists
the derivative term $J^a \, \pa \, J^b(w)$.
Recall that there is a relation from the footnote \ref{derinK}.
On the one hand,
we have the following result
\bea
N \Bigg[ -6  \, a_{17} +\frac{4 }{M} \, a_{11}+
\frac{8  }{M}\,  a_8
-\frac{4 (4-M^2) }{M^2} \, a_5
 -
\frac{  2 (k+M+N)}{k M} \, a_{13} \Bigg].
\label{reld30}
\eea
There are two contributions from $a_{17}$ term
as before. For the contributions from
$a_8$ and $a_{11}$, the previous relation (\ref{ffrel})
is used. Note that by combining the contributions
in the coefficient $(a_{13}-a_{12})$ and the coefficient
$a_{12}$, the final contribution from $a_{12}$ term
becomes zero. Then we do not have any contributions
from $a_{12}$ term in (\ref{reld30}).
On the other hand, 
this should be equal to
\bea
N \, d_{30}.
\label{d30con}
\eea
Note that
in (\ref{reld30}), the relation of (\ref{ffrel})
is used in 
the second, third and fourth terms of (\ref{reld30}).
Then from (\ref{reld30}) and (\ref{d30con}),
the coefficient can be determined 
\bea
d_{30} =
-\frac{ (k^2+3 k M+M^2+4) (k+2 N)
  (3 k+2 M+2 N)}{k M (k+M) (3 k+2 M)} \, a_1,
\label{d30}
\eea
which can be substituted into the (\ref{seven}).

Let us look at the $d_{32}$ term where we have
the derivative term $J^b \, \pa J^a(w)$ with the footnote \ref{derinK}.
We can collect the possible terms
as follows:
\bea
&& N \Bigg[ \frac{2  (M^2+4) }{M} \, a_3 -
\frac{4  (M^2-4)  }{M^2} \, a_5
+\frac{2   (2 k M+M^2+2 M N-4)}{M} \, a_8
\nonu \\
&&-\frac{2   (k+M+N)}{k M} \, a_{12}
-\frac{2   (2 k+M+N)}{k M} \, a_{13} + 24  \, a_{17} \Bigg].
\label{d32relation1}
\eea
There are two contributions from
both $a_3$ term and $a_8$ term
and the final result can be written as above.
The contribution from $a_5$ term
also appears in $i \, f^{a d e} \, d^{b c d}\,
(K^e \, K^c)_{pole-1}(w)$.
Note that the additional contribution from $a_{12}$ term
can be found in $\frac{1}{2} \, (K^a \, K^b)_{pole-1}(w)$.
From the three places of $a_{17}$ term, the final
result for this coefficient is given above.
On the other hand, there exists
\bea
N \, \Big( -  \, d_{32} +\frac{4 }{M} \, d_{33} \Big).
\label{d32relation2}
\eea
Then we arrive at the
following result, by using (\ref{d32relation1}) and
(\ref{d32relation2}) which are equal to each other,
\bea
d_{32} =\frac{ (k^3-2 k^2 M-3 k M^2+4 k-M^3+4 M)
  (k+2 N) (3 k+2 M+2 N)}{k M (k+M)^2 (3 k+2 M)} \, a_1.
\label{d32}
\eea
Then this coefficient can be substituted in (\ref{seven}).

For the $d_{39}$ term, we have
the derivative term $d^{a c e} \, d^{e b d} \, J^c \pa \, J^d(w)$.
We can collect the possible terms as follows:
\bea
N \Bigg[ -\frac{(8 k+M^2 N+4 M)}{M (2 k+M)} \,a_5
+ 2 \, a_8 -\frac{ (2 k+M+2 N)}{2 (2 k+M)} \, a_{13}
+  \, a_{11} \Bigg].
\label{reld39}
\eea
The two contributions from $a_{12}$
are cancelled each other.
Similarly, those from $a_{17}$
can be also cancelled.
After simplifying the contributions from
the $a_{5}$ term, the net result comes from
$i \,f^{a d e} \, d^{b c d} \, (K^e \, K^c)_{pole-1}$
as above.
This should be  equal to
\bea
N \, d_{39}.
\label{39}
\eea
Then from (\ref{reld39}) and (\ref{39})
by taking them to be equal to each other, we obtain
\bea
d_{39} =-\frac{(k^2+k M+4) (k+2 N) (3 k+2 M+2 N)}{4 k (k+M) (3 k+2 M)}
\, a_1,
\label{d39}
\eea
which can be substituted into the (\ref{seven}).

Similarly,
the corresponding terms for the $d_{42}$ term
which involves various different kind of coefficients by
considering the term $f^{a b c}\, J^c \, \pa \, J^{u(1)}(w)$
can be obtained 
\bea
&& N \, \sqrt{\frac{M+N}{M N}}\,\Bigg[ \frac{4  M  }{k}\,  a_3
+\frac{2  (M^2-4)  (k+N)}{k M}\,
 a_5+
\frac{4   (k+N) }{k}\,  a_8
\nonu \\
&&-\frac{2  M  }{k} \,  a_{11}
-\frac{  (k+N) }{k} \, a_{12}
+\frac{ N }{k} \,  a_{13}
+\frac{4  }{k} \,  a_{16}
+ \frac{2  (M^2+6)  }{k}
\,  a_{17} \, \Bigg].
\label{d42rel}
\eea
The $a_{3}$ term and $a_{11}$ term
can be obtained by changing the
ordering of the two operators.
After we simplify all the contributions from
the $a_{5}$ term, the final result comes from
$i \,f^{a d e} \, d^{b c d} \, (K^e \, K^c)_{pole-1}$
as above. Note that there are contributions from
various places corresponding to the $a_{12}$ and $a_{13}$ terms.
Then the above should be equal to
\bea
 \frac{2 N}{k}\, \sqrt{\frac{M+N}{M N}}\, \Bigg[ \frac{1}{4} \,
   \frac{(k^2-4)(2k+M)(k+2N)(3k+2M+2N)}{2k(k+M)(3k+2M)}
   \,   a_1 
+  d_{42} \Bigg].
   \label{d42rel1}
\eea
Then it is easy to obtain the following result from
(\ref{d42rel}) and (\ref{d42rel1})
\bea
d_{42} =-\frac{ (k-2) (k+2) M (k+2 N) (3 k+2 M+2 N)}{8 k (k+M)
  (3 k+2 M)} \, a_1.
\label{d42}
\eea
This can be substituted in (\ref{seven}).

For the final coefficient, we use a little different method.
It is straightforward to calculate the OPEs between $T(z)$
and each term of (\ref{seven}) respectively.
The third order pole of these (except the $d_{41}$ term) has the form
$i \, f^{a b c} K^c(w)$.
By requiring that the expression (\ref{seven})
should be a quasi primary operator,
there exists for the vanishing of the third order pole 
\bea
d_{30} - d_{32} + M\, d_{33} -  \frac{(M^2-4)}{M} \, d_{38}
+  \frac{(M^2-4)}{M} \, d_{39} + 4\, d_{42}=0.
\label{d30d42}
\eea
This (\ref{d30d42}) implies that we obtain the coefficient
$d_{38}$ by using (\ref{d30}), (\ref{d32}), (\ref{d33}), (\ref{d39}),
and (\ref{d42}) as follows:
\bea
d_{38} =-\frac{ (k^2+k M+4)
  (k+2 N) (3 k+2 M+2 N)}{4 k (k+M) (3 k+2 M)} \, a_1,
\label{d38}
\eea
which can be substituted in (\ref{seven}).
We can also check the above result by following previous method
after extracting the corresponding terms from the second order pole. 

Therefore, by substituting (\ref{d30}),
(\ref{d32}), (\ref{d33}), (\ref{d38}), (\ref{d39}),
(\ref{d41}), and (\ref{d42})
into the previous expression
(\ref{seven}) the known quasi spin-$3$ operator
can be written as
\bea
Q^{a b}(w) & \equiv & \frac{N (k+2N) (
3k +2M+2N)}{6 (k+M)(3k+2M)} \, a_1\, \Bigg[
    -\frac{6 (4+k^2+3 k M+M^2)}{k M N} \, J^a \, K^b  \nonu \\
& + &
\frac{6 (4k +k^3 + 4 M-2 k^2 M-3 k M^2 -M^3)}{
k(k+M) M N} \,
J^b \, K^a \nonu \\
& + & \frac{3(2k+M)^2}{2(k+M)N} \,
f^{a c e} \, f^{b d e} \,
J^c \, K^d
+  \frac{3(4k +k^3+4M)}{2k(k+M)N}  \,
d^{a c e} \, d^{b d e}  \, J^d \, K^c \nonu \\
& - &  \frac{3(4+k^2+k M)}{2 k N} \,
d^{a c e} \, d^{b d e}  \, J^c \, K^d
\nonu \\
& - & \frac{6k(2k+M)}{(k+M)(k+2M)N} \,
\de^{a b} \,
J^c \, K^c
 -  \frac{3(k^2-4)M}{4k N} \,
i \, f^{a b c} \, \pa K^c \Bigg](w).
\label{Qab}
\eea
By multiplying $f$ or $d$ symbols into (\ref{Qab}),
we obtain the primary operator having a single index
\footnote{Under the large $k$ limit,
  the coefficients in (\ref{Qab}) become
  $\frac{ k (\la^2-4)}{3 \la^2 M}\, a_1$, $-\frac{ k
    (\la^2-4)}{3 \la^2 M}\, a_1$,$-\frac{ k
    (\la^2-4)}{3 \la^2 }\, a_1$,$-\frac{ k
    (\la^2-4)}{12 \la^2 }\, a_1$,$\frac{ k
    (\la^2-4)}{12 \la^2 }\, a_1$,$\frac{ 
    2(\la^2-4)}{3 \la^2 }\, a_1$, and $\frac{ k
    (\la^2-4) M}{24 \la^2 }\, a_1$ respectively.}.
We observe that the OPEs between the operator (\ref{Qab})
and $J^{u(1)}(w)$ (or $J^{\alpha}(w)$) are regular
because this operator consists of the spin-$1,2$ currents.
We can check the primary condition for the spin-$3$ operator
having the two indices
\bea
T(z) \, Q^{a b}(w) = \frac{1}{(z-w)^2} \, 3 \, Q^{a b}(w) + \frac{1}{(z-w)}
\, \pa \, Q^{a b}(w) + \cdots.
\label{TQ}
\eea

Therefore, the second order pole can be described as
\bea
&&
(K^a \, P^b)_{pole-2}(w) =
\frac{1}{4} \,
   \frac{(k^2-4)(2k+M)(k+2N)(3k+2M+2N)}{2k(k+M)(3k+2M)}
   \, a_1 \, i\, f^{a b c} \, \pa \, K^c(w)
   \nonu \\
   &&  - \de^{a b} \, \frac{a_1}{b_1} \, W^{(3)}(w)
   +\frac{k(3k+2M)(2k+M+2N)}{(k+M)(2k+M)} \, d^{a b c} \, P^c(w)
   +  Q^{a b}(w),
\label{KP2}
\eea
where the spin-$3$ primary (\ref{TQ})
operator $Q^{ab}(w)$ is given by (\ref{Qab}).
Compared to the third order pole in (\ref{KP3}) where
there exists the term $f^{a b c} \, K^c(w)$, 
the second order pole in (\ref{KP2}) has both $d^{a b c}\, P^c(w)$
and $\delta^{ab} \, W(w)$ terms which are symmetric under
the interchange between the index $a$ and $b$
as well as some descendant.
We expect that
this alternating feature will appear through the whole singular
terms in the given OPEs.

\subsection{The first order pole and charged
  quasi primary spin-$4$ current}

Compared to other singular terms described in previous
subsections, the first order pole 
can be obtained by simple contraction between the operators.
We present this in Appendix $E$.
We expect that there exists a new quasi spin-$4$ current
in this singular term.
In the third order pole, the field content is given by
$i \, f^{ a b c}\, K^c(w)$.
Along the line of this behavior,
by introducing the following quantity
\bea
i \, f^{a b c} \, (K^a \, P^b)_{pole-1}(w) \equiv R^c(w),
\label{Rdef}
\eea
and subtracting the
corresponding
quantity from the descendants with a multiplication
of $i \, f^{a b c}$,
a new quasi spin-$4$ current is given by
\footnote{We have the nontrivial fourth order pole
  in the OPE between $T(z)$ and $\hat{R}^a(w)$
  which is given by $
  -\frac{288  (k-2) (k+2)^2 (k+6) (k+10)}{5 k (k+4) (3 k+8)}\, a_1 \,
  K^a(w)$ for $(N,M)=(5,4)$. By adding
 $i\, f^{a b c} \, (J^b \, \pa \, K^c -2 \pa \, J^b \, K^c-\frac{i}{10} \,
  f^{b c d} \, \pa^2 \, K^d)(w)$ into
  this new quasi primary spin-$4$ current
  and removing the fourth order pole above, we can make
  a primary spin-$4$ current at least for $(N,M)=(5,4)$.
  For generic $(N,M)$ case, we should
  find out the above fourth order pole
  for the general case. Then
we can easily fix the above relative coefficient we want to add above.
  See also the footnote
  \ref{quasiandprimary}. }
\bea
\hat{R}^c(w) & \equiv &
R^c(w) -\frac{1}{3}\, i\, f^{a b c} \, \pa \, Q^{a b}(w) 
\nonu \\
&+ & \frac{M(k^2-4)(2k+M)(k+2N)(3k+2M+2N)}{20k(k+M)(3k+2M)}
   \, a_1 \, \pa^2 \, K^c(w),
\label{Rhatab}
\eea
where $Q^{ab}(w)$ is given by (\ref{Qab}).
When the $f$ symbols meet the Kronecker delta or $d$ symbols
by two index contractions,
we get zero.
Let us emphasize that
this new quasi spin-$4$ current is completely determined
via the left hand side of (\ref{Rdef}) from Appendix $E$ and
the two terms of the right hand side of (\ref{Rhatab}).

Then the first order pole is given by
\bea
(K^a \, P^b)_{pole-1}(w) & = &
\frac{1}{20} \,
   \frac{(k^2-4)(2k+M)(k+2N)(3k+2M+2N)}{2k(k+M)(3k+2M)}
   \, a_1 \, i\, f^{a b c} \, \pa^2 \, K^c(w)
   \nonu \\
   & - &  
\frac{1}{3}
   \de^{a b} \, \frac{a_1}{b_1} \, \pa \, W^{(3)}(w)
   +\frac{1}{3} \,
   \frac{k(3k+2M)(2k+M+2N)}{(k+M)(2k+M)} \, d^{a b c} \, \pa \, P^c(w)
   \nonu \\
   & + & \frac{1}{3} \pa \, Q^{a b}(w)
   + R^{a b}(w),
\label{KPpole1}
\eea
where we introduce the
operator $R^{a b}(w)$ which is given by
the first order pole subtracted by the descendant terms.
Then how we can connect this with the above quasi spin-$4$ current?
From the relation
\bea
(K^a \, P^b)_{pole-1}(z)=
-\frac{1}{2M} \, i\, f^{a b c } \, R^c(w) + S^{a b}(w),
\label{KPpole1other}
\eea
and by equating (\ref{KPpole1}) and (\ref{KPpole1other}) each other,
we can write down the above $R^{a b}(w)$ in terms of
$R^c(w)$, $S^{a b}(w)$ and other known operators.
Note that the above behavior can be seen from
the first order pole of the OPE between the spin-$2$ current and
itself (\ref{fullKK}).
This (\ref{KPpole1other})
can be seen from the fixed $(N,M)=(5,4)$ case.
In other words,
\bea
R^{a b} = -\frac{1}{2M} \, i\, f^{a b c } \, \hat{R}^c(w)
+ \mbox{other terms}.
\label{Rab}
\eea
Therefore, the $R^{a b}(w)$ contains the previous quasi spin-$4$
current in (\ref{Rhatab})
and can be treated as similar quasi spin-$4$ current
with two indices. We can easily see that
the overall factor  $\frac{1}{2M}$ in (\ref{Rab})
can be checked by multiplying $i \, f^{a b d}$
into (\ref{KPpole1other}) and using (\ref{Rdef})
and (\ref{Rhatab}).
When we substitute (\ref{Rab}) into (\ref{KPpole1})
and use (\ref{Rhatab}), then the above ``other terms'' can be read off
explicitly. Note that the left hand side of (\ref{KPpole1})
is given by
Appendix $E$ in terms of coset realization.

Although the expression of $S^{a  b}(w)$
can be determined for general $N$ and $M$
by following the procedure we have described in the construction of
$Q^{a b}(w)$ in (\ref{Qab}), it will be
rather nontrivial and complicated due to the fact that
there are tensorial structures having five indices.
Instead we present them for fixed $(N, M)=(5,4)$ as follows:
\bea
&& S^{a b}(w) \Bigg|_{N=5,M=4} 
\equiv -\frac{1}{3}\, \frac{a_1}{b_1}\,
\de^{a b} \, \pa \, W^{(3)}(w) + \frac{k (k+7)}{(k+4)}\,
d^{a b c}\, \pa P^c(w) 
\nonu \\
&& +\frac{ (k+7)}{2 (k+2) (k+4)}\,
\Bigg[-(3 k+8)\,
d_{4AA2}^{d c b a}+i\,  k \,
(d_{4SA}^{c a d b}+
d_{4SA}^{c b d a}) \Bigg] \, J^c \, P^d(w)
\nonu \\
&& +\frac{(k+6) (k+10)}{(k+4)^2 (3 k+8)}\,
a_1 \,\Bigg[-\frac{3   (7 k-4)}{40 k  }\,
  d_{51}^{d b e c a} -\frac{3   (7 k-4)}{
    40 k }  \, d_{51}^{e  a d c b}\nonu \\
  && +\frac{3 
    (5 k^2+33 k+4)}{80 k  } \,
  d_{51}^{e d c b a} -\frac{3 
    (20 k^2+59 k+12)}{40 k  } \, d_{52}^{d b e c a}
  \nonu \\
  && -\frac{3  (10 k^2+21 k-12)}{
    40 k  }  \, d_{52}^{e  b d c a}
  -\frac{3 (15 k^2+23 k-36)}{80 k  }
  \, d_{52}^{e d c b a} \nonu \\
  && + \frac{3 i  (k+2)}{4  }\,
  f^{a b c} \, \de^{d e}+\frac{3 i (k+4)  }{8 }
  \, f^{a b e} \, \de^{d c} +\frac{3 i (k+4) }{k  }\,
  f^{a e d} \, \de^{b c} \Bigg] \, J^c \, J^d \, K^e(w)
\nonu\\
&&+ \frac{(k+6) (k+10)}{ (k+4)^2 (3 k+8)}\,
a_1 \, \Bigg[
  -\frac{3   (10 k^3+75 k^2+346 k+568)}{40 k
   } \, \pa \, J^a\, K^b \nonu \\
  && -\frac{3(k+4)^2 }{8 }\,
  J^a \, \pa \, K^b  +\frac{3 
    (10 k^3-20 k^2-149 k-132)}{40 k }\,
  \pa \, J^b\, K^a
   + \frac{3   (k^2-4 k-16)}{8 } \,
  J^b \, \pa \, K^a
  \nonu \\
  && + \frac{3   (20 k^3+165 k^2+277 k-44)}{
    40 k }\,
  f^{a c e} \, f^{b d e} \, \pa J^c \, K^d  + \frac{3  k (k+2) }{4  }
  \,
  f^{a c e} \, f^{b d e} \,  J^c \, \pa \, K^d
  \nonu \\
  && +  \frac{3  (10 k^3+5 k^2-78 k-104)}{
    40 k}\,
  d^{a c e} \, d^{e b d} \,  \pa \, J^d  \, K^c  +
\frac{3  k^2 }{8  } \,
  d^{a c e} \, d^{e b d} \,  \, J^d  \, \pa \, K^c
  \nonu \\
  && 
-\frac{3  (2 k^3+6 k^2-9 k-20)}{8 k  }
  \,
  d^{a c e} \, d^{e b d} \,   \pa \, J^c   \, K^d
  -\frac{3  k (k+4)}{8  }
\,
  d^{a c e} \, d^{e b d} \,    J^c  \, \pa  \, K^d
  \nonu \\
  &&-\frac{3   (45 k^3+233 k^2+268 k+32)}{
    40 k  (k+8) } \, \de^{a b}  \, \pa J^c
  \, K^c  
  -\frac{3  k^2 }{2  (k+8) }
  \, \de^{a b}  \,  J^c
  \, \pa \, K^c
  \Bigg](w).
\label{Sab}
\eea
It is rather nontrivial to extract this expression
without any unwanted terms like as $J^{u(1)}(w)$,
$J^{\alpha}(w)$, $J^{(\rho \bar{i})}(w)$, or $J^{(\bar{\si} j)}(w)$
explicitly.
The other unwanted spin-$1$ currents can be absorbed in
the new quasi primary current in (\ref{Rhatab}).
Note that the operator contents in (\ref{Sab})
consist of the spin-$1,2,3$ currents and some of the derivative terms
can be seen from the derivative of $Q^{a b}(w)$ in (\ref{Qab}).
This implies that the right hand side of (\ref{Sab})
depends on the adjoint operators (or singlet operator) living in $SU(M)$
\footnote{We can check the following properties.
  The $J^a \, W^{(3)}(w)$ is a primary spin-$4$ operator and the
  $J^a \, P^b(w) -\frac{1}{6} \, i\, f^{a b c} \, \pa \,
  P^c(w)$ is a primary operator. On the other hand,
  $J^a \, \pa \, K^b(w) -2 \pa \, J^a \, K^b(w)-\frac{i}{10} \,
  f^{a b c} \, \pa^2 \, K^c(w)$ is quasi primary spin-$4$
  operator and the fourth order pole in the OPE between
  the stress energy tensor and this operator is given by
  $-\frac{21}{5}\, i \, f^{a b c} \, K^c(w)$.
\label{quasiandprimary}}.

Moreover, the OPE between the spin-$1$ current
and the above quasi primary
spin-$4$ current for fixed $(N,M)=(5,4)$ is described by
\bea
J^a(z) \, \hat{R}^b(w) & = &
-\frac{1}{(z-w)^3} \, \Bigg[ \frac{28
  (k-2) (k+2)^2 (k+6) (k+10)}{5 k (k+4) (3 k+8)}\,
a_1 \Bigg]\,  i \, f^{a b c} \, K^c(w)
\nonu \\
&+&
\frac{1}{(z-w)^2} \,  \Bigg[
  \frac{ (k+2) (k+6) (k+10)}{(k+4)^2 (3 k+8)}\,
  \Bigg(-\frac{(k+4) (7 k+4)}{k }\,
 a_1\, J^a \, K^b \nonu \\
  &- &
 \frac{  (11 k^2+16 k-16)}{k  } 
 \, a_1\, J^b \, K^a +
 \frac{6  (k-4) (k+2)}{k }
 \, a_1 \,
 f^{a c e}\, f^{b d e }  \, J^c \, K^d \nonu \\
 &-& \frac{ (5 k^2-8 k-16)}{k }
 \,
 a_1 \,d^{a c e}\, d^{e b d}\, J^d \, K^c
 \nonu \\
 &-& \frac{(k+4)^2}{k }
 \, a_1 \, d^{a c e} \, d^{e b d} \, J^c \, K^d
 - \frac{6  k^2 }{(k+8) }
 \, a_1 \, \de^{a b} \, J^c \, K^c \nonu \\
 &+& \frac{ (k-2) (k+4) (7 k+24)}{5 k }
 \, a_1 \, i\, f^{a b c} \, \pa \, K^c \Bigg) \nonu \\
 &-& \frac{4 k (k+7) (3 k+8)}{(k+2) (k+4)}\, d^{a b c} P^c
 \Bigg](w)  + 
\frac{1}{(z-w)} \, i \,
f^{a b c} \, \hat{R}^c(w) +\cdots.
\label{JRhat}
\eea
We observe that there exists a factor $(k+10)$
which comes from the factor $(k+2N)$.
The first order pole is what we  have expected.  
Contrary to the spin-$2,3$ currents,
there are more singular terms in addition to
the first order pole in (\ref{JRhat}).
In principle, the above calculation can be done for any $N$ and $M$
but it will take time to complete this computation.

\subsection{The final OPE}

In summary, we present the OPE between
the charged spin-$2$ current and the charged spin-$3$ current
as follows:
\bea
K^a(z) \, P^b(w) & = &
\frac{1}{(z-w)^3} \,
\Bigg[\frac{(k^2-4)(2k+M)(k+2N)(3k+2M+2N)}{2k(k+M)(3k+2M)}
  \Bigg]
\, a_1 \, i\, f^{a b c} \, K^c(w)
\nonu \\
&+& \frac{1}{(z-w)^2} \, \Bigg[
  \frac{1}{4} \,
   \frac{(k^2-4)(2k+M)(k+2N)(3k+2M+2N)}{2k(k+M)(3k+2M)}
   \, a_1 \, i\, f^{a b c} \, \pa \, K^c
   \nonu \\
   & + & Q^{a b} - \de^{a b} \, \frac{a_1}{b_1} \, W^{(3)}
+\frac{k(3k+2M)(2k+M+2N)}{(k+M)(2k+M)} \, d^{a b c} \, P^c
    \Bigg](w) 
\nonu \\
&+& \frac{1}{(z-w)} \, \Bigg[
 \frac{1}{20} \,
   \frac{(k^2-4)(2k+M)(k+2N)(3k+2M+2N)}{2k(k+M)(3k+2M)}
   \, a_1 \, i\, f^{a b c} \, \pa^2 \, K^c
   \nonu \\
   & + & \frac{1}{3} \pa \, Q^{a b} -
\frac{1}{3}
   \de^{a b} \, \frac{a_1}{b_1} \, \pa \, W^{(3)}
   +\frac{1}{3} \,
   \frac{k(3k+2M)(2k+M+2N)}{(k+M)(2k+M)} \, d^{a b c} \, \pa \, P^c
   \nonu \\
   &+& R^{a b} \Bigg](w) + \cdots.
\label{KPfinal}
\eea
In the right hand side of (\ref{KPfinal}), there exists
an overall factor $a_1$ \footnote{The structure constant in the third
  order becomes $-\frac{k^2(\la^2-4)}{3\la^2} \, a_1$
  when we take the infinity limit of $k$.
  The one appearing in charged spin-$3$ current of
  the second order pole becomes $\frac{3 k}{\la}$. }.
Although
the explicit form for 
the quasi primary spin-$4$ current $\hat{R}^c(w)$ is determined
via the known currents and coset operators,
the explicit form for the $R^{a b}$ for general $(N,M)$
is not known but it is known for $(N,M)=(5,4)$
because we do not know how Appendix (\ref{KPfirstorder})
can be written in terms of the known currents.
Their operator contents are known but
the relative coefficients are known for  $(N,M)=(5,4)$.
Or if we interpret $R^{a b}$ itself as
the whole new charged quasi primary spin-$4$ current
(without splitting the quasi primary spin-$4$ current
with a single free index and others), 
we do not worry about the fact that
this is written in terms of coset realization \footnote{
For the first order pole of the OPE in (\ref{fullKK}),
we can treat the sum of four quasi primary spin-$3$ operators
(after subtracting the descendant terms)
and a single primary spin-$3$ current as the  whole
single quasi primary spin-$3$ current having two free indices.
Then we do not need to specify the above four quasi primary
spin-$3$ currents in terms of multiple products between the
known currents.}.

\section{ The OPE between the charged  spin-$3$
  current and itself
with $(N,M)=(5,4)$}


\subsection{The sixth, fifth, fourth order poles}

For the sixth order pole we expect to have the
Kronecker delta $\de^{ab}$ term.
For the fifth order pole, there exists $i \, f ^{a b c} \, J^c(w)$ term.
For the fourth order pole, there are symmetric
$\de^{ a b}$ and $d^{a b c}$ tensor terms  and other symmetric
tensor terms in (\ref{fullKK}) as well as the descendant term. 

\subsection{The third order pole}

We can take the operator contents
in the first order pole of (\ref{fullKK})
at the third order pole of this OPE.
In other words, in addition to the descendant terms,
there are $ i \, f^{a b c} \, P^c(w)$
 and the quasi primary spin-$3$  operator 
 including the stress energy tensor.
Compared to (\ref{fullKK}),
the other two kinds of quasi primary spin-$3$ operators
do not appear in this OPE.

\subsection{The second and first order poles}

Then we obtain the OPE as follows:
\bea
&& P^a(z) \, P^b(w)  = 
\frac{1}{(z-w)^6}\,
\Bigg[\frac{15(k^2-4)(k^2-1)(k+6)(k+10)(2k+9)}{2k(k+4)(3k+8)}\Bigg]
\, a_1^2 \, \de^{a b}
\nonu \\
&& +
\frac{1}{(z-w)^5}\,
\Bigg[\frac{15(k^2-4)(k^2-1)(k+6)(k+10)(2k+9)}{2k^2
  (k+4)(3k+8)}\Bigg]
\, a_1^2 \, i \, f^{a b c} \, J^c(w)
\nonu \\
&& + 
\frac{1}{(z-w)^4} \, \Bigg[ \frac{3(k^2-4)(k+5)(k+6)(k+9)(k+10)}{4k(k+4)} \,
  a_1^2 \, \de^{a b} \, T \nonu \\
  &&+  \frac{3(k^2-4)(k+6)(k+7)(k+10)}{4(k+4)^2} \,
  a_1^2\, d^{a b c } \, K^c \nonu \\
  &&+ \frac{1}{2} \,
\frac{15(k^2-4)(k^2-1)(k+6)(k+10)(2k+9)}{2k^2
  (k+4)(3k+8)}\, a_1^2\,
  i\, f^{a b c} \, \pa \, J^c
  \nonu \\
  &&+ \frac{(k+6)(k+10)}{(k+4)(3k+8)}\, \Bigg( \frac{15(k-2)(2k+9)}{8} \,
  a_1^2 \, d^{a b c d}_{4SS1}  - 
  \frac{15(k^2-4)(2k+9)}{8k} \,
  a_1^2 \, 
  d^{a b c d}_{4SS2} \nonu \\
  && -  \frac{3(k-2)(180+191k+44k^2+3k^3)}{
  8k} \, a_1^2 \,  \de^{a b} \, \de^{c d}
   -  \frac{15(k^2-4)(2k+9)}{2k^2 } \, a_1^2
  \,  \de^{a c} \, \de^{b d}
  \Bigg) \nonu \\
  && \times\, \frac{1}{2} \, (J^c \, J^d + J^d \, J^c)
  \Bigg](w) \nonu \\
&&+ 
\frac{1}{(z-w)^3} \, \Bigg[
\frac{1}{2} \, \frac{3(k^2-4)(k+5)(k+6)(k+9)(k+10)}{4k(k+4)} \,
  a_1^2 \, \de^{a b} \, \pa \, T \nonu \\
  &&+  \frac{1}{2} \, \frac{3(k^2-4)(k+6)(k+7)(k+10)}{4(k+4)^2} \,
  a_1^2\, d^{a b c } \, \pa \, K^c \nonu \\
  &&+  
  \frac{1}{6} \,
\frac{15(k^2-4)(k^2-1)(k+6)(k+10)(2k+9)}{2k^2
  (k+4)(3k+8)}\, a_1^2\,
  i\, f^{a b c} \, \pa^2 \, J^c
  \nonu \\
  &&+ \frac{1}{2} \,
\frac{(k+6)(k+10)}{(k+4)(3k+8)}\,
  \Bigg( \frac{15(k-2)(2k+9)}{8} \,
  a_1^2 \, d^{a b c d}_{4SS1} - 
  \frac{15(k^2-4)(2k+9)}{8k} \,
  a_1^2 \, 
  d^{a b c d}_{4SS2} \nonu \\
  && -  \frac{3(k-2)(180+191k+44k^2+3k^3)}{
  8k} \, a_1^2 \,  \de^{a b} \, \de^{c d}
   -  \frac{15(k^2-4)(2k+9)}{2k^2} \, a_1^2
  \,  \de^{a c} \, \de^{b d}
  \Bigg) \nonu \\
  && \times\,  \frac{1}{2} \, \pa \, (J^c \, J^d + J^d \, J^c)
  \nonu \\
  && + \frac{ (15 k^6+278 k^5+1648 k^4+2208 k^3-10480 k^2
    -37088 k-34560)}{2 k (k+4)^2 (3 k+8)}
  \, a_1\, i \, f^{a b c} \, P^c \nonu \\
  && +  \frac{3  (k^2-4)  (k+5) (k+6) (k+9) (k+10)}{4 k^2 (k+4)}
  \, a_1^2 \, (T \, J^a -\frac{1}{2}\,
  \pa^2 \, J^a)\Bigg](w) \nonu \\
&& + {\cal O}\Bigg(\frac{1}{(z-w)^2}\Bigg)
+ \cdots.
\label{papb}
\eea
%
In the second order pole, there exist spin-$4$ quasi primary operators
in addition to the descendant terms as usual. 
We expect that there will be symmetric terms,
$\de^{a b} \, W^{(4)}(w)$ where the neutral
primary spin-$4$ current will be
presented in next section and $d^{a b c}\, \hat{R}^c(w)$
by recalling the second order pole of (\ref{KPfinal}).
We observe that in the second order pole 
there exists a  term $
J^{\alpha=1}\,  J^{\alpha=1}\,  J^{\alpha=22}\,  J^{\alpha=23}(w)$
which is one of the terms in  the neutral
primary spin-$4$ current $W^{(4)}(w)$ for the equal indices
$a=b$. 
So far we do not obtain the explicit form for the second order pole
due to the fact that there are two many candidates with various
tensorial structures at this singular terms.
       We expect that there is a new primary field of spin-$5$
in the first order pole.

\section{ The OPE between the uncharged higher spin-$3$
  current and itself}


\subsection{For fixed  $(N,M)=(5,4)$ case}


\subsubsection{The sixth, fifth, fourth and third order poles}

We expect that the highest order pole contains
the central term.
We observe that this contains the factor $(k+10)$
which is given by $(k+2N)$ for general $N$.
There will be no fifth order pole because we are considering
the OPE between the spin-$3$ current and itself.
The fourth order pole should contain the spin-$2$ current as usual.
It turns out that
there is also the quadratic term $J^a \, J^a(w)$.
Then the third order pole
should contain the descendant terms.

\subsubsection{The second and  first order poles
with the presence of uncharged primary spin-$4$ current}

The second order pole
can provide us to have a new  primary current of spin-$4$.
It turns out that
the final OPE for fixed $(N,M)=(5,4)$ is given by
\bea
W^{(3)}(z) \, W^{(3)}(w) & = &
\frac{1}{(z-w)^6} \frac{\hat{c}}{3}
+ \frac{1}{(z-w)^4} 2 \, \hat{T} (w)
+\frac{1}{(z-w)^3} \pa \, \hat{T} (w)
\nonu \\
&+& \frac{1}{(z-w)^2} \Bigg[ \frac{3}{10}\, \pa^2 \, \hat{T} +
  \frac{16 (k+4) (k+5) (k+9)}{3 (37 k^3+216 k^2+337 k+510)} \,
  \Big(\hat{T} \, \hat{T} -
  \frac{3}{10} \, \pa^2 \, \hat{T} \Big) \nonu \\
  & + &  W^{(4)} \Bigg](w)
\nonu \\
&+&
\frac{1}{(z-w)} \Bigg[
 \frac{1}{15}\, \pa^3 \, \hat{T} +
 \frac{1}{2} \,\frac{16 (k+4) (k+5) (k+9)}{3 (37 k^3+216 k^2+337 k+510)}\,
\pa \, \Big(\hat{T} \, \hat{T} -
\frac{3}{10} \, \pa^2 \, \hat{T} \Big)
\nonu \\
& + & \frac{1}{2}\, \pa \,  W^{(4)} \Bigg](w)
 +  \cdots,
\label{spin3spin3-1}
\eea
where
the $b_1^2$ of the overall constant in the neutral primary
spin-$3$ current
$W^{(3)}$ is fixed as follows:
\bea
b_1^2 = \frac{8k(k+8)}{27(k^2-4)(k+5)^2(k+6)(k+9)^2(k+10)}.
\label{b1b1}
\eea
In (\ref{b1b1}), the requirement we impose is
that the central term of (\ref{spin3spin3-1}) should be equal to
$\frac{\hat{c}}{3}$ with (\ref{chatThat}).
As described before, there exists a $J^a \, J^a(w)$ term in the
fourth order pole.
Note that the modified central charge and stress energy tensor
are given by
\bea
\hat{c} & \equiv & c- \frac{15 k }{(k+4)}= \frac{20(k^2-1)(2k+9)}{(k+4)(k+5)(k+9)},
\nonu \\
\hat{T} & \equiv & T- \frac{1}{2(k+4)} \, J^a \, J^a. 
\label{chatThat}
\eea
We can easily see that the OPE between the spin-$1$ current and
the modified stress energy tensor is regular 
\bea
J^a(z) \, \hat{T}(w) =0 + \cdots.
\label{Regular}
\eea
We can calculate the OPE between
the stress energy tensor and the  $J^a \, J^a(w)$ term
and this leads to the central charge $\frac{15 k }{(k+4)}$.
Due to the fact that there is a relation in (\ref{Regular}),
we obtain the modified central charge is given by (\ref{chatThat}).
Therefore, the spin-$1$ current is decoupled from the modified
stress energy tensor according to (\ref{Regular}).
See also \cite{AP1812} where the $U(1)$ spin-$1$ current is
decoupled
from the stress energy tensor, spin-$3,4$ currents in the specific
model.

We have explicit form for the primary spin-$4$ current \footnote{
  Then we have
  $T(z) \, W^{(4)}(w) =\frac{1}{(z-w)^2}\, 4 W^{(4)}(w) +
  \frac{1}{(z-w)}\, \pa \, W^{(4)}(w) +\cdots $.}
as follows:
\bea
W^{(4)}(w) & = &
-\frac{9  (k+2) (k+5) (k+9) (53 k^4+800 k^3+3409 k^2+3078 k-3600)}{
  2 k (k+8) (37 k^3+216 k^2+337 k+510)}
\, b_1^2 \nonu \\
&\times & J^{\alpha=1}\,  J^{\alpha=1}\,  J^{\alpha=1}\,  J^{\alpha=1}(w)
+\mbox{other 9330 terms},
\label{w4}
\eea
together with (\ref{b1b1}).
We can also check that the regularity between the spin-$1$
current and this spin-$4$ current (\ref{w4})
\bea
J^a(z) \, W^{(4)}(w) = 0 + \cdots. 
\label{regular1}
\eea
This implies that the spin-$1$ current is decoupled.
We will observe in next section that
 the regularity between the spin-$1$
current and this spin-$3$ current
\bea
J^a(z) \, W^{(3)}(w) = 0 + \cdots. 
\label{regular2}
\eea
This fact can be seen from the closure of the OPE between
$K^a(z)$ and $W^{(3)}(w)$.

Therefore, we have the spin-$2,3,4$ currents $\hat{T}$, $W^{(3)}$
and $W^{(4)}$,
having the regularity
behavior in (\ref{Regular}), (\ref{regular1}) and (\ref{regular2}).
The OPE between the spin-$2$ current and itself
takes the standard form with modified central charge. 
The (quasi)primary condition under the stress energy tensor
is preserved when we modify the stress energy tensor
because the $J^a \, J^a(z)$ term does not spoil the spin-$3,4$
currents according to the regularity.

\subsection{For general  $(N,M)$ case}

The coefficient appearing in the
the quasi spin-$4$ operator
$(\hat{T} \, \hat{T} -\frac{3}{10} \, \pa^2 \,
\hat{T})$
is fractional function of $k$
and both numerator and denominator 
are polynomials of $k$. The highest power is given by
three. We can express this coefficient in terms of
the central charge. It turns out that
this is equal to the well known quantity
$\frac{32}{(5\hat{c} +22)}$ for fixed $N$ and $M$
with (\ref{chatThat}).
Because the structure constant depends only on the central charge,
we expect that when we change the different values
of $N$ and $M$, the OPE of (\ref{spin3spin3-1})
still satisfies together with the corresponding
central charge. The structure constants do not change
and are given by function of central charge as above. 
Therefore, we obtain the general OPE for arbitrary
$N$ and $M$ by realizing 
modified central charge written in terms of $N$, $M$ and $k$.  

We claim that the OPE between the neutral spin-$3$ current and itself
is described as
\bea
W^{(3)}(z) \, W^{(3)}(w) & = &
\frac{1}{(z-w)^6} \frac{\hat{c}}{3}
+ \frac{1}{(z-w)^4} 2 \, \hat{T} (w)
 +\frac{1}{(z-w)^3} \pa \, \hat{T} (w)
\nonu \\
&+& \frac{1}{(z-w)^2} \Bigg[ \frac{3}{10}\, \pa^2 \, \hat{T} +
  \frac{32}{(5\hat{c} +22)} \Big(\hat{T} \, \hat{T} -
  \frac{3}{10} \, \pa^2 \, \hat{T} \Big) +  W^{(4)} \Bigg](w)
\nonu \\
&+&
\frac{1}{(z-w)} \Bigg[
 \frac{1}{15}\, \pa^3 \, \hat{T} +
 \frac{1}{2} \,
 \frac{32}{(5\hat{c} +22)} \pa \, \Big(\hat{T} \, \hat{T} -
 \frac{3}{10} \, \pa^2 \, \hat{T} \Big) +
 \frac{1}{2}\, \pa \,  W^{(4)} \Bigg](w)
\nonu \\
& + & \cdots.
\label{Spin3Spin3}
\eea
where
the modified central charge and modified stress
energy tensor for generic $(N,M)$ are given by
\bea
\hat{c} & = & c-\frac{k (M^2-1)}{(k+M)} =
\frac{(k^2-1)M N (2k +M+N)}{(k+M)(k+N)(k+M+N)},
\nonu \\
\hat{T}(w) & = & 
T(w)- \frac{1}{2(k+M)} \, J^a \, J^a(w). 
\label{modified}
\eea
The central charge in (\ref{Spin3Spin3}) is fixed by
manipulating the overall constant $b_1^2$ in
the spin-$3$ current (\ref{W}).
Once we fix the structure constant in the fourth order pole
as two, then the corresponding descendant terms with known
coefficients are determined automatically.
Moreover, the first order pole can be determined
from the information of the second order pole.
Because the additional term in the
modified central charge in (\ref{modified}) under the infinity limit of
$k$ contributes to $-(M^2-1)$ which can be ignored,
the modified central charge
behaves as $ \hat{c} \rightarrow  M \, (1-\lambda^2) \, k$
where $\lambda \equiv \frac{k}{(k+N)}$ \cite{CH1812}.
The coefficient of the quadratic $J^a \, J^a(w)$ in (\ref{modified})
becomes $-\frac{1}{2 k}$ under the infinity limit of $k$.
We expect from the experience of \cite{Ahn1111,AK1308}
that the neutral primary spin-$4$ current contains
$d^{a b e}\, d^{c d e}\, J^a \, J^b \, J^c \, J^d(z)$
as well as other terms.

\section{ The OPE between the charged (higher) spin
  currents and the uncharged higher spin-$3$
  current}


\subsection{  The OPE  $J^a(z) \, W^{(3)}(w)$}

We can calculate the OPE between $J^a(z)$ and $W^{(3)}(w)$.
The third order pole is given by
$\Big[2(k+M) \, a_{10}+ N \, (a_{12}+a_{13})\Big]\, J^a(w)$
which vanishes by imposing the condition (\ref{interbvalue}).
Furthermore it turns out, under the condition of (\ref{interbvalue}),  that
the nontrivial second order pole is
\bea
J^a(z) \, W^{(3)}(w)  & = & \frac{1}{(z-w)^2} \,
\Bigg[ \frac{N}{2(2k+M)} \, (-3(k+N) \, (k+2N) \, b_1 + M \, (k+2M) \,
  b_7)
  \nonu \\
  & + & \frac{M}{2} \, (3(k+M) \, b_2 + N \, b_7)\Bigg]
\, d^{a b c } \, J^b \, J^c(w) + \cdots. 
\label{JWpole2}
\eea
According to the discussion of next subsection,
the coefficients $b_2$ and $b_7$ can be determined completely
in terms of $b_1$ and leads to the vanishing of
second order pole in (\ref{JWpole2}).
Therefore, the spin-$1$ current is decoupled. In other words,
\bea
J^a(z) \, W^{(3)}(w)  & = & 0  + \cdots. 
\label{regularjw}
\eea
In addition to the modified stress energy tensor (\ref{modified})
with (\ref{Regular}) and the spin-$4$ current (\ref{w4})
with (\ref{regular1}),
this neutral spin-$3$ current with (\ref{regularjw})
belongs to the generators
of $W$ algebra.

\subsection{  The OPE  $K^a(z) \, W^{(3)}(w)$}

For the calculation of $b_7$ term in the second order pole
of the OPE $K^a(z) \, W^{(3)}(w)$, the following identity
\bea
\mbox{Tr} (t^a \, t^b \, t^c \, t^d) = \frac{1}{M} \, \de^{a b}\, \de^{cd}
+ \frac{1}{4} \, (i \, f + d)^{a b e} \, (i \, f +d)^{e c d}
\label{traceabcd}
\eea
is used. 
It is rather nontrivial to calculate the OPEs between
the composite operators (evaluated at $z$) appearing
in the first order pole in the first equation of (\ref{KJJ})
and $\de_{\rho \bar{\si}}\, t^b_{j \bar{i}}\, J^{(\bar{\si} j)}(w)$.

We focus on the particular singular terms in the second order pole.
It is rather nontrivial to calculate the contributions from
$b_7$ and $b_8$ terms. 
We can collect $J^a \, J^{u(1)} \, J^{u(1)}(w)$ term coming from $b_5, b_7$
and $b_8$ terms
as follows:
\bea
&& 2M \frac{2 N}{k} \sqrt{\frac{M+N}{M N}} b_5
-\frac{4}{k M} (k+N)(M+N) b_7 + \frac{4}{k} N (2k +M+2N)
\sqrt{\frac{M+N}{M N}} b_8 \nonu \\
&& = C_{K^a \, W}^{P^a} \, a_4,
\label{JJu1Ju1}
\eea
where the corresponding coefficient of $P^a(w)$ is given by $a_4$
and the coefficient of $P^a(w)$ is denoted by $C_{K^a \, W}^{P^a}$
we should determine.
For the $d^{a b c} \, J^b \, J^c \, J^{u(1)}(w)$ term coming from $b_2, b_5, b_7$
and $b_8$ terms
we have
\bea
&& 3 M \frac{2N}{k}  \sqrt{\frac{M+N}{M N}} b_2 - 2M \frac{N}{(2k+M)} b_5
+ \frac{2N(4k^2+ 2 k M + 4 k N + M N)}{k(2k+M)}  \sqrt{\frac{M+N}{M N}} b_7
\nonu \\
&& -  2N \frac{(2k+M+2N)}{(2k+M)} \, b_8 = C_{K^a \, W}^{P^a} \, a_9,
\label{JJJu1}
\eea
where the corresponding coefficient of $P^a(w)$ is given by $a_9$
in the right hand side.
Moreover,
the $d^{a b c} \, d^{c d e} \, J^b \, J^d \, J^e(w)$ term
leads to the following relation
\bea
- 3 M \, \frac{N}{(2k+M)} \, b_2 - N \, \frac{(2k+M+2N)}{(2k+M)} \,
b_7
= C_{K^a \, W}^{P^a} \, \frac{3}{2} \, a_{17},
\label{JJJ}
\eea
where the corresponding coefficient of $P^a(w)$ is $\frac{3}{2} a_{17}$. 
By solving the equations (\ref{JJu1Ju1}), (\ref{JJJu1}) and (\ref{JJJ})
together with (\ref{avalues}) and (\ref{interbvalue}), we obtain
\bea
b_2  & = &  -\frac{N(k+N)(k+2N)}{M(k+M)(k+2M)} \, b_1,
\qquad
b_7 = \frac{3(k+N)(k+2N)}{M(k+2M)} \, b_1,
\nonu \\
C_{K^a \, W}^{P^a} & = & - \frac{12(3k+2M)(k+N)(k+M+N)}{M(k+2M) } \, \frac{b_1}{a_1}. 
\label{b2b7X}
\eea
Then all the coefficients in the neutral spin-$3$ current are
completely fixed.
See Appendix $F$.

From the term of
$ i\, f^{a b c}\, \de_{\rho \bar{\si}}\, t^c_{j \bar{i}} \,
J^b \, J^{u(1)}\, J^{(\rho\bar{i})}\, J^{\bar{\si} j}(w)$
associated with $a_7$ term of (\ref{spin3exp}),
the following relation satisfies
\bea
2 \, a_7 \, C_{K^a \, W}^{J^b \,P^c}  = 4\, \Bigg[ b_5 +
\sqrt{\frac{M+N}{M N}}\, b_7-b_8 \Bigg].
\label{rreal}
\eea
Then by substituting (\ref{avalues}), (\ref{interbvalue})
and (\ref{b2b7X}) into (\ref{rreal}),
we obtain the structure constant
\bea
C_{K^a \, W}^{J^b \,P^c} = \frac{24
  (k+N) (k+M+N)}{ M(k+2M)}\, \frac{b_1}{a_1}.
\label{stru}
\eea
Also other terms can be checked.
See Appendix $G$.

Therefore, we have the following OPE between the charged spin-$2$
current and the uncharged spin-$3$ current
\bea
K^a(z) \, W^{(3)}(w) & = & -\frac{1}{(z-w)^2} \,
\Bigg[
  \frac{12(3k+2M)(k+N)(k+M+N)}{M(k+2M) } \Bigg]
\, \frac{b_1}{a_1} \, P^a(w)
\nonu \\
& + &
\frac{1}{(z-w)} \Bigg[ -\frac{1}{3}\,
  \frac{12(3k+2M)(k+N)(k+M+N)}{M(k+2M) } \, \frac{b_1}{a_1} \, \pa \, P^a
  \nonu \\
  & + &   \frac{24   (k+N) (k+M+N)}{ M(k+2M)}\, \frac{b_1}{a_1} \,
\Big(i \, f^{a b c} \, J^b \, P^c +\frac{M}{3} \pa \, P^a\, \Big) \, \Bigg](w) \nonu \\
& + & \cdots,
\label{KW}
\eea
where the relation (\ref{stru}) is used.
Note that the last line in (\ref{KW}) is a primary operator
written in terms of the known spin-$1$ and spin-$3$ currents. 
Compared to the one in (\ref{KPfinal}),
the OPE structure is rather simple because
in this case, there exists only one free index.
Under the large $k$ limit, the structure constant
in the second order pole becomes $\frac{36k^2}{\la^2 M}$
and the one in the last line of (\ref{KW}) leads to
$\frac{24 k}{\la^2 M}$.

\subsection{ The OPE  $P^a(z) \, W^{(3)}(w)$ with $(N,M)=(5,4)$ }


\subsubsection{The sixth, fifth, fourth
  and third order poles}

Because the free index of this OPE
is given by the index $a$, there will  no singular terms
in sixth and fifth order poles.
The nonzero singular terms appear in the fourth 
order pole. The natural candidate is given by the charged spin-$2$
current $K^a(w)$.
In the third order pole, 
there will be a quasi charged spin-$3$ operator
in addition to the descendant term.
It turns out that 
there exists a $i\, f^{a b c}\, J^b \, K^c(w)$ term with
derivative term which is a primary.

\subsubsection{The second and first order poles}

The quasi charged spin-$4$ current can also arise
and the composite operators between the spin-$1$ operator
and the spin-$3$ current with appropriate tensor structures occur.
We can consider the derivative terms with free index $a$
without any difficulty.

We summarize the OPE as follows:
\bea
&& P^a(z) \, W^{(3)}(w)  = 
\nonu \\
&& -  \frac{1}{(z-w)^4} \,
\Bigg[\frac{9(k-2)(k+2)^2(k+5)(k+6)(k+9)(k+10)}{4 k (k+4)(k+8)}\Bigg]
\, a_1 \, b_1\, K^a(w)
\nonu \\
&& + 
\frac{1}{(z-w)^3} \, \Bigg[
- \frac{1}{2} \, \frac{9(k-2)(k+2)^2(k+5)(k+6)(k+9)(k+10)}{4 k (k+4)(k+8)}
\, a_1 \, b_1\, \pa \, K^a \nonu \\
&& + 
 \frac{9(k-2)(k+2)(k+5)(k+6)(k+9)(k+10)}{8 k (k+4)(k+8)}
\, a_1 \, b_1\, \nonu \\
& & \times  \Big( i \, f^{a b c} \, J^b \, K^c + \frac{M}{2}  \, \pa K^a \Big)
\Bigg](w)
\nonu \\
&& + 
\frac{1}{(z-w)^2} \, \Bigg[
- \frac{3}{20} \, \frac{9(k-2)(k+2)^2(k+5)(k+6)(k+9)(k+10)}{4 k (k+4)(k+8)}
\, a_1 \, b_1\, \pa^2 \, K^a \nonu \\
&& + 
\frac{1}{2} \,
 \frac{9(k-2)(k+2)(k+5)(k+6)(k+9)(k+10)}{8 k (k+4)(k+8)}
\, a_1 \, b_1\, \nonu \\
&& \times  \pa \,
\Big( i\, f^{a b c} \, J^b \, K^c + \frac{M}{2}  \, \pa K^a \Big)
+ R_2^a \, \Bigg](w)
+  {\cal O}\Bigg(\frac{1}{(z-w)}\Bigg)+
\cdots.
\label{PW}
\eea
Note that there is a $(k+10)$ factor.
A quasi primary spin-$4$ takes the form 
\bea
R_2^a(w) & \equiv & \frac{ 3(k+5) (k+9)}{(k+8)} \,
b_1\, \Bigg[ \frac{  (k+4)}{8 }\,
 R^a(w)  -\frac{  (k+4) }{k }\,
\frac{a_1}{b_1}\, J^a \ W^{(3)}(w) \nonu \\
& + & \frac{  k  (k+7) (3 k+8)}{2 (k+2) (k+4)}
\, d^{a b c} \, J^b\, P^c(w) \Bigg]
\nonu \\
& + & \frac{9(k+2)(k+5) (k+6) (k+9) (k+10)}{(k+4) (k+8)(3 k+8)}\,
 a_1\, b_1 \,
\Bigg[
  \frac{  k  }{8 (k+4) } \, d_{4SS2}^{b c d a}\, J^b \, J^c\, K^d
\nonu \\
& + &
\frac{    (k^2+16)}{16 k (k+4)  }\,
d_{4AA1}^{b a c d}\ J^b \, J^c \, K^d
+  \frac{   (k^2+6 k+16)}{4 (k+4) (k+8) }
\,  \de^{b a}\, \de^{c d} \, J^b \, J^c \,K^d
\nonu \\
& + & \frac{1   }{8  }\, 
\de^{b c} \, \de^{d a} \, J^b \, J^c \, K^d
+  \frac{  (k-2)   (k^2+3 k-8)}{20 k }\, \pa^2 \, K^a\nonu \\
& - & \frac{   (k-2) (k+8)  }{16 k  }
\, i\, f^{a b c}  \,J^b \, \pa \, K^c -  \frac{ 
  (k^2+k-16)}{8 k  }
\, i\, f^{a b c}  \, \pa \, J^b  \, K^c \Bigg](w),
\label{r2a}
\eea
where $R^a(w)$ is given by (\ref{Rdef}).
Again by using the relation (\ref{Rhatab}),
we can rewrite the above in terms of quasi primary spin-$4$ current.
Compared to the previous OPE between the spin-$3$ current and itself
(\ref{papb}), the OPE structure is rather simple.
We expect that
in the first order pole there will be  no new (quasi)primary field.
Although the construction in (\ref{KPfinal}) provides the information
on the quasi primary spin-$4$ current, due to the presence of
free two indices, we should multiply the $f$ symbols into the
first order pole. On the other hands, the construction in (\ref{PW})
is rather complicated because the spins of the left hand side are
given by three and three. Nonetheless, due to the one single
free index, once we have determined the second order pole, then
the quasi primary spin-$4$ current is determined without manipulating
further. After subtracting the descendant terms, we are left with
the quasi primary spin-$4$ current.

\section{ Conclusions and outlook}

In particular,
we have constructed i) the OPE between the charged spin-$2$ current
and itself in (\ref{fullKK}) with (\ref{KKcoeff}),
ii) the OPE between the charged spin-$2$ current and the
charged spin-$3$ current in (\ref{KPfinal}) where the first order pole
is known for $(N,M)=(5,4)$ case by rearranging it in terms of
the known operators,
iii) the OPE between the neutral spin-$3$ current and itself
(\ref{Spin3Spin3}) where the neutral primary spin-$4$ current
is known for $(N,M)=(5,4)$ and iv) the OPE
between the charged spin-$2$ current and the neutral spin-$3$
current in (\ref{KW}).

In doing this, we have determined the charged
quasi primary spin-$4$ current
in (\ref{Rhatab}) together with
(\ref{Rdef}) and Appendix (\ref{KPfirstorder})
in terms of coset realization completely.
In the OPE between the charged spin-$3$ current
and the neutral spin-$3$ current for fixed $(N,M)=(5,4)$
values, we have checked that the above
 charged quasi primary spin-$4$ current
(\ref{Rhatab}) occurs at the second order pole of this OPE.
 We have some evidence for the presence of the above
 neutral primary spin-$4$ current in the second order pole in the
 OPE between the charged spin-$3$ current and itself
 for fixed $(N,M)=(5,4)$ by focusing on the particular nontrivial
 term.

 Under the presence of the charged higher spin currents, 
the algebra obtained from the whole charged and neutral higher spin
currents leads to the one in an extension of \cite{GG1011}.
The algebra coming from the neutral ones is closed. Its extension
is closed and the right hand side contains
 the whole charged and neutral higher spin
currents in general.

We list the possible open problems  along the line of this paper
as follows:

\begin{itemize}
  
\item More OPEs

  So far, the charged spin-$2,3,4$ currents
  and the neutral spin-$3$ current
  are known in terms of coset realization.
  It is an open problem to determine
  the neutral spin-$4$ current in terms of coset realization
  for generic $(N,M)$. Moreover, some of the OPEs we have presented
  in this paper
  do not have their complete expressions. In doing this, the
  new quasi primary spin-$5$ current will be determined.
  In the bulk theory side, it is an open problem to construct
  an extension of the higher spin algebra studied in
  \cite{AK2009,AKK1910} for general $M$
  by adding the $SO(2 N M)$ factor in
  the numerator of the coset (\ref{coset}).
  It is better to oberve how the case $M=2$ and the case $M=4$,
  where the nontrivial $SU(M)$ invariant tensors can occur,
  appear explicitly.
  

\item Three point functions
  
Because the charged spin-$2,3,4$ currents
  and the neutral spin-$3$ current
  are known explicitly, it is natural to
  ask what are the three-point functions
  by evaluating the zero mode eigenvalue equations of these
  currents in the large $N$ limit.
  The relevant primary states in the coset (\ref{coset})
  are given by $(\Lambda_{N+M}; \Lambda_N, m)$
  where $\Lambda_L$ represents the highest weight of
  $SU(L)$ and $m$ is the $U(1)$ charge \cite{CH1906}.
  Recall that the previous relevant works are given in
  \cite{GH,CY,AKP,Ahn1111,AK1308} and we will keep track of
  the nonsinglet parts of the construction.
  The nontrivial part is to identify the $SU(M)$ adjoint indices
  in the three point functions explicitly.
  
\item Orthogonal group
  
  So far we have considered the special unitary group
  in the coset model. We can apply the present results
  for the unitary group to the orthogonal group \cite{CHU1906-1}
  where they decompose the $SU(M)$ generators into $\frac{M(M-1)}{2}$
  antisymmetric matrices and $(M^2-1)-\frac{M(M-1)}{2}$ traceless
  symmetric matrices. For the former, we do have spin-$1$ current
  and for the latter, we can associate with the spin-$2$ current.
  Then the nontrivial OPE between the spin-$1$ current and the spin-$2$
  current will give us the nontrivial structure constant whose indices
  are mixed together at the
  first order pole. This will be an extension of \cite{Ahn1106,GV}.
  We need to classify the various invariant tensors in this
  context correctly. 
  
\item Supersymmetric case
  
  By the additional $SO(2 N M)$ factor, which leads to
  $N M$ complex fermions, in the numerator of
  (\ref{coset}), the ${\cal N}=2$ supersymmetric model is studied
  in \cite{CH1906} where the spin contents are given by
  one $U(1)$ spin-$1$ current, two $(M^2-1)$ spin-$1$ currents,
  $2M^2$ spin-$s$ currents ($s=2,3, \cdots, n$),
  $M^2$ spin-$(n+1)$ currents and $2M^2$ spin-$(s-\frac{1}{2})$
  currents ($s=2, 3, \cdots, (n+1)$). Note that
  the standard $U(1)$ spin-$1$ current,
  two spin-$\frac{3}{2}$ currents
  and spin-$2$ stress energy tensor of
  ${\cal N}=2$ superconformal algebra can
  be seen from the above spin contents.
  It is natural to observe how the previous works in
  \cite{Ahn1206,Ahn1208} can be generalized in this enlarged model.
  Furthermore, for the particular level
  $k=N$ or $k=N+M$ \cite{CHR1406},
  we expect to have the supersymmetric models 
  and it is an open problem to observe how an extension of
  \cite{Ahn1211,Ahn1305} arises. See also the relevant work in
  \cite{AK1607} for different supersymmetry
  and there are some partial lists on the
  supersymmetric cases in
  \cite{CG1203,Henneaux:2012ny,
Hanaki:2012yf,Candu:2012tr,
Creutzig:2012xb,Gaberdiel:2013vva,Gaberdiel:2014vca,
Datta:2017ert,
Eberhardt:2018plx,Castroetal}.
  Due to the complex fermions, the (higher) spin currents
  will contain the bosonic currents as well as these complex
  fermions.
  Moreover, it is known that under the superalgebra
  description on \cite{BBSS}, we have similar coset construction.
  Then it is an open problem to consider the coset construction
  \cite{CH1906}
  where the numerator is given by the superalgebra.
  
  
\end{itemize}

\vspace{.7cm}

\centerline{\bf Acknowledgments}

We would like to thank  C. Peng for the general discussion
on the higher spin square \cite{GG1406,GG1501,GG1512} and
Y. Hikida for the discussions on his papers
\cite{CHR1306,CH1812,CH1906,CHR1406}. 
This work was supported by
the National Research Foundation of Korea(NRF) grant
funded by the Korea government(MSIT)(No. 2020R1F1A1066893).
CA acknowledges warm hospitality from 
the School of  Liberal Arts (and Institute of Convergence Fundamental
Studies), Seoul National University of Science and Technology.

\newpage

\appendix

\renewcommand{\theequation}{\Alph{section}\mbox{.}\arabic{equation}}

\section{ An $SU(M)$ invariant tensors in terms of Kronecker
  delta, $f$ and $d$ symbols }

Let us present the various $SU(M)$ invariant tensors
in terms of $f$ and $d$ symbols of rank $3$
\bea
d_{4SS1}^{a b c d} & = & \frac{4}{M} \, \de^{a b} \, \de^{c d} + d^{a b e} \,
d^{e c d},
\nonu \\
d_{4SS2}^{a b c d} & = & \frac{2}{M} \, \de^{a d} \, \de^{b c}
+\frac{2}{M} \, \de^{a c} \, \de^{b d} - \frac{1}{2} \,f^{a c e} \,
f^{e b d} + \frac{i}{2}  \, f^{a c e} \, d^{e b d} + \frac{i}{2} \,
 d^{a c e} \, f^{e b d}+ \frac{1}{2} \, d^{a c e} \,
d^{e b d}
\nonu \\
&- & \frac{1}{2} \,f^{b c e} \,
f^{e a d} + \frac{i}{2}  \, f^{b c e} \, d^{e a d} + \frac{i}{2} \,
 d^{b c e} \, f^{e a d}+ \frac{1}{2} \, d^{b c e} \,
d^{e a d},
\nonu \\
d_{4SA}^{a b c d} & = & d^{a b e}\, f^{e c d},
\qquad
d_{4AA1}^{a b c d}  =  f^{a b e} \, f^{e cd},
\nonu \\
d_{4AA2}^{a b c d} & = & \frac{2}{M} \, \de^{a c} \, \de^{b d}-
\frac{2}{M} \, \de^{a d} \, \de^{b c} -\frac{1}{2} \, f^{a c e} \, f^{e b d} +
\frac{1}{2} \, f^{a d e} \, f^{e b c} + \frac{i}{2} \, f^{a c e} \, d^{e b d}
\nonu \\
&-& \frac{i}{2} \, f^{a d e} \, d^{e b c} +
\frac{i}{2} \, d^{a c e} \, f^{e b d} - \frac{i}{2} \, d^{a d e} \, f^{e b c}
 +
 \frac{1}{2} \, d^{a c e} \, d^{e b d} - \frac{1}{2} \, d^{a d e} \, d^{e b c},
 \nonu \\
 d_{51}^{a b c d e} & = &
 \frac{2}{M} \, \de^{f c} \, \de^{d e}
+ 
\frac{2}{M} \, \de^{f e} \, \de^{c d}
+
\frac{2}{M} \, \de^{f d} \, \de^{c e} + \frac{i}{2} \, f^{f c g} \, d^{g d e}
+ 
\frac{1}{2} \, d^{f c g} \, d^{g d e}
\nonu \\
& + & \frac{i}{2} \, f^{f e g} \, d^{g c d}
+ 
\frac{1}{2} \, d^{f e g} \, d^{g c d}  + \frac{i}{2} \, f^{f d g} \, d^{g c e}
+ 
\frac{1}{2} \, d^{f d g} \, d^{g c e},
\nonu \\
d_{52}^{a b c d e} & = & d_{51}^{a b c d e} + i \, f^{c b f} \, \Big(
\frac{1}{M} \de^{a f} \, \de^{d e} + \frac{1}{4} (i f +d)^{a f g}(i f +d)^{g d e}\Big) \nonu \\
& - & i \, f^{c a f} \, \Big(
\frac{1}{M} \de^{b f} \, \de^{d e} + \frac{1}{4}
(i f +d)^{b f g}(i f +d)^{g d e}\Big)\nonu \\
&+& i \, f^{c b f} \, \Big(
\frac{1}{M} \de^{a f} \, \de^{e d} +
\frac{1}{4} (i f +d)^{a f g}(i f +d)^{g e d}\Big)
\nonu \\
& - & i \, f^{c a f} \, \Big(
\frac{1}{M} \de^{b f} \, \de^{e d} + \frac{1}{4}
(i f +d)^{b f g}(i f +d)^{g e d}\Big)\nonu \\
&+& i \, f^{e b f} \, \Big(
\frac{1}{M} \de^{a f} \, \de^{c d} +
\frac{1}{4} (i f +d)^{a f g}(i f +d)^{g c d}\Big)
\nonu \\
&-& i \, f^{e a f} \, \Big(
\frac{1}{M} \de^{b f} \, \de^{c d} +
\frac{1}{4} (i f +d)^{b f g}(i f +d)^{g c d}\Big)\nonu \\
&+& i \, f^{d b f} \, \Big(
\frac{1}{M} \de^{a f} \, \de^{c e} +
\frac{1}{4} (i f +d)^{a f g}(i f +d)^{g c e}\Big)
\nonu \\
&-& i \, f^{d a f} \, \Big(
\frac{1}{M} \de^{b f} \, \de^{c e} +
\frac{1}{4} (i f +d)^{b f g}(i f +d)^{g c e}\Big)
\nonu \\
&+& i \, f^{d b f} \, \Big(
\frac{1}{M} \de^{a f} \, \de^{e c} +
\frac{1}{4} (i f +d)^{a f g}(i f +d)^{g e c}\Big)
\nonu \\
&-& i \, f^{d a f} \, \Big(
\frac{1}{M} \de^{b f} \, \de^{e c} +
\frac{1}{4} (i f +d)^{b f g}(i f +d)^{g e c}\Big)
\nonu \\
&+& i \, f^{e b f} \, \Big(
\frac{1}{M} \de^{a f} \, \de^{d c} +
\frac{1}{4} (i f +d)^{a f g}(i f +d)^{g d c}\Big)
\nonu \\
&-& i \, f^{e a f} \, \Big(
\frac{1}{M} \de^{b f} \, \de^{d c} +
\frac{1}{4} (i f +d)^{b f g}(i f +d)^{g d c}\Big).
\label{tensor}
\eea
We can further simplify these
relations
when they are multiplied by some composite
operators having symmetric or antisymmetric properties in the indices.

\section{ The first order pole in the OPE between the charged spin-$2$
current and itself}

\subsection{ The substitution of charged spin-$2$ current}

In this Appendix, we simplify (\ref{KKpole1}) by rearranging
the operators
and substituting the spin-$2$ current (\ref{spin2expression})
and obtain 
\bea
&& i \, f^{a b c}\, N \, \pa \, K^c(w) =
i \, f^{a b c}\, N \, \pa \,
\Bigg[
\de_{\rho \bar{\si}} \,
t^c_{j\bar{i}} \, (J^{(\rho \bar{i})} \,  J^{(\bar{\si} j)} +
J^{(\bar{\si} j)} \,  J^{(\rho \bar{i})})  
-\frac{N}{(M+2k)} \, d^{c f g} \, J^f\, J^g
\nonu \\
&& +   \frac{2N}{k} \sqrt{\frac{M+N}{M N}} \, J^c  \,  J^{u(1)}
\Bigg](w),
\nonu \\
&& i \, f^{a c e} \, d^{b c d} \, K^e \, J^d(w) =
i \, f^{a c e} \, d^{b c d} \, J^d \, \Bigg[
\de_{\rho \bar{\si}} \,
t^e_{j\bar{i}} \, (J^{(\rho \bar{i})} \,  J^{(\bar{\si} j)} +
J^{(\bar{\si} j)} \,  J^{(\rho \bar{i})})  
-\frac{N}{(M+2k)} \, d^{e f g} \, J^f\, J^g
\nonu \\
&& +   \frac{2N}{k} \sqrt{\frac{M+N}{M N}} \, J^e  \,  J^{u(1)}
\Bigg](w)+ M \, d^{a b c} \, \pa \,
\Bigg[
\de_{\rho \bar{\si}} \,
t^c_{j\bar{i}} \, (J^{(\rho \bar{i})} \,  J^{(\bar{\si} j)} +
J^{(\bar{\si} j)} \,  J^{(\rho \bar{i})})  
\nonu \\
&& -  \frac{N}{(M+2k)} \, d^{c f g} \, J^f\, J^g
+   \frac{2N}{k} \sqrt{\frac{M+N}{M N}} \, J^c  \,  J^{u(1)}
\Bigg](w),
\nonu \\
&& i\, f^{a d e} \, d^{b c d} \, J^c \, K^e  = 
i\, f^{a d e} \, d^{b c d} \, J^c \, \Bigg[
\de_{\rho \bar{\si}} \,
t^e_{j\bar{i}} \, (J^{(\rho \bar{i})} \,  J^{(\bar{\si} j)} +
J^{(\bar{\si} j)} \,  J^{(\rho \bar{i})})  
-\frac{N}{(M+2k)} \, d^{e f g} \, J^f\, J^g
\nonu \\
&& +   \frac{2N}{k} \sqrt{\frac{M+N}{M N}} \, J^e  \,  J^{u(1)}
\Bigg](w),
\nonu \\
&& i \, f^{a b c} \,
K^c \, J^{u(1)}(w)  = i \, f^{a b c} \,
 J^{u(1)} \Bigg[
\de_{\rho \bar{\si}} \,
t^c_{j\bar{i}} \, (J^{(\rho \bar{i})} \,  J^{(\bar{\si} j)} +
J^{(\bar{\si} j)} \,  J^{(\rho \bar{i})})  
\nonu \\
&& -  \frac{N}{(M+2k)} \, d^{c f g} \, J^f\, J^g
+   \frac{2N}{k} \sqrt{\frac{M+N}{M N}} \, J^c  \,  J^{u(1)}
\Bigg](w),
 \nonu \\
&& \de_{\rho \bar{\si}} \,
t^b_{j \bar{i}} \,
\de^{k \bar{i}} \, t^a_{k \bar{l}} \,
((J^{u(1)} \, J^{(\rho \bar{l})}) J^{(\bar{\si} j)})(w)
= \Bigg[ \frac{k N}{2} \, \de^{a b} \, \pa^2 \, J^{u(1)}
\nonu \\
&&-  \sqrt{\frac{M+N}{M N}} \, (
\frac{1}{M} \,\de^{a b}\, \de_{j\bar{i}}  +
\frac{1}{2} \, (i f +d )^{ b a c}
\, t^c_{j \bar{i}} )
\, \de_{\rho \bar{\si}} \, \pa \,
J^{(\rho \bar{i})} \,  J^{(\bar{\si} j)}
\nonu \\
&&+ N \,  \sqrt{\frac{M+N}{M N}} \,
\pa \, J^{u(1)} \, J^{u(1)}
+ \frac{i}{2} \, N\, f^{a b c}\, J^c \, \pa \, J^{u(1)}
-\frac{N}{2} \, d^{a b c} \, J^c\, \pa \, J^{u(1)}
\nonu \\
&&+ \de_{\rho \bar{\si}}\, (\frac{1}{M}\, \de^{b a}\, \de_{j \bar{i}}
+ \frac{1}{2}\, (i f +d )^{b a c}\, t^c_{j \bar{i}}) \,
J^{u(1)} \, J^{(\rho \bar{i})} \,  J^{(\bar{\si} j)} \Bigg](w),
\nonu \\
&& \de_{\rho \bar{\si}} \,
t^b_{j \bar{i}} \,((J^a\, J^{(\rho\bar{i})})\, J^{(\bar{\si} j)})(w)
=
(
\frac{1}{M} \,\de^{a b}\, \de_{j\bar{i}}  +
\frac{1}{2} \, (i f +d )^{a b  c}
\, t^c_{j \bar{i}} )
\, \de_{\rho \bar{\si}} \, \pa \,
J^{(\rho \bar{i})} \,  J^{(\bar{\si} j)}
\nonu \\
&& - N\, \pa \, J^a\, J^b +\de_{\rho \bar{\si}} \,
t^b_{j \bar{i}} \,
J^a\, J^{(\rho\bar{i})} \, J^{(\bar{\si} j)} \Bigg](w),
  \nonu \\
&& \de_{\rho \bar{\si}} \,
t^b_{j \bar{i}} \,
( i \, f -\frac{(2k+M+2N)}{(2k+M)} \, d )^{a c d}\,
\de^{k \bar{i}} \, t^d_{k \bar{l}} \,
((J^c \, J^{(\rho \bar{l})}) \, J^{(\bar{\si} j)})(w)=
\Bigg[-\frac{1}{2} \, k \,N \, i \, f^{a b c} \, \pa^2 \, J^c
  \nonu \\
  && - \frac{k N(2k+M+2N)}{2(2k+M)}\, d^{a b c} \, \pa^2 \, J^c
  -\frac{2(-4k-2M-4N+M^2 N)}{M^2(2k+M)}\,
  \de^{a b}\, \de_{\rho \bar{\si}} \, \de_{j \bar{i}}\, \pa \,
J^{(\rho \bar{i})} \,  J^{(\bar{\si} j)}
\nonu \\
&&+ \frac{(2N+M+2k)}{(2k+M)} \, \frac{2}{M} \,
 d^{a b c}\, \de_{\rho \bar{\si}} \, t^c_{j \bar{i}}\, \pa \,
J^{(\rho \bar{i})} \,  J^{(\bar{\si} j)}-
 \sqrt{\frac{M+N}{M N}} \, N \, f^{a b c} \, \pa \, J^c \, J^{u(1)}
 \nonu \\
 && -\frac{(2N+M+2k)}{(2k+M)} \,\sqrt{\frac{M+N}{M N}} \,
 N\, d^{a b c} \, \pa \, J^c \, J^{u(1)}
 \nonu \\
 && -\frac{N}{2} \,
 ( i \, f -\frac{(2k+M+2N)}{(2k+M)} \, d )^{a c d}\,
 (i f +d)^{e b d} \, \pa \, J^{c}\, J^{e}
 \nonu \\
 &&+ ( i \, f -\frac{(2k+M+2N)}{(2k+M)} \, d )^{a c d}\,
(
\frac{1}{M} \,\de^{ b d }\, \de_{j\bar{i}}  +
\frac{1}{2} \, (i f +d )^{ b d  e}
\, t^e_{j \bar{i}} )
\, \de_{\rho \bar{\si}} \, J^c\, J^{(\rho \bar{i})} \,  J^{(\bar{\si} j)}
  \Bigg](w),
\nonu \\
&&
\de_{\rho \bar{\si}} \,
t^b_{j \bar{i}} \,
\de^{k \bar{i}} \, \de_{\si_1 \bar{\si_1}}\, (t^{\al})^{\bar{\si_1} \rho} \,
t^a_{k \bar{l}} \, ((J^{\al} \,
J^{(\si_1 \bar{l})}) \, J^{(\bar{\si} j)})(w) =
\nonu \\
&& \Bigg[ (t^{\al})^{\si \bar{\rho}} \,
  J^{\al} \, J^{(\si \bar{i})} \, J^{(\bar{\rho} j)}
(
\frac{1}{M} \,\de^{ a b }\, \de_{j\bar{i}}  +
\frac{1}{2} \, (i f +d )^{ a b  c}
\, t^c_{j \bar{i}} )
\nonu \\
&& - \frac{(N^2-1)}{N} \,
 (
\frac{1}{M} \,\de^{ a b }\, \de_{j\bar{i}}  +
\frac{1}{2} \, (i f +d )^{  b a c}
\, t^c_{j \bar{i}} )\, \de_{\rho \bar{\si}} \, \pa \,
J^{(\rho \bar{i})} \,  J^{(\bar{\si} j)} +
\de^{a b} \, \pa \, J^{\alpha}\, J^{\alpha}
\Bigg](w),
\nonu \\
&& \de_{\rho \bar{\si}} \,
t^b_{j \bar{i}} \,  \de_{k \bar{l}} \, (t^a)^{\bar{l} j} \, 
 J^{(\rho \bar{i})}\,
\pa \, J^{(\bar{\si} k)}(w)  = \de_{\rho \bar{\si}} \,(
\frac{1}{M} \,\de^{ a b }\, \de_{j\bar{i}}  +
\frac{1}{2} \, (i f +d )^{ a b  c}
\, t^c_{j \bar{i}} )\,  J^{(\rho \bar{i})}\,
\pa \, J^{(\bar{\si} i)}(w),
\nonu \\
&& \de_{\rho \bar{\si}} \,
t^b_{j \bar{i}} \, \de^{j \bar{l}} \, t^a_{k \bar{l}} \,
J^{(\rho \bar{i})}\, J^{u(1)} \, J^{(\bar{\si} k)}(w)
=\Bigg[\de_{\rho \bar{\si}} \,(
\frac{1}{M} \,\de^{ a b }\, \de_{j\bar{i}}  +
\frac{1}{2} \, (i f +d )^{ a b  c}
\, t^c_{j \bar{i}} )\, J^{u(1)} \,
J^{(\rho \bar{i})}\, J^{(\bar{\si} j)}
\nonu \\
&&- \sqrt{\frac{M+N}{M N}} \,\de_{\rho \bar{\si}} \,(
\frac{1}{M} \,\de^{ a b }\, \de_{j\bar{i}}  +
\frac{1}{2} \, (i f +d )^{ a b  c}
\, t^c_{j \bar{i}} )\, \pa \, J^{(\rho \bar{i})} \,  J^{(\bar{\si} j)}
\Bigg](w),
\nonu \\
&& \de_{\rho \bar{\si}} \,
t^b_{j \bar{i}} \,  J^{(\rho\bar{i})} \, J^a \, J^{(\bar{\si} j)}(w) =
\Bigg[\de_{\rho \bar{\si}} \,
  t^b_{j \bar{i}} \, J^a \, J^{(\rho\bar{i})} \,  J^{(\bar{\si} j)}\nonu \\
  && +
 \de_{\rho \bar{\si}} \,(
\frac{1}{M} \,\de^{ a b }\, \de_{j\bar{i}}  +
\frac{1}{2} \, (i f +d )^{  b a c}
\, t^c_{j \bar{i}} )\,
\, \pa \, J^{(\rho \bar{i})} \,  J^{(\bar{\si} j)}
  \Bigg](w),
\nonu \\
&& \de_{\rho \bar{\si}} \,
t^b_{j \bar{i}} \,( i \, f + \frac{(2k+M+2N)}{(2k+M)} \, d \Big)^{a c d}\,
\de^{j \bar{j_1}}\, t^d_{k \bar{j_1}} \,
J^{(\rho \bar{i})}\, J^c \, J^{(\bar{\si} k)}(w)=
\nonu \\
&& \Bigg[
  ( i \, f + \frac{(2k+M+2N)}{(2k+M)} \, d )^{a c d}\,
   \de_{\rho \bar{\si}} \,
(
\frac{1}{M} \,\de^{ d b }\, \de_{j\bar{i}}  +
\frac{1}{2} \, (i f +d )^{ d b  e}
\, t^e_{j \bar{i}} )\,
  J^c \, J^{(\rho \bar{i})}\, J^{(\bar{\si} i)}
  \nonu \\
  &&+
    ( i \, f + \frac{(2k+M+2N)}{(2k+M)} \, d )^{a c d}\,
  \de_{\rho \bar{\si}} \,
  \nonu \\
  && \times
  (\frac{1}{M} \, \de^{b c} \,t_{j \bar{i}}^d +
  \frac{1}{2M} \, \de^{d e} \, (i \, f +
d)^{b c e}\,
     \de_{j \bar{i}} 
+ \frac{1}{4} (i \, f +d)^{b c e} \, (i \, f + d)^{d e f} t_{j \bar{i}}^f)  
\, \pa \, J^{(\rho \bar{i})} \,  J^{(\bar{\si} j)}
\Bigg](w),
\nonu \\
&&
\de_{\rho \bar{\si}} \,
t^b_{j \bar{i}} \,
\de^{j \bar{l}} \, \de_{\si_1 \bar{\si_1}}\, (t^{\al})^{\bar{\si} \si_1} \,
t^a_{k \bar{l}} \, J^{(\rho \bar{i})}\, J^{\al} \, J^{(\bar{\si_1} k)}(w) =
\nonu \\
&& \Bigg[
 (t^{\al})^{\rho \bar{\si}} \,(
\frac{1}{M} \,\de^{ a b }\, \de_{j\bar{i}}  +
\frac{1}{2} \, (i f +d )^{ a b  c}
\, t^c_{j \bar{i}} )\,
J^{\al} \,J^{(\rho \bar{i})}\,  J^{(\bar{\si} k)} \nonu \\
&& - 
(N-\frac{1}{N})\, 
(
\frac{1}{M} \,\de^{ a b }\, \de_{j\bar{i}}  +
\frac{1}{2} \, (i f +d )^{ a b  c}
\, t^c_{j \bar{i}} )\,  \de_{\rho \bar{\si}} \, \de_{j \bar{i}} \,
\pa \, J^{(\rho \bar{i})} \,  J^{(\bar{\si} j)}
\Bigg](w).
\label{appbpoleone}
\eea
The quadratic terms in the Kronecker delta, $f$ or $d$ symbols
appearing in the second relation from the last
in Appendix (\ref{appbpoleone}) can be simplified further.

\subsection{ The adjoint spin-$1$ dependent terms
in the first order pole}

Now we collect the adjoint spin-$1$ dependent terms only
from Appendix (\ref{appbpoleone}) and (\ref{KKpole1})
in the first order pole as follows:
\bea
&& i \, f^{a b c}\, N \, \pa \,
\Bigg[
N\, \pa \, J^c  
-\frac{N}{(M+2k)} \, d^{c f g} \, J^f\, J^g
\Bigg](w)
\nonu \\
&& -\frac{N}{(M+2k)} \Bigg(
\, i \, f^{a c e} \, d^{b c d} \, J^d \, \Bigg[ N\, \pa \, J^e 
-\frac{N}{(M+2k)} \, d^{e f g} \, J^f\, J^g
\Bigg]\nonu \\
&& + M \, d^{a b c} \, \pa \,
\Bigg[  N\, \pa \, J^c
 -  \frac{N}{(M+2k)} \, d^{c f g} \, J^f\, J^g
\Bigg] \Bigg)(w)
\nonu \\
&&  -\frac{N}{(M+2k)} \Bigg( i\, f^{a d e} \, d^{b c d} \, J^c \, \Bigg[
  N\, \pa \, J^e
-\frac{N}{(M+2k)} \, d^{e f g} \, J^f\, J^g
\Bigg] \Bigg)(w)
\nonu \\
&& +\frac{4}{k M} \, (k+M+N) \, (-N)\, \pa \, J^a \, J^b(w)
-2\, \Bigg[-\frac{1}{2} \, k \,N \, i \, f^{a b c} \, \pa^2 \, J^c
  \nonu \\
  && - \frac{k N(2k+M+2N)}{2(2k+M)}\, d^{a b c} \, \pa^2 \, J^c
 \nonu \\
 && -\frac{N}{2} \,
 ( i \, f -\frac{(2k+M+2N)}{(2k+M)} \, d )^{a c d}\,
 (i f +d)^{e b d} \, \pa \, J^{c}\, J^{e}
  \Bigg](w),
\label{Jdep}
\eea
which should be equal to the terms of 
the first order pole in (\ref{fullKK}) by putting all the other
spin-$1$ currents to zero.
The cubic terms in Appendix (\ref{Jdep}) will participate in the various
places in the first order pole of (\ref{fullKK}).
In the calculations of 
(\ref{c53c72}), (\ref{c52rel}) and (\ref{c51rel}),
the above relation Appendix (\ref{Jdep}) is
used.

\subsection{ The  first order pole}

Eventually we obtain the first order pole as follows:
\bea
&& K^a(z) \, K^b(w)\Bigg|_{\frac{1}{(z-w)}}=
-\frac{2 N}{(M+2k)} \,i\, f^{a c e} \, d^{b c d} \, J^d \, K^e(w)
-\frac{M N}{(M+2k)} \, d^{a b c}\, \pa \, K^c(w)\nonu \\
&& +
N \,i\, f^{a b c} \, \pa \, K^c(w)
+\frac{2 N}{k}\, \sqrt{\frac{M+N}{M N}}
i\, f^{a b c}\, J^{u(1)} \, K^c(w)
+4(k+N) \, \de_{\rho \bar{\si}} \,
(\frac{1}{M} \, \de^{a b} \, \de_{j \bar{i}}
\nonu \\
&& + \frac{1}{2} (i \, f +d)^{a b c} \, t^c_{j \bar{i}} )\,
J^{(\rho \bar{i})} \,\pa \, J^{(\bar{\si} j)}(w)
\nonu \\
&& -2 \, (k+N) \, N \,\sqrt{\frac{M+N}{M N}}\,
\de^{a b} \, \pa^2 \, J^{u(1)}(w)
+\frac{4 (k+N)}{M} \, \de^{a b} \,
\de_{\rho \bar{\si}}\, \de_{j \bar{i}} \,
\pa \, J^{(\rho \bar{i})}  \, J^{(\bar{\si} j)}(w)
\nonu \\
&& -2(k+3 N) \, i\, f^{a b c}\, t^c_{j \bar{i}}\, \de_{\rho \bar{\si}}\,
\pa \, J^{(\rho \bar{i})}  \, J^{(\bar{\si} j)}(w)
+ 2(k+N) \,  d^{a b c}\, t^c_{j \bar{i}}\, \de_{\rho \bar{\si}}\,
\pa \, J^{(\rho \bar{i})}  \, J^{(\bar{\si} j)}(w)
\nonu \\
&& -\frac{4 \, (k+N) }{k}\,
\frac{M+N}{M } \, \de^{a b} \, \pa \, J^{u(1)}\,
J^{u(1)}(w) -\frac{2 N (k+N) }{ k} \, \sqrt{\frac{M+N}{M N}}
\, i\, f^{a b c} \, \pa \, J^{u(1)} \, J^c(w)
\nonu \\
&& + \frac{2 N (k+N)}{ k}\,  \sqrt{\frac{M+N}{M N}}\,
 d^{a b c} \, \pa \, J^{u(1)} \, J^c(w)
 \nonu \\
 &&+\frac{4 (k+N)}{k}\,  \sqrt{\frac{M+N}{M N}} \, i\,
 f^{a b c}\,  t^c_{j \bar{i}}\, \de_{\rho \bar{\si}}
 \, J^{u(1)}\,  J^{(\rho \bar{i})}  \, J^{(\bar{\si} j)}(w)
 \nonu \\
 && -\frac{4 \, N \, (k+M+N)}{k M}\, \pa \, J^a\, J^b(w)
 + k \, N\, i\, f^{a b c} \, \pa^2 \, J^c(w)
 +2 N \sqrt{\frac{M+N}{M N}}\, i\, f^{a b c}\,
 J^{u(1)} \, \pa \, J^c(w)
 \nonu \\
 &&+\frac{2 N  (2 k+M+2 N)}{2 k+M}
   \,
\sqrt{\frac{M+N}{M N}} \, d^{a b c} \, J^{u(1)}\, \pa \, J^c(w)
+ 4 \, i\, f^{a b c} \, t^{\alpha}_{\rho \bar{\si}} \, t^{c}_{j \bar{i}}\,
J^{\alpha} \, J^{(\rho \bar{i})}  \,
J^{(\bar{\si} j)}(w)
\nonu \\
&& -4 \, \de^{a b} \, \pa \, J^{\al} \, J^{\al}(w)
+ \frac{k N (2 k+M+2 N)}{(2 k+M)}\,
d^{a b c} \, \pa^2 \, J^c(w)
\nonu \\
&&- 4 \Bigg( \frac{i}{M} \, f^{a c b}\, \de_{j \bar{i}}\,
 +\frac{i}{2}\, f^{a c d} \, d^{b  d e} \,
t^{e}_{j \bar{i}} -\frac{(2 k+M+2 N)}{2 (2 k+M)}\,
d^{a c d} \, i\, f^{b  d e} \,
t^{e}_{j \bar{i}}\, \Bigg) \de_{\rho \bar{\si}}\, 
J^{c} \, J^{(\rho \bar{i})}  \,
J^{(\bar{\si} j)}(w) \nonu \\
&& + N \, ( i\, f -
\frac{(2 k+M+2 N)}{(2 k+M)} \, d )^{a c d}
(i \, f + d)^{e b d} \, \pa \, J^c \, J^e(w).
\label{pole1KK}
\eea
It is rather nontrivial to rewrite this
in terms of the known currents as well as
the charged spin-$3$ current.
This can be written in terms of
various quasi primary operators and the
charged spin-$3$ current as well as the descendant terms
in (\ref{fullKK}).
The expression in Appendix (\ref{pole1KK})
will be used in the second order pole of
the OPE between the charged spin-$2$ current and the
charged spin-$3$ current.

\section{ The structure constants in the infinity limit of $k$ of section
$3$}

The structure constants appearing in (\ref{fullKK})
under the infinity limit of $k$ become
\bea
c_1 & \rightarrow & \frac{4(1-\la^2)}{\la^2} \, k^3, \qquad
c_2 \rightarrow \frac{2(1-\la^2)}{\la^2} \, k^2, \qquad
\frac{a_{1,CH} \, c_1}{ c} \rightarrow \frac{4}{\la^2  M} \, k^2,
\nonu \\
c_6 & \rightarrow & \frac{1}{\la} \, k, \qquad
c_{31} \rightarrow  -\frac{4}{\la^2 M} \, k, \qquad
c_{32} \rightarrow  \frac{(1-\la^2)}{2 \la^2 } \, k,
\nonu \\
c_{33} & \rightarrow & \frac{(1-\la^2)}{2 \la^2} \, k, \qquad
c_{34} \rightarrow  -\frac{2(1-\la^2)}{\la^2 }, \qquad
c_2 \, a_{3,CH}  \rightarrow  \frac{8}{ \la^2 M} \, k,
\nonu \\
c_{41} & \rightarrow & \frac{(1-\la^2)}{12 \la^2} \, k, \qquad
c_{43} \rightarrow  -\frac{(1-\la^2)\, M}{2 \la^2 } \, \frac{1}{k}, \qquad
c_{51}        \rightarrow  -\frac{4 \, i}{ \la^2 M},
\nonu \\
c_{52}        & \rightarrow & \frac{(1-\la^2)}{ 18 \la^2 },
\qquad
c_{53}        \rightarrow  -\frac{(1-\la^2)}{ 6 \la^2 },
\qquad
c_{72}        \rightarrow  -\frac{1}{  \la },
\qquad
c_{73}        \rightarrow  \frac{i}{  \la },
\label{struc1}
\eea
together with
\bea
a_{1, CH} \rightarrow 1, \qquad a_{2,CH} \rightarrow \frac{1}{6}, \qquad
a_{3,CH} \rightarrow  \frac{4}{ (1- \la^2) M} \frac{1}{k},
\qquad
c \rightarrow (1-\la^2) \, M \, k.
\label{struc2}
\eea
It is not clear whether there are some relations between these
structure constants Appendix (\ref{struc1}) and
Appendix (\ref{struc2}) and
the ones in the free field realization given in
Appendix $H$.
Because the structure constants are given by the three
parameters, we can take any limits among these.
For example, the infinity limit of $N$ can be taken for fixed
$\lambda$ and $M$.

\section{ The second order pole in the OPE $K^a(z) \, P^b(w)$}

The second order pole of
the OPE between the charged spin-$2$ current and the
charged spin-$3$ current can be described as 
\bea
&& K^a(z) \, P^b(w)\Bigg|_{\frac{1}{(z-w)^2}}  = 
\Bigg[ \frac{2}{M}\, (2k+N) \, \de^{a b}\, \de_{j \bar{i}}
\, t^{\alpha}_{\rho \bar{\si}} \, J^{\alpha} \, J^{(\rho \bar{i})}\,
J^{(\bar{\si} j)} -2 k \, \de^{a b} \, J^{\alpha} \, \pa \, J^{\alpha}
\nonu \\
&& + \frac{2k(2k+M+N)}{(2k+M)} \,
d^{a b c} \, t^{c}_{j \bar{i}} \, t^{\alpha}_{\rho \bar{\si}}
\, J^{\alpha} \, J^{(\rho \bar{i})}\,
J^{(\bar{\si} j)} \nonu \\
&& - \frac{2}{k} \, (2k +N) \, \sqrt{\frac{M+N}{M N}}\,
\de^{a b} \, J^{\alpha} \, J^{\alpha} \, J^{u(1)}
+i \, f^{\alpha \beta \gamma} \, \de^{a b} \, J^{\alpha}\,
J^{\beta} \, J^{\gamma}
\nonu \\
&& +  \frac{2(2k+M+N)}{(2k+M)} \, d^{a b c} \,
J^{\alpha} \, J^{\alpha} \, J^c  -
 d^{\alpha \beta \gamma} \, \de^{a b} \, J^{\alpha}\,
J^{\beta} \, J^{\gamma}
\Bigg](w)\, a_1
\nonu \\
&& + \Big( 2 M \, K^a \, J^b - 2 \, f^{a c d} \, f^{d b e}
\, J^c \, K^e \Big)(w)\, a_3 
\nonu \\
&& + \Bigg[ i \, f^{a d e} \, d^{b c d} \, (K^e \, K^c)_{pole-1}
+ d^{b c d} \, J^d \, (K^a \, K^c)_{pole-2} \nonu \\
&& - \frac{2 N M}{k} \, \sqrt{\frac{M+N}{M N}}\, d^{a b c}
\, K^c \, J^{u(1)} + \frac{N}{(2k+M)} \, \Big(
  i \, f^{a b c} \, (M^2-4) \, \pa \, K^c \nonu \\
  && -  \frac{8(4-M^2)}{M^2} \, (J^b \, K^a - J^a \, K^b)
  -\frac{(8-M^2)}{M} \, d^{a c e} \, d^{b d e} \,
  ( J^d \, K^c \nonu \\
  && -   J^c \, K^d) + M \, d^{a b e} \, d^{e c d}\,
  J^c \, K^d + M \, d^{a c e} \, d^{b d e} \, J^d
  \, K^c
  \Big) \Bigg](w)\, a_5 
\nonu \\
&& + \Bigg[ J^{u(1)} \, (K^a \, K^b)_{pole-2} +
\frac{N M}{(2k+M)} \, d^{a b c} \, J^{u(1)} \, K^c \Bigg](w)\, a_7 
\nonu \\
&& + \Big( 2(2 k+ 2 N+M ) \, J^b \, K^a + 2 \, i \,
f^{a b c} \, (k+N) \, \pa \, K^c \nonu \\
&& + 2 \, f^{a b c} \, f^{c d e} \, J^d \, K^e \Big)(w)\, a_{8} 
\nonu \\
&& +  M \, d^{a b c} \, J^{u(1)} \, K^c(w) \, a_9 
\nonu \\
&& -  f^{a c d} \, f^{b c e} \, K^d \, J^e(w) \, a_{11} 
\nonu \\
&& +
\Big( \frac{2 (k^2-1) (2 k+M+N)}{k (2 k+M)} \,
\de_{\rho \bar{\si}}\,
(\frac{2}{M} \, \de^{ a b } \, \de_{j \bar{i}}
+ d^{ a b c} \, t^c_{j \bar{i}})
J^{(\rho \bar{i})} \, \pa \,  J^{(\bar{\si} j)}
\nonu \\
&& + 
\de_{\rho \bar{\si}} \, t^b_{j \bar{i}}\,
((K^a \, J^{(\rho \bar{i})})_{pole-1} \, \pa \, J^{(\bar{\si} j)})_{pole-1}
\nonu \\
&& +  \de_{\rho \bar{\si}} \, t^b_{j \bar{i}}\,
 J^{(\rho \bar{i})} \, (K^a \, \pa\ J^{(\bar{\si} j)})_{pole-2}+
N \, i \, f^{a b c} \, \pa \, K^c \Big)(w)\, (a_{13}-a_{12}) 
\nonu \\
&& + \Bigg[ \frac{1}{2}\, \pa \, (K^a \, K^b)_{pole-2}+
\frac{1}{2}  \, (K^a \, K^b)_{pole-1} \nonu \\
&& - 
\frac{N}{k}\, \sqrt{\frac{M+N}{M N}} \, i \,
f^{a b c} \, K^c \, J^{u(1)} +
\frac{N}{2(2k+M)} \, \Big( M \, d^{a b c} \, \pa \,
K^c \nonu \\
&& + 
i \, f^{a d e} \, d^{b d f} \, K^e \, J^f 
+ i \, f^{a d e} \, d^{b  f d } \, J^f \, K^e \Big)
\Bigg](w)\, a_{12} 
\nonu \\
&& + 2 i \, f^{a b c} \, \pa \, K^c(w) \, a_{16} 
\nonu \\
&& + \Bigg[ \frac{6(4-M^2)}{M^2} \, J^a \, K^b -\frac{6}{M}
\, f^{a b c} \, f^{c d e}\, J^d \, K^e +
3(8-\frac{16}{M^2}) \, J^b \, K^a \nonu \\
&& - \frac{6}{M} \, f^{a c e} \, f^{b d e} \, J^c \, K^d+
\frac{M}{4} \, i \, f^{c d e} \, d^{a b c}\,
J^e \, K^d -
\frac{M}{4} \, i \, f^{a b c} \, d^{c d e}\,
J^e \, K^d \nonu \\
&& + \frac{M}{4} \, i \, f^{e b d} \, d^{a c e}\,
J^d \, K^c -\frac{M}{4} \, i \, f^{a c e} \, d^{e b d}\,
J^d \, K^c + \frac{6(4-M^2)}{M^2} \, \de^{a b} \, J^c \,
K^c \nonu \\
&& + 3 ( M -\frac{M}{4})  \, d^{a c e} \, d^{e b d}\,
J^d \, K^c+
\frac{6}{M} \, d^{a c e} \, d^{e b d}\,
J^c \, K^d \nonu \\
&& +  \frac{6}{M}  \, d^{a b e} \, d^{e c d}\,
J^c \, K^d
+(M^2+6)\, i \, f^{a b c}\, \pa \, K^c  \Bigg](w)\, a_{17}.
\label{pole2KP}
\eea
We will further simplify these expressions
in next subsections.

\subsection{ The $a_5$ terms of the second order pole }

In the $a_5$ terms,
we have the following results
\bea
&& i \, f^{a d e} \, d^{b c d} \, (K^e \, K^c)_{pole-1}(w) =
\frac{8 (4-M^2)  N}{M^2 (2 k+M)}
\, J^b \, K^a(w) -
\frac{8 (4-M^2)  N}{M^2 (2 k+M)}
\, J^a \, K^b(w)
\nonu \\
&& +\frac{(8-M^2)  N}{ M (2 k+M)}\,
d^{a c e} \, d^{e b d} \, J^d \, K^c(w)-
\frac{(8-M^2)  N}{ M (2 k+M)} \,
d^{a c e} \, d^{e b d} \, J^c \, K^d(w)
\nonu \\
&& -\frac{M  N}{ (2 k+M)} \,
d^{a b e} \, d^{e c d} \, J^c \, K^d(w)
-\frac{(M^2-4)  N}{ (2 k+M)}\,
i \, f^{a b c} \, \pa \, K^c(w)
\nonu \\
&& + 2 M (k+N) \, d^{a b c} \,
\de_{\rho \bar{\si}} \, t^c_{j \bar{i}}\,
J^{(\rho \bar{i})}  \,  \pa \, J^{(\bar{\si} j)}(w) 
\nonu \\
&& + \frac{2 M N }{k}\,\sqrt{\frac{ M+N}{M N}}\,
d^{a b c} \, J^{u(1)}\, K^c(w)
+
\frac{2}{M} (M^2-4) (k+N)
\,i \, f^{a b c} \,
\de_{\rho \bar{\si}} \, t^c_{j \bar{i}}\,
J^{(\rho \bar{i})}  \,  \pa \, J^{(\bar{\si} j)}(w) 
\nonu \\
&& -2 M (k+3 N)\,
\, d^{a b c} \,
\de_{\rho \bar{\si}} \, t^c_{j \bar{i}}\,
\pa \, J^{(\rho \bar{i})}   \, J^{(\bar{\si} j)}(w) 
\nonu \\
&& +\frac{2}{M} (M^2-4) (k+N) \,i \, f^{a b c} \,
\de_{\rho \bar{\si}} \, t^c_{j \bar{i}}\,
\pa \, J^{(\rho \bar{i})}   \, J^{(\bar{\si} j)}(w) 
\nonu \\
&& - \frac{2 M N (k+N) }{ k}\,
\sqrt{\frac{M+N}{M N}}
\, d^{a b c} \, \pa \, J^{u(1)}\, J^c(w)
\nonu \\
&& +\frac{2 (M^2-4) N (k+N) }{ k M}\,
\sqrt{\frac{M+N}{M N}}\, i \, f^{a b c} \, \pa \, J^{u(1)}\, J^c(w)
\nonu \\
&& +\frac{4 M (k+N) }{k}\,\sqrt{\frac{M+N}{M N}}\,
d^{a b c} \,
\de_{\rho \bar{\si}} \, t^c_{j \bar{i}}\,
J^{u(1)} \, J^{(\rho \bar{i})}   \, J^{(\bar{\si} j)}(w) 
\nonu \\
&&+\frac{N (4 k^2-k M^2 N-6 k M+4 k N-4 M^2-4 M N)}{k M (2 k+M)}
\, i \, f^{a e c} \, d^{b e d} \, \pa \, J^c \, J^d(w)
\nonu \\
&&+ k M N \, d^{a b c} \, \pa^2 \, J^c(w)
+ 2 M N \sqrt{\frac{M+N}{M N}}\,
d^{a b c} \, J^{u(1)} \, \pa \, J^c(w)
\nonu \\
&&+ \frac{2 (M^2-4) N  (2 k+M+2 N)}{M (2 k+M)}\,
\sqrt{\frac{M+N}{M N}}\,
i \, f^{a b c} \, J^{u(1)} \, \pa \, J^c(w)
\nonu \\
&& + \frac{M N^2}{2 k+M}\,
 i \, d^{a  c e} \, f^{e d b} \, \pa \, J^c \, J^d(w)
 \nonu \\
 &&+ \frac{4(4-M^2) N}{M^2}\, \pa \, J^a \, J^b(w)
 - \frac{4(4-M^2) N}{M^2}\, \pa \, J^b \, J^a(w)
 \nonu \\
 && +N (\frac{M (2 k+M+2 N)}{2 (2 k+M)}
 +\frac{(8-M^2)}{2 M})\,
 d^{a  c e} \, d^{e d b} \, \pa \, J^c \, J^d(w)
 \nonu \\
 && +(-\frac{M (2 k+M+2 N)}{2 (2 k+M)}-
 \frac{8-M^2}{2 M}) N\,
 d^{a  d e} \, d^{e c b} \, \pa \, J^c \, J^d(w)
 +
 \frac{M N^2}{(2 k+M)} \,
 d^{a  b e} \, d^{e c d} \, \pa \, J^c \, J^d(w)
 \nonu \\
 && +\frac{2 N  (2 k+M+2 N)}{M (2 k+M)}\,
 i \, f^{a  b e} \, d^{e c d} \, \pa \, J^c \, J^d(w)
 -\frac{2 N  (2 k+M+2 N)}{M (2 k+M)}\,
  i \, f^{a d  e} \, d^{e c b} \, \pa \, J^c \, J^d(w)
  \nonu \\
  && + 4 M \,
t^{\al}_{\rho \bar{\si}} \, t^c_{j \bar{i}}\,
J^{\al} \, J^{(\rho \bar{i})}   \, J^{(\bar{\si} j)}(w)
+\frac{k (M^2-4) N (2 k+M+2 N)}{M (2 k+M)}\,
i \, f^{a b c} \, \pa^2 \, J^c(w)
\nonu \\
&& + M N \, d^{a b c} \, \pa \, K^c(w)
\nonu \\
&&+ 4 \, \de_{\rho \bar{\si}} \, \de_{j \bar{i}}\,
d^{a b c} \, J^{c} \, J^{(\rho \bar{i})}   \, J^{(\bar{\si} j)}(w)
- \frac{8(4-M^2)}{M^2}\,  t^b_{j \bar{i}}\,
\de_{\rho \bar{\si}} \,
 J^{a} \, J^{(\rho \bar{i})}   \, J^{(\bar{\si} j)}(w)
 \nonu \\
 &&- \frac{8(4-M^2)}{M^2}\,  t^a_{j \bar{i}}\,
\de_{\rho \bar{\si}} \,
J^{b} \, J^{(\rho \bar{i})}   \, J^{(\bar{\si} j)}(w)
\nonu \\
&& -4 (\frac{M (2 k+M+2 N)}{4 (2 k+M)}+\frac{(8-M^2)}{4 M})\,
 \de_{\rho \bar{\si}} \, t^{d}_{j \bar{i}}\,
d^{a c e} \, d^{e d b} \, J^{c} \, J^{(\rho \bar{i})}   \, J^{(\bar{\si} j)}(w)
\nonu \\
&& + 4\, (\frac{M (2 k+M+2 N)}{
  4 (2 k+M)}+\frac{(8-M^2)}{4 M})\,
 \de_{\rho \bar{\si}} \, t^{d}_{j \bar{i}}\,
d^{a d e} \, d^{e c b} \, J^{c} \, J^{(\rho \bar{i})}   \, J^{(\bar{\si} j)}(w)
\nonu \\
&& -\frac{2 M N}{2 k+M}\,
 \de_{\rho \bar{\si}} \, t^{d}_{j \bar{i}}\,
d^{a b e} \, d^{e c d} \, J^{c} \, J^{(\rho \bar{i})}   \, J^{(\bar{\si} j)}(w).
\label{a5one}
\eea
We should 
analyze these complicated results in order to rewrite
them in terms of the known currents.
Moreover, we have the following expression
\bea
&& d^{b c d} \, J^d \, (K^a \, K^c)_{pole-2}(w) =
-\frac{4 (k+N) (M+N)}{k M}\,
d^{ a b c} \, J^{u(1)} \, J^{u(1)} \, J^c(w)
\nonu \\
&& -4 \, d^{ a b c} \, J^{\al} \, J^{\al} \, J^c(w)
- \frac{2 N^2 }{k} \,
\sqrt{\frac{M+N}{M N}}\,
i\,  f^{ a e c}\, d^{b e d} \, J^{d}\, J^{c} \, J^{u(1)}(w)
\nonu \\
&& + \frac{2  N (4 k^2+2 k M+4 k N+M N)}{
  k (2 k+M)} \, \sqrt{\frac{M+N}{M N}}\,
 d^{ a e c}\, d^{b e d} \, J^{d}\, J^{c} \, J^{u(1)}(w)
 \nonu \\
 && - \frac{4 N (k+M+N)}{k M}\,
 d^{ b c d} \, J^d \, J^a \, J^c(w)
 \nonu \\
 && + N\, d^{b f g}\, (i \, f -\frac{(2N+M+2k)}{(M+2k)})^{a c d}
 \, (i \, f+ d)^{e f d}\,
  J^g \, J^c \, J^e(w)
  \nonu \\
  && + \frac{8 (k+N)}{M}\, d^{a b c}\, 
\de_{\rho \bar{\si}} \, \de_{j \bar{i}}\,
J^{c} \, J^{(\rho \bar{i})}   \, J^{(\bar{\si} j)}(w)
-2 N \, i \, f^{a c e} \, d^{b c d}\,
\de_{\rho \bar{\si}} \, t^e_{j \bar{i}}\,
J^{d} \, J^{(\rho \bar{i})}   \, J^{(\bar{\si} j)}(w)
\nonu \\
&&+\frac{2 (4 k^2+2 k M+4 k N+M N)}{(2 k+M)} \,
 d^{b c d} \, d^{a c e}\,
\de_{\rho \bar{\si}} \, t^e_{j \bar{i}}\,
J^{d} \, J^{(\rho \bar{i})}   \, J^{(\bar{\si} j)}(w)
\nonu \\
&& -4\, N \, (k+N)\, \sqrt{\frac{M+N}{M N}}\,
d^{a b c} \, J^c \, \pa \, J^{u(1)}(w)
+ 2 k \, N\,
i\, d^{b c d}\, f^{a c e} \, J^d \, \pa \, J^{e}(w)
\nonu \\
&& \frac{2 k N (2 k+M+2 N)}{(2 k+M)}\,
d^{b c d}\, d^{a c e} \, J^d \, \pa \, J^{e}(w)
- \frac{M N}{(2 k+M)} \,
d^{b c d}\, d^{a c e} \, J^d  \, K^{e}(w)
\nonu \\
&& + N \, i \, d^{b c d}\, f^{a c e} \,
J^d  \, K^{e}(w).
\label{a5two}
\eea
Then by substituting Appendix (\ref{a5one}) and Appendix (\ref{a5two})
into Appendix (\ref{pole2KP}), we obtain the corresponding
$a_5$ terms explicitly.
This is necessary step we should do in order to obtain the
final result in (\ref{KPfinal}).

\subsection{ The $(a_{13}-a_{12})$ terms of the second order pole }

In particular, 
in order to
calculate the $(a_{13}-a_{12})$ terms in Appendix (\ref{pole2KP})
we should calculate the following first order poles
which can be obtained in the OPEs 
between the first order poles at the point $z$ in the OPE 
between $K^a$ and $ J^{(\rho \bar{i})}$
with $\de_{\rho \bar{\si}} \, t^b_{j \bar{i}}$
and the current $ J^{(\bar{\si} j)}(w)$
\bea
&& \de_{\rho \bar{\si}} \,
t^b_{j \bar{i}} \, \de^{k \bar{i}}  \, t^a_{k \bar{l}} \,   J^{u(1)}
\, J^{(\rho \bar{l})}
(z) \, J^{(\bar{\si} j)}(w) \Bigg|_{\frac{1}{(z-w)}}
=
k \, N \, \de^{a b} \, \pa \, J^{u(1)}(w)
\nonu \\
&& + \frac{1}{2}\, \sqrt{\frac{M+N}{M N}} \, i \, f^{a b c} \,
\de_{\rho \bar{\si}} \, t^c_{j \bar{i}}\,
J^{(\rho \bar{i})}  \,  J^{(\bar{\si} j)}(w)
 - \frac{1}{2}\, \sqrt{\frac{M+N}{M N}}  \, d^{a b c} \,
\de_{\rho \bar{\si}} \, t^c_{j \bar{i}}\,
J^{(\rho \bar{i})}  \,  J^{(\bar{\si} j)}(w)
\nonu \\
&& - \frac{1}{M}\, \sqrt{\frac{M+N}{M N}}  \, \de^{a b} \,
\de_{\rho \bar{\si}} \, \de_{j \bar{i}}\,
J^{(\rho \bar{i})}  \,  J^{(\bar{\si} j) }(w)
+ \sqrt{\frac{M+N}{M N}}  \, N \, \de^{a b} \, J^{u(1)} \, J^{u(1)}(w)
\nonu \\
&&
+\frac{N}{2}\, i \, f^{a b c} \, J^{u(1)} \, J^c(w)
-\frac{N}{2} \, d^{a b c} \, J^{u(1)} \, J^c(w),
\nonu \\
&& \de_{\rho \bar{\si}} \,
t^b_{j \bar{i}} \,   J^a \, J^{(\rho \bar{i})}(z) \, \, J^{(\bar{\si} j)}(w) \Bigg|_{\frac{1}{(z-w)}}
= \frac{1}{2}\,  i \, f^{a b c} \,
\de_{\rho \bar{\si}} \, t^c_{j \bar{i}}\,
J^{(\rho \bar{i})}  \,  J^{(\bar{\si} j)}(w)
\nonu \\
&& +\frac{1}{2} \, d^{a b c} \,
\de_{\rho \bar{\si}} \, t^c_{j \bar{i}}\,
J^{(\rho \bar{i})}  \,  J^{(\bar{\si} j)}(w)
+ \frac{1}{M} \, \de^{a b} \,
\de_{\rho \bar{\si}} \, \de_{j \bar{i}}\,
J^{(\rho \bar{i})}  \,  J^{(\bar{\si} j) }(w) - N \, J^a \, J^b(w),
\nonu \\
&& \de_{\rho \bar{\si}} \,
t^b_{j \bar{i}} \,  (i \, f -\frac{(2k+
    M+2N)}{
    (2k+M)} \, d)^{a c d} \, \de^{k_1 \bar{i}} \, t^d_{k_1 \bar{j_1}} \, J^c \,
   J^{(\rho \bar{j_1})}(z) \, \, J^{(\bar{\si} j)}(w) \Bigg|_{\frac{1}{(z-w)}}
 = 
 \nonu \\
 && -\frac{(2N+M+2k)}{(M+2k)}\, \frac{2}{M}\,
  d^{a b c} \,
\de_{\rho \bar{\si}} \, t^c_{j \bar{i}}\,
J^{(\rho \bar{i})}  \,  J^{(\bar{\si} j)}(w)-
\de^{a b} \,
\de_{\rho \bar{\si}} \, \de_{j \bar{i}}\,
J^{(\rho \bar{i})}  \,  J^{(\bar{\si} j) }(w)
\nonu \\
&& + \frac{(M^2-4) (2 k+M+2 N)}{M^2 (2 k+M)}\,
\de^{a b} \,
\de_{\rho \bar{\si}} \, \de_{j \bar{i}}\,
J^{(\rho \bar{i})}  \,  J^{(\bar{\si} j) }(w)-
\frac{2 N (2 k+M+2 N)}{M (2 k+M)}\, d^{a b c} \, \pa \, J^c(w)
\nonu \\
&& - \frac{2 N  (-4 k+M^2 N-2 M-4 N)}
      {M (2 k+M)} \, \sqrt{\frac{M+N}{M N}}
      \, \de^{a b} \,\pa \, J^{u(1)}(w)
      \nonu \\
      && +
\frac{2 N (2 k+M+2 N)}{M (2 k+M)}\, d^{a b c} \, \pa \, J^c(w)
\nonu \\
&& - (i \, f -\frac{(2k+
    M+2N)}{
  (2k+M)} \, d)^{a c d} \,
\Bigg( \sqrt{\frac{M+N}{M N}}\, N\, \de^{b d} \, J^{c} \, J^{u(1)}
-\frac{N}{2} \, (i \, f + d)^{e b d} \, J^{c}\, J^{e} \nonu \\
&& + \frac{1}{2}\, \sqrt{\frac{M+N}{M N}}\, N\,
(i \, f + d)^{b d c}\, \pa \, J^{u(1)} + k \, N\, \de^{b d}\,
\pa \, J^{c}
\Bigg)(w),
\nonu \\
&& \de_{\rho \bar{\si}} \,
t^b_{j \bar{i}} \,  \de^{k_1 \bar{i}} \, \de_{\si_2 \bar{\si_1}} \, (t^{\al})^{\bar{\si_1} \rho}
   \, t^a_{k_1 \bar{j_1}} \, J^{\al} \, J^{(\si_2 \bar{j_1})}(z) \,
   \, J^{(\bar{\si} j)}(w) \Bigg|_{\frac{1}{(z-w)}}
 =-\frac{(N^2-1)}{M N} \, \de^{a b} \,
 \Bigg(
 \de_{\rho \bar{\si}} \, \de_{j \bar{i}}\,
 J^{(\rho \bar{i})}  \,  J^{(\bar{\si} j) }
 \nonu \\
&& - \sqrt{\frac{M+N}{M N}}\, N\,M\, \pa \, J^{u(1)}\Bigg)(w)
-\frac{(N^2-1)}{2N} \, (i \, f + d)^{b a c}
 \Bigg( t^c_{j \bar{i}} \,  \de_{j \bar{i}} \,
 J^{(\rho \bar{i})}  \,  J^{(\bar{\si} j) } + N\, \pa \, J^c \Bigg)(w)
\label{fourfirst}
 \\
 && + \de^{a b} \, J^{\al}\, J^{\al}(w) - \sqrt{\frac{M+N}{M N}}\,
 (N^2-1) \, \de^{a b} \, \pa \, J^{u(1)}(w)
 +\frac{1}{2}\, (N^2-1)(- i \, f + d)^{a b c} \, \pa \, J^c(w).
 \nonu
 \eea
 Note that
 the first order pole of
 the OPE between $\de_{k\bar{l}} \, (t^a)^{\bar{i} k} \,
 \pa \, J^{(\rho \bar{l})}(z)$
 and $ J^{(\bar{\si} j)}(w)$ is zero.
Then it is obvious to obtain the corresponding
$(a_{13}-a_{12})$ terms simply  by differentiating these
Appendix (\ref{fourfirst}) with respect
to the variable $w$.
We can easily calculate the second order poles of the OPE
between $K^a$ and $\pa\, J^{(\bar{\si} j)}$ from (\ref{KJJ})
by differentiating the first relation with respect to the variable
$w$ consisting of five terms.

Finally, we obtain the
$(a_{13}-a_{12})$ terms as follows:
\bea
&& \frac{2 (k^2-1) (2 k+M+N)}{k (2 k+M)} \,
\de_{\rho \bar{\si}}\, (\frac{1}{M} \, \de^{ b a} \, \de_{j \bar{i}}(w)
+ \frac{1}{2} (i \, f +d)^{ b a c} \, t^c_{j \bar{i}} )\,
J^{(\rho \bar{i})}  \,  \pa \, J^{(\bar{\si} j) }(w)
\nonu \\
&& +
\frac{2 (4 k^3+2 k^2 M+3 k^2 N+k M N-2 k-M-N)}{k (2 k+M)}
\nonu \\
& & \times  \de_{\rho \bar{\si}}
\, (\frac{1}{M} \, \de^{ a b} \, \de_{j \bar{i}}
+ \frac{1}{2} (i \, f +d)^{  a b c} \, t^c_{j \bar{i}} )\,
J^{(\rho \bar{i})}  \,  \pa \, J^{(\bar{\si} j) }(w)
\nonu \\
&& -2 N (k+N) \sqrt{\frac{M+N}{M N}}\, \pa^2 \, J^{u(1)}(w)
-N\, \,i\, f^{a b c}  \,  t^c_{j \bar{i}} \, \de_{\rho \bar{\si}}
\pa \, (J^{(\rho \bar{i})}   \, J^{(\bar{\si} j) })(w)
\nonu \\
&& +\frac{(2 k^2 N+k M N+4 k+2 M+2 N)}{k (2 k+M)}
\,  d^{a b c}  \,  t^c_{j \bar{i}} \, \de_{\rho \bar{\si}}\, 
\pa \, (J^{(\rho \bar{i})}   \, J^{(\bar{\si} j) })(w)
\nonu \\
&& + \frac{4 (k^2 N+k M N+2 k+M+N)}{k M (2 k+M)}
\, \de^{a b} \,  \de_{\rho \bar{\si}}\, \de_{j \bar{i}}\,
\pa \, (J^{(\rho \bar{i})}   \, J^{(\bar{\si} j) })(w)
\nonu \\
&& -\frac{2 (k+N) (M+N)}{k M} \,  \de^{a b} \,
\pa \, ( J^{u(1)} \, J^{u(1)})(w)
-\frac{N^2 }{k}\,\sqrt{\frac{M+N}{M N}}\,
i\, f^{a b c} \, \pa \, ( J^{u(1)} \, J^c)(w)
\nonu \\
& & + \frac{N  (4 k^2+2 k M+4 k N+M N)}{k (2 k+M)}\,
\sqrt{\frac{M+N}{M N}}\,
 d^{a b c} \, \pa \, ( J^{u(1)} \, J^c)(w)
 \nonu \\
 && -\frac{2 N  (k+M+N)}{k M}\,
 \pa \, (J^a \, J^b)(w) +
 \frac{k N (2 k+M+2 N)}{(2 k+M)}\, d^{a b c} \, \pa^2 \,
 J^c(w)
 \nonu \\
 && +\frac{N}{2} \,
(i \, f -\frac{(2k+
    M+2N)}{
  (2k+M)} \, d)^{a c d} \,
 \, (i \, f + d)^{e b d} \, \pa \,( J^{c}\, J^{e})(w)
+ k \, N\, i\, f^{a b c} \, \pa^2 \,
 J^c(w)
 \nonu \\
 && -2 \, \de^{a b} \, \pa \, (J^{\al} \, J^{\al})(w)
 \nonu \\
 && + \frac{2}{k} \, (k+N)\,
 \sqrt{\frac{M+N}{M N}}\,
\Bigg[\de_{\rho \bar{\si}} \,(
\frac{1}{M} \,\de^{ a b }\, \de_{j\bar{i}}  +
\frac{1}{2} \, (i f +d )^{ a b  c}
\, t^c_{j \bar{i}} )\, J^{u(1)} \,
J^{(\rho \bar{i})}\, J^{(\bar{\si} j)}
\nonu \\
&&- \sqrt{\frac{M+N}{M N}} \,
\de_{\rho \bar{\si}} \,(
\frac{1}{M} \,\de^{ a b }\, \de_{j\bar{i}}  +
\frac{1}{2} \, (i f +d )^{ a b  c}
\, t^c_{j \bar{i}} )\, \pa \, J^{(\rho \bar{i})} \,  J^{(\bar{\si} j)}
\Bigg](w)
\nonu \\
&&- \frac{2}{ k M} \, (k+M+N) \,
\Bigg[\de_{\rho \bar{\si}} \,
  t^b_{j \bar{i}} \, J^a \, J^{(\rho\bar{i})} \,  J^{(\bar{\si} j)}\nonu \\
  && +
 \de_{\rho \bar{\si}} \,(
\frac{1}{M} \,\de^{ a b }\, \de_{j\bar{i}}  +
\frac{1}{2} \, (i f +d )^{  b a c}
\, t^c_{j \bar{i}} )\,
\, \pa \, J^{(\rho \bar{i})} \,  J^{(\bar{\si} j)}
  \Bigg](w)
\nonu \\
&& -
\Bigg[
  ( i \, f + \frac{(2k+M+2N)}{(2k+M)} \, d \Big)^{a c d}\,
   \de_{\rho \bar{\si}} \,
(
\frac{1}{M} \,\de^{ d b }\, \de_{j\bar{i}}  +
\frac{1}{2} \, (i f +d )^{ d b  e}
\, t^e_{j \bar{i}} )\,
  J^c \, J^{(\rho \bar{i})}\, J^{(\bar{\si} i)}
  \nonu \\
  &&+
    ( i \, f + \frac{(2k+M+2N)}{(2k+M)} \, d \Big)^{a c d}\,
  \de_{\rho \bar{\si}} \,
  \nonu \\
  && \times
  (\frac{1}{M} \, \de^{b c} \,t_{j \bar{i}}^d +
  \frac{1}{2M} \, \de^{d e} \, (i \, f +
d)^{b c e}\,
     \de_{j \bar{i}} 
+ \frac{1}{4} (i \, f +d)^{b c e} \, (i \, f + d)^{d e f} t_{j \bar{i}}^f)  
\, \pa \, J^{(\rho \bar{i})} \,  J^{(\bar{\si} j)}
\Bigg](w)
\nonu
\\
&& + 2 \,
\Bigg[
 (t^{\al})^{\rho \bar{\si}} \,(
\frac{1}{M} \,\de^{ a b }\, \de_{j\bar{i}}  +
\frac{1}{2} \, (i f +d )^{ a b  c}
\, t^c_{j \bar{i}} )\,
J^{\al} \,J^{(\rho \bar{i})}\,  J^{(\bar{\si} k)}
\label{a13a12}
\\
&& - 
(N-\frac{1}{N})\, 
(
\frac{1}{M} \,\de^{ a b }\, \de_{j\bar{i}}  +
\frac{1}{2} \, (i f +d )^{ a b  c}
\, t^c_{j \bar{i}} )\,  \de_{\rho \bar{\si}} \, \de_{j \bar{i}} \,
\pa \, J^{(\rho \bar{i})} \,  J^{(\bar{\si} j)}
\Bigg](w)
+ N \, i\, f^{a b c} \, \pa \, K^c(w),
\nonu
\eea
where the last four relations in Appendix (\ref{appbpoleone})
are used here in Appendix (\ref{a13a12}).

Therefore, we will obtain
the final second order pole by collecting all the
relevant terms from
Appendix (\ref{a5one}), Appendix (\ref{a5two})
and Appendix (\ref{a13a12}) explicitly.
It seems that they have rather complicated coset operators.
However, the second order pole can be written in simple form
as the one in (\ref{KPfinal}).

\subsection{ The
 relations between the remaining
  coefficients of $W^{(3)}(w)$ in the second order pole}

In (\ref{pole2W}), we have identified the coefficient of $W^{(3)}(w)$
in the second order pole of the OPE between $K^a(z) \, P^b(w)$
and we present the remaining terms of the neutral spin-$3$ current
including the $b_3$ term as follows:
\bea
&& \frac{a_1 }{b_1}\, b_3-\frac{4 (k+N) (M+N)}{k M}\,  a_7=0,
\nonu\\
&& \frac{a_1}{b_1}\,  b_4-\frac{2  (2 k+N) }{k}
\, \sqrt{\frac{M+N}{M N}}\, a_1-
4 \, a_7=0,
\nonu \\
&& 2\, \frac{a_1}{b_1}\,  b_6 +
\frac{2 (2 k+N)}{M}\,  a_1+\frac{2 }{M}\, (a_{13}-a_{12})=0,
\nonu \\
&& 2\, \frac{a_1 }{b_1}\,  b_8+
\frac{8  (k+N)}{M}\, a_7+\frac{2  (k+N)
  }{k M}\, \sqrt{\frac{M+N}{M N}}\, (a_{13}-a_{12})=0,
\nonu \\
&&   \frac{a_1 }{b_1}\,  b_9-
M \, \frac{a_1}{b_1}\, b_6-(2  k+N) \, a_1 -
4 a_{13}-2 a_{12}=0,
\nonu \\
&& \frac{a_1 }{b_1}\,  b_{11}
-M \, N\, \frac{a_1}{b_1}\,  \sqrt{\frac{M+N}{M N}}\, b_8
-4  N (k+N) \sqrt{\frac{M+N}{M N}}\, a_7
\nonu \\
&& -\frac{4(k+N)}{k M}\, (M+N)\,
a_{13}
-\frac{2  (k+N) (M+N)}{k M}\, a_{12}=0,
\nonu \\
&&  \frac{a_1}{b_1}\, b_{12}+\frac{6 (k+N)}{M}\, a_{12} 
=0,
\nonu \\
&& \frac{a_1}{b_1}\, b_{13}+\frac{6 (k+N)}{M}\, a_{13} 
=0,\nonu\\
&&
-\frac{ M N }{2}\, \frac{a_1}{b_1}\,
\sqrt{\frac{M+N}{M N}}\,  b_{13}+
\frac{a_1 }{b_1}\, b_{14}-2  N (k+N) \sqrt{\frac{M+N}{M N}}\,
a_{13}
\nonu \\
&& -
(k+N) \sqrt{\frac{M+N}{M N}}
\,  a_{12}
=0.
\label{relation1}
\eea
It is easy to observe that 
the above relations Appendix (\ref{relation1})
are satisfied by substituting (\ref{avalues}),
(\ref{interbvalue}) and (\ref{b2b7X}).

\subsection{ The relations between the remaining
  coefficients of $P^b(w)$ in the second order pole }

In (\ref{pole2P}), the structure of
charged spin-$3$ current in the second order pole of
the OPE $K^a(z) \, P^b(w)$ is found and
we present the remaining terms in the charged spin-$3$ current
including the $a_2$ term as follows:
\bea
&&-\Bigg[\frac{ k (3 k+2 M) (2 k+M+2 N)}{(k+M) (2 k+M)}\,
  \Bigg] \, a_1+
\frac{ 2 k (2 k+M+N)}{(2 k+M)}\, a_1+4  M\, a_5- a_{12}+a_{13}=0,
\nonu \\
&& \frac{ 2 (2 k+M+N)}{(2 k+M)}\, a_1-
\Bigg[ \frac{ k (3 k+2 M) (2 k+M+2 N)}{(k+M) (2 k+M)}\,
  \Bigg] \, a_2-4 \, a_5=0,
\nonu \\
&& \frac{2  N  (4 k^2+2 k M+4 k N+M N)}{k (2 k+M)}\,\sqrt{\frac{M+N}{M N}} \, a_7-\Bigg[\frac{ k (3 k+2 M) (2 k+M+2 N)}{(k+M) (2 k+M)}\, \Bigg]\,
a_4
\nonu \\
&& -
\frac{4  (k+N) (M+N)}{k M}\, a_5+
\frac{2  M }{k}\, \sqrt{\frac{N (M+N)}{M}}\, a_9=0,
\nonu\\
&& \frac{2 (4 k^2+2 k M+4 k N+M N)}{2 k+M}\, a_7+
\frac{4  M (k+N) }{k}\, \sqrt{\frac{M+N}{M N}}\, a_5\nonu \\
&& -2 \Bigg[
\frac{ k (3 k+2 M) (2 k+M+2 N)}{(k+M) (2 k+M)}\Bigg]\, a_7+
\frac{  (k+N)}{ k}\,  \sqrt{\frac{M+N}{M N}}
\, (a_{13}-a_{12})+2 \, M\,
a_9 =0,
\nonu \\
&& -2  M (k+2 N)\, a_5+
\frac{ (6 k^2+3 k M+6 k N+2 M N)}{2 k+M}\, a_{12}\nonu \\
&& -
\Bigg[\frac{k (3 k+2 M) (2 k+M+2 N)}{(k+M) (2 k+M)}\Bigg] \, a_{12}=0,
\nonu \\
&& 2  M (k+2 N)\, a_5 +
\frac{(6 k^2+3 k M+6 k N+2 M N)}{2 k+M}\, a_{13}
\nonu \\
&&- \Bigg[ \frac{ k (3 k+2 M) (2 k+M+2 N)}{
    (k+M) (2 k+M)}\Bigg] \, a_{13}=0,
\nonu \\
&& -2  N \sqrt{\frac{M+N}{M N}} \, (2 k+M+2 N) \, a_5+
\frac{ N (k+N)}{k}\, \sqrt{\frac{M+N}{M N}} \, a_{12}\nonu \\
&& +
\frac{ N 
  (4 k^2+2 k M+4 k N+M N)}{k (2 k+M)}\, \sqrt{\frac{M+N}{M N}}
\, a_{13}
\nonu \\
&& + \Bigg[\frac{   k (3 k+2 M) (2 k+M+2 N)}{(k+M) (2 k+M)}\Bigg] \,
(\sqrt{\frac{M+N}{M N}} \, M N\, a_8 - a_{14})=0,
\nonu \\
&&\frac{2  M N (k+N) }{k} \, \sqrt{\frac{M+N}{M N}} \, a_5
+ \frac{ N (4 k^2+2 k M+4 k N+M N)}{2 k+M}\,
a_7 \nonu \\
&& + M \, N\, a_9 +\frac{ N (2 k+M+2 N)}{2 k+M} \,
\sqrt{\frac{M+N}{M N}} \, a_{12}\nonu \\
&& +
\frac{ N  (4 k^2+2 k M+4 k N+M N)}{k (2 k+M)}\,
\sqrt{\frac{M+N}{M N}}
\, a_{13}
\nonu \\
&& -\Bigg[\frac{  k (3 k+2 M) (2 k+M+2 N)}{(k+M) (2 k+M)}\Bigg]
\, (N\, a_7
-a_{15})=0.
\label{relation2}
\eea
The above relations Appendix (\ref{relation2})
are satisfied by substituting (\ref{avalues}).

\section{ The first order pole  in the OPE $K^a(z) \, P^b(w)$}

The first order pole of
the OPE between the charged spin-$2$ current and the
charged spin-$3$ current can be obtained  
\bea
&& K^a(z) \, P^b(w)\Bigg|_{\frac{1}{(z-w)}}  = 
\,     
t^{\alpha}_{\rho \bar{\si}} \, t^b_{j \bar{i}} \,
 J^{\alpha} \, \Bigg[
 ((K^a \, J^{(\rho \bar{i})})_{pole-1} \, J^{(\bar{\si} j)})
 +
 J^{(\rho \bar{i})} \, (K^a \, J^{(\bar{\si} j)})_{pole-1} \Bigg](w)
  \, a_1
  \nonu \\
 & &+ i \, f^{a b c} \, J^{\alpha} \, J^{\alpha} \, J^c(w) \, a_2
  \nonu \\
  & &+ \Big(
  i \, f^{a c d} \, K^{d} \, J^{c} \, J^b
  + i \, f^{a c e} \, J^{c} \, K^{e} \, J^b
+ i \, f^{a b e} \, J^{c} \, J^{c} \, K^e
  \Big)(w) \, a_3
  \nonu \\
  &&+   i \, f^{a b c} \, K^{c} \, J^{u(1)} \, J^{u(1)}(w) \, a_4
  \nonu \\
  &&+ \Bigg[ i \, f^{a d e} \, d^{b c d} \, K^e \, K^c +
    d^{b c d} \, J^d \, ( K^a \, K^c )_{pole-1} +
    \nonu \\
   & &+ \frac{N}{(2k+M)} \Big( i \, f^{a d g} \,
    d^{c e f} \, d^{b c d} \, K^g \, J^e \, J^f  +
     i \, f^{a g e} \,
    d^{b c d} \, d^{c g f} \, J ^d \, K^e \, J^f
    \Big) \nonu \\
   & &+   i \, f^{a g e} \,
    d^{b c d} \, d^{c f g} \, J ^d \, J^f \, K^e \nonu \\
   & & -  \frac{2 N}{k} \, \sqrt{\frac{M+N}{M N}} \Big(
   i \, f^{a d e} \,
   d^{b c d}  \, K^e \, J^c \, J^{u(1)} +
 i \, f^{a c e} \,
   d^{b c d}  \, J^d \, K^e \, J^{u(1)} 
   \Big)\Bigg](w) \, a_5
  \nonu \\
  && + \Bigg[ J^{u(1)} \, ( K^a \, K^b )_{pole-1} 
    -\frac{2 N}{k} \, \sqrt{\frac{M+N}{M N}} \, i \, f^{a b c}
    \, J^{u(1)} \, K^c \, J^{u(1)}
    \nonu \\
 && + \frac{N}{(2k+M)} \Big( i \, f^{a d e} \,
    d^{b d f}  \, J^{u(1)} \, K^e \, J^f  +
     i \, f^{a d e} \,
    d^{b f d}  \, J ^{u(1)} \, J^f \, K^e
    \Big)    
    \Bigg](w) \, a_7
  \nonu \\
 & &+ \Bigg[ 2 \, i \, f^{a b c} \, (k +M+N) \, K^c \, T
    + 2 \, (k+M+N) \, J^b \, \pa \, K^a \nonu \\
  &  &- 2 \, i \, f^{a c d} \, J^b \, J^c \, K^d -
    2 M \, J^b \pa \, K^a - i \, f^{a b d}\, K^d \, J^c \, J^c
    \nonu \\
    & &+   i\, f^{a b c} \, \frac{M}{(k+N)} \, K^c \, J^{\alpha}\,
    J^{\alpha} + i \, 
    f^{a b c} \,  \frac{(M+N)}{k} \,
    K^c \, J^{u(1)} \, J^{u(1)} \Bigg](w) \, a_8
  \nonu \\
 & &+ \Big(  i \, f^{a c e} \, d^{b c d}  \, J^{u(1)} \, K^e \, J^d
+ i \, f^{a d e} \, d^{b c d} \, J^{u(1)} \, J^c \, K^e
    \Big)(w) \, a_9
  \nonu \\
  & &+ \Big( - f^{a c d} \, f^{b c e} \, \pa \,
  K^d \, J^e + f^{a c d} \, f^{b c e} \, \pa \, J^e \, K^d
  \Big)(w) \, a_{11}
  \nonu \\
  & &+ \de_{\rho \bar{\si}} \, t^b_{j \bar{i}} \,
  \Big( ((K^a \, J^{(\rho \bar{i})})_{pole-1} \, \pa \,
  J^{(\bar{\si} j}) + J^{(\rho \bar{i})}\,
 (K^a \, \pa \, J^{(\bar{\si} j)})_{pole-1} \Big)(w) \, (a_{13}-a_{12})
  \nonu \\
  & &+ \Bigg[ \frac{1}{2} \, \pa \, ( K^a \, K^b )_{pole-1}
    - \frac{ N}{k} \, \sqrt{\frac{M+N}{M N}} \,
    i \, f^{a b c} \, \pa \, (K^c \, J^{u(1)}) \nonu \\
  &  &+
    \frac{N}{(2k+M)} \Big( i \, f^{a d e} \,
    d^{b d f}  \, \pa \, (K^e \, J^f)  +
     i \, f^{a d e} \,
    d^{b f d}  \, \pa \, (J^f \, K^e)
    \Big)    
    \Bigg](w) \, a_{12}
  \nonu \\
  & &+ i \, f^{a b c} \, \pa^2 \, K^c(w) \, a_{16}
  \nonu \\
  & & +  \Big( \frac{2}{M} \, \de^{b c} \de^{d e} +
  \frac{2}{M} \, \de^{b e} \de^{c d} +
  \frac{2}{M} \, \de^{b d} \de^{c e} +
  \frac{i}{2} \, f^{b c f} \, d^{f d e}+
  \frac{1}{2} \, d^{b c f} \, d^{f d e}
  \nonu \\
  && +
\frac{i}{2} \, f^{b e f} \, d^{f c d}+
  \frac{1}{2} \, d^{b e f} \, d^{f c d}
  +
\frac{i}{2} \, f^{b d f} \, d^{f c e}+
  \frac{1}{2} \, d^{b d f} \, d^{f c e}
  \Big) 
  \nonu \\
   & & \times  \Big(  i \, f^{a c f} \,
  \, K^{f} \, J^d \, J^e +
 i \, f^{a d f} \,
 \, J^{c} \, K^f \, J^e +
  i \, f^{a e f} \,
  \, J^{c} \, J^d \, K^f 
  \Big)(w)\, a_{17}.
  \label{KPfirstorder}
\eea
In this case, from Appendix (\ref{KPfirstorder}),
we do not have to consider the additional contractions
between the operators because we are focusing on
the first order pole.
 We can also further simplify the above expressions by changing
 the ordering of operators appropriately.
 Then the above expression Appendix (\ref{KPfirstorder})
 plays an important role of a new quasi primary spin-$4$ current
 $\hat{R}^c(w)$
 by multiplying $i \, f^{a b c}$ together with (\ref{Rdef}) and
 (\ref{Rhatab}).
 
\section{ The second  order pole
  in the OPE  $K^a(z) \, W^{(3)}(w)$}

The second order pole of
the OPE between the charged spin-$2$ current and the
neutral spin-$3$ current can be described as 
\bea
&& K^a(z) \, W^{(3)}(w)\Bigg|_{\frac{1}{(z-w)^2}}  =
3M\, d^{a b c} \, J^b \, K^c(w) \, b_2-
2 M\, J^{u(1)} \, K^a(w) \, b_5
\nonu \\
&& +\Big( 4(2k+M+N) \, t^{\al}_{\rho, \bar{\si}} \, t^a_{j \bar{i}} \,
J^{\al}\,J^{(\rho \bar{i})}  \,  J^{(\bar{\si} j)}+
\frac{4}{k}\, (2k +M+N) \, J^a \, J^{\al}\, J^{\al}\Big)(w) \, b_6
\nonu \\
&& +\Big( i \, f^{a b c} \, (K^c \, K^b)_{pole-1}
+ \frac{M N}{(2k+M)} \, d^{a b c} \, (K^b \, J^c +
J^c \, K^b) -\frac{ 4 M N}{k}\, \sqrt{\frac{M+N}{M N}}\,
K^a \, J^{u(1)}
\nonu \\
&&+ J^b \, (K^a \, K^b)_{pole-2} +\frac{M N}{(2k+M)}\,
d^{a b c } \, J^c \, K^b
\Big)(w) \, b_7 + 2(2k +2N+M)\, J^{u(1)}\, K^a(w) \, b_8
\nonu \\
&&
+
\Big( \frac{2 (k^2-1) (2 k+M+N)}{k (2 k+M)} \,
\de_{\rho \bar{\si}} \, \de^{k \bar{i}}\, t^a_{k \bar{l}}
J^{(\rho \bar{l})} \, \pa \,  J^{(\bar{\si} j)}
\nonu \\
&& + 
\de_{\rho \bar{\si}} \, \de_{j \bar{i}}\,
((K^a \, J^{(\rho \bar{i})})_{pole-1} \, \pa \, J^{(\bar{\si} j)})_{pole-1}
\nonu \\
&& +  \de_{\rho \bar{\si}} \, \de_{j \bar{i}}\,
J^{(\rho \bar{i})} \, (K^a \, \pa\ J^{(\bar{\si} j)})_{pole-2}
\Big)(w)\, (b_{13}-b_{12}) 
\nonu \\
&& + \Big( 3(k +M+N)\, \pa \, K^a -\frac{1}{2}(
2 M \, \pa \, K^a + i \, f^{a b c} \, (K^c \, J^b +
J^b \, K^c)) \Big)(w) \, b_{12}.
\label{k3wpole2}
\eea
Moreover, in order to calculate the $(b_{13}-b_{12})$ terms
the following relations can be used by considering
the similar relations as in Appendix (\ref{fourfirst})
\bea
&& \de_{\rho \bar{\si}} \,
\de_{j \bar{i}} \, \de^{k \bar{i}}  \, t^a_{k \bar{l}} \,   J^{u(1)}
\, J^{(\rho \bar{l})}
(z) \, J^{(\bar{\si} j)}(w) \Bigg|_{\frac{1}{(z-w)}}
= -  \sqrt{\frac{M+N}{M N}}  \,
\de_{\rho \bar{\si}} \, t^a_{j \bar{i}}\,
J^{(\rho \bar{i})}  \,  J^{(\bar{\si} j)}(w)
\nonu \\
&& -N  \, J^{u(1)} \, J^a(w),
\nonu \\
&& \de_{\rho \bar{\si}} \,
\de_{j \bar{i}} \,   J^a \, J^{(\rho \bar{i})}(z) \, \, J^{(\bar{\si} j)}(w) \Bigg|_{\frac{1}{(z-w)}}
= 
\de_{\rho \bar{\si}} \, t^a_{j \bar{i}}\,
J^{(\rho \bar{i})}  \,  J^{(\bar{\si} j)}(w)
\nonu \\
&& +k \, M \, N \, \pa \, J^a -  \sqrt{\frac{M+N}{M N}}\, M\, N\,
J^a \, J^{u(1)}(w),
\nonu \\
&& \de_{\rho \bar{\si}} \,
\de_{j \bar{i}} \,  (i \, f -\frac{(2k+
    M+2N)}{
    (2k+M)} \, d)^{a c d} \, \de^{k_1 \bar{i}} \, t^d_{k_1 \bar{j_1}} \, J^c \,
   J^{(\rho \bar{j_1})}(z) \, \, J^{(\bar{\si} j)}(w) \Bigg|_{\frac{1}{(z-w)}}
 = 
 \nonu \\
 && \frac{2 (2 k M^2-4 k+M^3+M^2 N-2 M-4 N)}{M (2 k+M)}\,
\de_{\rho \bar{\si}} \, t^a_{j \bar{i}}\,
J^{(\rho \bar{i})}  \,  J^{(\bar{\si} j)}(w)
\nonu \\
&& - (i \, f -\frac{(2k+
    M+2N)}{
  (2k+M)} \,d)^{a b c} \, N
\, J^{b}\, J^{c}(w),
\label{fourfirst1}
 \\
&& \de_{\rho \bar{\si}} \,
\de_{j \bar{i}} \,  \de^{k_1 \bar{i}} \, \de_{\si_2 \bar{\si_1}} \, (t^{\al})^{\bar{\si_1} \rho}
   \, t^a_{k_1 \bar{j_1}} \, J^{\al} \, J^{(\si_2 \bar{j_1})}(z) \,
   \, J^{(\bar{\si} j)}(w) \Bigg|_{\frac{1}{(z-w)}}
   = -\frac{(N^2-1)}{ N}  \,
 t^a_{\rho \bar{\si}} \, \de_{j \bar{i}}\,
 J^{(\rho \bar{i})}  \,  J^{(\bar{\si} j) }(w).
 \nonu
 \eea
 Then
 we obtain the
 $\de_{\rho \bar{\si}} \, \de_{j \bar{i}}\,
 ((K^a \, J^{(\rho \bar{i})})_{pole-1} \, \pa \, J^{(\bar{\si} j)})_{pole-1}$
in Appendix (\ref{k3wpole2})
 by taking the appropriate derivatives
 in Appendix (\ref{fourfirst1}).
 By simplifying Appendix (\ref{k3wpole2}), we will end up with
 the second order pole in (\ref{KW}).

\section{ The first  order pole
  in the OPE  $K^a(z) \, W^{(3)}(w)$}

In the $b_6$ term of $W^{(3)}$ in (\ref{W}),
after moving the second factor to the left
in the second term,  there exists a $\pa \, J^{\alpha}$
term.
But the OPE with $K^a(z)$ does not contribute to
nonzero expression.
In the $b_7$ term of $W^{(3)}$, there exists
a $\pa \, J^b$ term by  moving the second factor to the left
in the second term. We should include this contribution also.
In the $b_8$ term of $W^{(3)}$,
 there exists
a $\pa \, J^{u(1)}$ term by  moving the second factor to the left
in the second term.
But the OPE with $K^a(z)$ does not contribute to
nonzero expression.
Finally,
in the $b_{13}$ term of $W^{(3)}$,
 there exists
 a $\pa^2 \, J^{u(1)}$ term, by  moving the second factor to the
 left, which does not contribute to the nonzero result.
 It is easy to observe that the OPEs between
 $K^a(z)$ and other terms in the $W^{(3)}(w)$
 vanish.
 
 Therefore, we summarize the first order pole as follows:
\bea
&& K^a(z) \, W^{(3)}(w)\Bigg|_{\frac{1}{(z-w)}}  = 
\Big( i \, f^{a b e } \, K^e \, J^c\, J^d +
 i \, f^{a c e } \, J^b \, K^e\, J^d 
+ i \, f^{a d e } \, J^b \, J^c\, K^e 
 \Big)(w)\, d^{b c d} \, b_2
 \nonu \\
 & & + i \, f^{a b c} \, J^{u(1)} \, \Big( K^c \, J^b+
 J^b \, K^c\Big)(w) \, b_5
 \label{KWpoleone} \\
 && + 2 \,  \Big( t^{\alpha}_{\rho \bar{\si}} \, \de_{j \bar{i}} \,
 J^{\alpha} \, [
 ((K^a \, J^{(\rho \bar{i})})_{pole-1} \, J^{(\bar{\si} j)})
 +
 J^{(\rho \bar{i})} \, (K^a \, J^{(\bar{\si} j)})_{pole-1}]
 \Big)(w) \, b_6 
 \nonu \\
 && + \Big( i \, f^{a b c} \, N\, K^c \, \pa \, J^b +
  i \, f^{a b c} \, N \, J^b \, \pa \, K^c
  + 2 \, i\, f^{a b c} \,  t^{b}_{j \bar{i}} \,
  \de_{\rho \bar{\si}} \, K^c \,  J^{(\rho \bar{i})} \,
   J^{(\bar{\si} j)}
   \nonu \\
   && +  2 \,  t^{b}_{j \bar{i}} \,
   \de_{\rho \bar{\si}} \,J^b \, [
     ((K^a \, J^{(\rho \bar{i})})_{pole-1} \, J^{(\bar{\si} j)})+
  J^{(\rho \bar{i})} \, (K^a \, J^{(\bar{\si} j)})_{pole-1}  ]\Big)(w) \, b_7
   \nonu \\
   && + 2  \de_{j \bar{i}} \,
   \de_{\rho \bar{\si}} \, J^{u(1)} \,  \Big(
 ((K^a \, J^{(\rho \bar{i})})_{pole-1} \, J^{(\bar{\si} j)})+
  J^{(\rho \bar{i})} \, (K^a \, J^{(\bar{\si} j)})_{pole-1}
   \Big)(w) \, b_8
   \nonu \\
   && + \de_{j \bar{i}} \,
   \de_{\rho \bar{\si}}  \,  \Big(
 ((K^a \, \pa \, J^{(\rho \bar{i})})_{pole-1} \, J^{(\bar{\si} j)})+
  \pa \, J^{(\rho \bar{i})} \, (K^a \, J^{(\bar{\si} j)})_{pole-1}
   \Big)(w) \, b_{12}
   \nonu \\
   && +\de_{j \bar{i}} \,
   \de_{\rho \bar{\si}}  \,  \Big(
 ((K^a \,  J^{(\rho \bar{i})})_{pole-1} \, \pa \, J^{(\bar{\si} j)})+
   J^{(\rho \bar{i})} \, (K^a \, \pa \, J^{(\bar{\si} j)})_{pole-1}
   \Big)(w) \, b_{13},
   \nonu
\eea
where the four quantities in the above are given, from
(\ref{KJJ}),  by 
\bea
&&  K^a(z) \, J^{(\rho \bar{i})}(w) \Bigg|_{\frac{1}{(z-w)}}\equiv
(K^a \, J^{(\rho \bar{i})})_{pole-1}  =  \nonu \\
&& 
2(k+N) \de_{k \bar{j}} \, (t^{a})^{\bar{i} k} \,
  \pa \, J^{(\rho \bar{j})}(w) -\frac{2(k+N)}{k}  \sqrt{\frac{M+N}{M N}}
   \de^{k \bar{i}}  t^a_{k \bar{j}}    J^{u(1)}    J^{(\rho \bar{j})}(w)
  \nonu \\
  && + \frac{2}{k M}(k+M+N) \, J^a \, J^{(\rho \bar{i})}(w)-(i \, f -\frac{(2k+
    M+2N)}{
    (2k+M)} \, d)^{a b c} \, \de^{k \bar{i}} \, t^c_{k \bar{j}} \, J^b \,
  J^{(\rho \bar{j})}(w) \nonu \\
  & &- 2 \de^{k \bar{i}} \, \de_{\si \bar{\si_1}} \, (t^{\al})^{\bar{\si_1} \rho}
  \, t^a_{k \bar{j}} \, J^{\al} \, J^{(\si \bar{j})}(w),
  \nonu \\
&&  (K^a \, J^{(\bar{\si} j)})_{pole-1}  = 
 2(k+N) \, \de_{k \bar{k_1}} \, (t^{a})^{\bar{k_1} j} \,
  \pa \, J^{(\bar{\si} k)}(w) +\frac{2(k+N)}{k}  \sqrt{\frac{M+N}{M N}}
   \de^{j \bar{l}}  t^a_{k \bar{l}}  J^{u(1)}  J^{(\bar{\si} k)}(w)
  \nonu \\
  && - \frac{2}{k M}(k+M+N) \, J^a \, J^{(\bar{\si} j)}(w)-(i \, f +
  \frac{(2k+M+2N)}{
    (2k+M)} \, d)^{a b c} \, \de^{j \bar{j_1}} \, t^c_{k \bar{j_1}} \, J^b \,
  J^{(\bar{\si} k)}(w) \nonu \\
  &&+ 2 \de^{j \bar{l}} \, \de_{\tau \bar{\si_1}} \,
  (t^{\al})^{\bar{\si} \tau}
  \, t^a_{k \bar{l}} \, J^{\al} \, J^{(\bar{\si_1} k)}(w)  
  \nonu \\
&& (K^a \, \pa \, J^{(\rho \bar{i})})_{pole-1}  =
\pa \Bigg[ 2(k+N) \de_{k \bar{j}} \, (t^{a})^{\bar{i} k} \,
  \pa \, J^{(\rho \bar{j})} -\frac{2(k+N)}{k}  \sqrt{\frac{M+N}{M N}}
   \de^{k \bar{i}}  t^a_{k \bar{j}}    J^{u(1)}    J^{(\rho \bar{j})}
  \nonu \\
  &&+ \frac{2}{k M}(k+M+N) \, J^a \, J^{(\rho \bar{i})}-(i \, f -\frac{(2k+
    M+2N)}{
    (2k+M)} \, d)^{a b c} \, \de^{k \bar{i}} \, t^c_{k \bar{j}} \, J^b \,
  J^{(\rho \bar{j})} \nonu \\
  &&- 2 \de^{k \bar{i}} \, \de_{\si \bar{\si_1}} \, (t^{\al})^{\bar{\si_1} \rho}
  \, t^a_{k \bar{j}} \, J^{\al} \, J^{(\si \bar{j})} \Bigg](w),
  \nonu \\
 && (K^a \, \pa \, J^{(\bar{\si} j)})_{pole-1}  = 
 \pa \, \Bigg[ 2(k+N) \, \de_{k \bar{k_1}} \, (t^{a})^{\bar{k_1} j} \,
  \pa \, J^{(\bar{\si} k)} +\frac{2(k+N)}{k}  \sqrt{\frac{M+N}{M N}}
   \de^{j \bar{l}}  t^a_{k \bar{l}}  J^{u(1)}  J^{(\bar{\si} k)}
  \nonu \\
  & &- \frac{2}{k M}(k+M+N) \, J^a \, J^{(\bar{\si} j)}-(i \, f +
  \frac{(2k+M+2N)}{
    (2k+M)} \, d)^{a b c} \, \de^{j \bar{j_1}} \, t^c_{k \bar{j_1}} \, J^b \,
  J^{(\bar{\si} k)} \nonu \\
 & &+ 2 \de^{j \bar{l}} \, \de_{\tau \bar{\si_1}} \,
  (t^{\al})^{\bar{\si} \tau}
  \, t^a_{k \bar{l}} \, J^{\al} \, J^{(\bar{\si_1} k)} \Bigg](w).
  \label{simplepoleone}
\eea
Note that we should be careful about the
normal ordering \cite{BBSS} in the composite operators
containing the first and the third of Appendix
(\ref{simplepoleone}) 
in Appendix (\ref{KWpoleone}).
For any operators, $A,B,C$, we have the relation
$((AB)C)(w)= ([(AB),C])(w)+ (C(AB))(w)$
where the bracket stands for the normal ordering
between the operators.

\section{ Relevant free field realization}

In this section, the
free field realization in \cite{OS}
is reviewed and we comment on its relevance in the context of
previous sections.

\subsection{Free field construction}

The generators of $W_{1+\infty}$ algebra \cite{PRS}
are given by $V^i(z)$ of spin-$(i+2)$ and
the generator of affine $SU(M)$ algebra
is given by the spin-$1$ current $4 \, q\, W^{-1,a}(z)$
where $a=1,2,\cdots, (M^2-1)$ and $q$ is a parameter and
is fixed by $q =\frac{1}{4}$.
We follow the notation of \cite{OS} except that
their $N$ corresponds to our $M$.
Moreover the additional generators are given by
$W^{i,a}(z)$ of spin-$(i+2)$ transforming as the adjoint
representation of $SU(M)$.
The realization of the algebra is represented by
bilinear free fermions.
The complex free fermions satisfy the following OPE
\bea
\bar{\psi}^{\alpha}(z) \, \psi^{\beta}(w) =\frac{1}{(z-w)}\,
\de^{\alpha \, \beta} + \cdots,
\label{freeOPE}
\eea
where $\alpha, \beta = 1, 2, \cdots, M$.
Then the level $k=1$ realization is given by the following
forms \footnote{In the calculation of OPEs between the free fields
  in this section we will consider the $M=4$ case in order to
  see the structure of the algebra and the general
calculation for arbitrary $M$ can be done by hand.}
\bea
V^j(z) & = &
\frac{2^{j-1} (j+1)!}{(2 j+1)!!}\,
q^j \, \sum _{r=0}^{j+1} \, (-1)^r
\, \left(\begin{array}{c}
  j+1 \\
  r\\
\end{array}
\right)^2 \,
 \pa^{j-r+1} \, \bar{\psi}^{\alpha} \,
\pa^r \, \psi^{\alpha}(z),
\nonu \\
W^{j,a }(z) & = &
\frac{2^{j-1} (j+1)!}{(2 j+1)!!}\,
q^j \, \sum _{r=0}^{j+1} \, (-1)^r
\, \left(\begin{array}{c}
  j+1 \\
  r\\
\end{array}
\right)^2 \,
 \pa^{j-r+1} \, \bar{\psi}^{\alpha} \,
\pa^r \, t^a_{\alpha \, \beta}\, \psi^{\beta}(z).
\label{VW}
\eea
Then we can check that 
the stress energy tensor $V^0(z)$ has the central charge
$c=M \, k$ and it becomes $c=M$
by using the fundamental relation in
Appendices
(\ref{freeOPE}) and (\ref{VW}). The spin-$3$ operator $V^1(z)$
is a quasi primary operator and has the fourth order pole
$V^{-1}(w)$ in the OPE with the above stress energy tensor.
Moreover, there is a quasi primary spin-$4$ $V^2(z)$ operator.

We can check that
there exist two primary spin-$1, 2$ operators
$W^{-1,a}(z)$ and $W^{0,a}(z)$. For the
quasi primary spin-$3$ operator $W^{1,a}(z)$,
the OPE between the stress energy tensor $V^0(z)$ and
$W^{1,a}(w)$ has nonzero fourth order pole
$W^{-1,a}(w)$.

We consider two cases as follows:

$\bullet$ The OPEs between the nonsinglet currents and singlet current

When we calculate the OPE between $W^{-1,a}(z)$
and $V^1(w)$, we observe that the second order pole contains
the spin-$2$ operator $W^{0,a}(w)$.
This can be compared to the previous result in (\ref{JWpole2})
  which eventually becomes zero. 
The OPE between $W^{0,a}(z)$
and $V^1(w)$ implies that the nonzero singular terms are
given by  $W^{-1,a}(w)$ in the fourth order pole,
$ 3\, W^{1,a}(w)$ in the second order pole and
$\pa \, W^{1,a}(w)$ in the first order pole.
We can compare this with the one in (\ref{KW}) and realize that
there are common linear terms in the (quasi) spin-$3$ operator.
We can further calculate the OPE between the spin-$3$ operator
 $W^{1,a}(z)$
and other spin-$3$ operator  $V^1(w)$. It turns out that
there are $4 \, W^{0,a}(w)$, $2 \, \pa \, W^{0,a}(w)$,
$\frac{3}{5} \,\pa^2\,  W^{0,a}(w)+4 \, W^{2,a}(w)$, and
$\frac{2}{15} \,\pa^3\,  W^{0,a}(w)+2 \, \pa \, W^{2,a}(w)$
in the fourth, third, second and first order poles respectively.
Again by comparing with (\ref{PW}), we observe that
the same linear terms occur in both cases.

$\bullet$ The OPEs between the nonsinglet currents

The OPE between the spin-$1$ current and
the charged spin-$2$ current gives us
the nonzero second order pole which is given by
$(\frac{1}{2}\, d^{a b c} \, W^{-1,c}+\frac{1}{4} \, \de^{a b}\,
V^{-1})(w)$.
Moreover, the first order pole gives
$ i\, f^{a b c} \, W^{0,c}(w)$.
Note that there exists a spin-$1$ singlet current.
This can be compared to the relation (\ref{JK}).
Similarly, the OPE between the charged spin-$2$ current
and itself leads to
$\frac{1}{2}\, \de^{a b}$  in the fourth order pole,
$\frac{1}{2}\, i\, f^{a b c}\,
W^{-1,c}(w)$  in the third order pole, $( \frac{1}{4}\, i\, f^{a b c}\,
\pa \, W^{-1,c} + \frac{1}{2}\, \de^{a b} \, V^0+
d^{a b c} \, W^{0,c})(w)$ in the second order pole
and $(\frac{1}{12}\, i\, f^{a b c}\,
\pa^2 \, W^{-1,c} + \frac{1}{4}\, \de^{a b} \, \pa \, V^0+
\frac{1}{2} \, d^{a b c} \, \pa \, W^{0,c}+
i\, f^{a b c}\, W^{1,c})(w)$ in 
the first order pole respectively.
Now we observe that when we compare this with (\ref{fullKK}),
both cases share the common linear terms at each singular term.
The OPE between the spin-$1$ current
and  the charged spin-$3$ current leads to
$\frac{1}{3}\, f^{a b c}\, W^{-1,c}(w)$,
$(-\frac{1}{6}\, f^{a b c}\, W^{-1,c}+ d^{a b c} \, W^{0,c})(w)$
and $i \, f^{a b c}\, W^{1,c}(w)$
in the third, the second and the first order poles
respectively. Note that there are nonzero singular terms
in the third order and second order poles when we compare
with the one in (\ref{JP}).

We can also check the OPE
between the charged spin-$2$ current
and  the charged spin-$3$ current. It turns out that
the fourth order pole is
$(\frac{1}{4}\, \de^{a b}\, V^{-1}-\frac{i}{2}\, d^{a b c}
\, W^{-1,c})(w)$, the third order pole is
$\frac{4}{3}\, i \, f^{a b c} \, W^{0,c}(w)$,
the second order pole is
$(\frac{3}{4} \, \de^{a b} \, V^1-\frac{3 i}{2}
\, d^{a b c}
\, W^{1,c}+\frac{1}{3}\, i \, f^{a b c} \, \pa \, W^{0,c})(w)$
and the first order pole is
$(\frac{1}{4} \, \de^{a b} \, \pa \, V^1-\frac{ i}{2}
\, d^{a b c}
\, \pa\, W^{1,c}+
\frac{1}{15}\, i \, f^{a b c} \, \pa^2 \, W^{0,c}+
i \, f^{a b c}\, W^{2,c})(w)$.
In this case, some of the linear terms of this OPE
occur in the (\ref{KPfinal}).
Note the presence of a spin-$1$ singlet current.
Finally, the OPE between
the charged spin-$3$ current and itself provides
the following singular terms \footnote{
  The sixth order pole is proportional to
  the Kronecker delta symbols and is given by
  $\frac{2}{3}\, \de^{a b}$.
  The fifth order pole contains the spin-$1$ current and
  is $\frac{2}{3}\,
i\,f^{a b c}  \, W^{-1,c}(w)$.
The fourth order pole contains the two spin-$2$ currents
as well as the descendant terms and is given by
$(\frac{1}{3}\,
i\,f^{a b c} \, \pa \, W^{-1,c}+ \de^{a b} \, V^0 + 2 \,
d^{a b c} \, W^{0,c})(w)$.
The third order pole has a spin-$3$ current as well as various
descendant terms and is given by
$(\frac{1}{9}\,
i\,f^{a b c} \, \pa^2 \, W^{-1,c}+ \frac{1}{2}\,
\de^{a b} \, \pa \, V^0 +  
d^{a b c} \, \pa \, W^{0,c}+ \frac{8}{3}\, i\, f^{a b c} \,
W^{1,c})(w)$. The second order pole contains the two kinds of
spin-$4$ currents and the descendant terms and  is
$(\frac{1}{36}\,
i\,f^{a b c} \, \pa^3 \, W^{-1,c}+ \frac{3}{20}\,
\de^{a b} \, \pa^2 \, V^0 +  
\frac{3}{10} \,
d^{a b c} \, \pa^2 \, W^{0,c}+ \frac{4}{3}\, i\, f^{a b c} \,
\pa \, W^{1,c} +\de^{a b}\, V^2 + 2 \, d^{a b c} \, W^{2,c})(w)$.
Finally, the first order pole contains a spin-$5$ current besides
the various descendant terms and is given by
$(\frac{1}{180}\,
i\,f^{a b c} \, \pa^4 \, W^{-1,c}+ \frac{1}{30}\,
\de^{a b} \, \pa^3 \, V^0 +  
\frac{1}{15} \,
d^{a b c} \, \pa^3 \, W^{0,c}+ \frac{8}{21}\, i\, f^{a b c} \,
\pa^2 \, W^{1,c} +  d^{a b c} \, \pa \, W^{2,c}+\frac{1}{2}\,
\de^{a b}\, \pa \, V^2+
i \, f^{a b c}\, W^{3,c})(w)$.
Note that the relative coefficients appearing in the descendant terms
are fixed automatically.}.

Therefore, we observe that
the presence of a neutral spin-$1$ current
with Kronecker delta symbols appears in the
OPEs between the nonsinglet currents
where the sum of spins of the left hand side is given by
odd integer numbers. Although there are
some higher order terms which do not appear in
the coset realization,
we observe that by simply ignoring the above
uncharged spin-$1$ current, all the linear terms in the free
field realization arise in the coset realization.
One of the lessons from the free field realization is to expect how
the new quasi primary operators  arise in the specific singular terms
of the given OPEs.
From this fact we can rearrange each singular term in the
coset realization by expecting that there should be a new quasi
primary operator at that singular term. If we do not expect
a new quasi primary operator, then we should manage to rewrite
each singular term in terms of the multiple product of
known currents.

\subsection{ After decoupling the neutral spin-$1$ current}

We can construct the nonsinglet and singlet operators
which do not have any singular terms in the OPEs
with the above neutral spin-$1$ current
and present them as follows:
\bea
\hat{V}^0(z) & = & V^0(z) -\frac{1}{8}\, V^{-1}\, V^{-1}(z),
\nonu \\
\hat{V}^1(z)  & = & V^1(z)-\frac{1}{2} \, V^{-1}\, V^0(z) +
\frac{1}{24} \, V^{-1}\, V^{-1}\, V^{-1}(z),
\nonu \\
\hat{W}^{-1,a}(z) & = & W^{-1,a}(z),
\nonu \\
\hat{W}^{0,a}(z) & = & W^{0,a}(z) -\frac{1}{4} \, V^{-1}\, W^{-1,a}(z),
\nonu \\
\hat{W}^{1,a}(z) & = & W^{1,a}(z) -\frac{1}{2} \, V^{-1}\, W^{0,a}(z)+
\frac{1}{16}\,  V^{-1}\, V^{-1}\, W^{-1,a}(z)
\nonu \\
& - & \frac{3}{10} \, \hat{V}^0 \,  W^{-1,a}(z)+
\frac{3}{20}\,  \pa^2 \, W^{-1,a}(z).
\label{firstred}
\eea
The central charge is given by 
$c=M-1=3$. The singlet and nonsinglet operators are primary
under the new stress energy tensor.
Due to the nonlinear terms in the right hand side,
we expect that the algebra between these operators
leads to the nonlinear terms in the right hand side of the OPEs.
Note that the OPE between
the charged spin-$1$ current and
the neutral spin-$1$ current is regular.
The charged and uncharged spin-$4$ currents can be
determined similarly.

Moreover, the above operators
in Appendix (\ref{firstred}) can be written in terms of
$W^{-1,a}(z)$ as follows:
\bea
\hat{V}^0(z) & = & \frac{1}{10}\, W^{-1,a}\, W^{-1,a}(z),
\nonu \\
\hat{V}^1(z) & = &
\frac{1}{90}\, d^{a b c} \, W^{-1,a} \, W^{-1,b}\, W^{-1,c}(z),
\nonu \\
\hat{W}^{-1,a}(z) & = & W^{-1,a}(z),
\nonu \\
\hat{W}^{0,a}(z) & = &   \frac{1}{12} \,
d^{a b c} \, W^{-1,b}\, W^{-1,c}(z),
\nonu \\
\hat{W}^{1,a}(z) & = &
-\frac{2}{25} \, W^{-1,b} \, W^{-1,b}\, W^{-1,a}(z)-
\frac{1}{75} \,  \pa^2 \, W^{-1,a}(z) \nonu \\
& - & \frac{9}{100} \,
i \, f^{a b c} \, \pa \, W^{-1,b} \, W^{-1,c}(z).
\label{secondred}
\eea
Note that
in terms of the complex free fermions,
the above operators in Appendix (\ref{secondred}) contain
the quartic,  
the sextic,
the quadratic,
the quartic,  
and the sextic terms in the free fermions respectively.
The OPE between the spin-$1$ current and
the charged spin-$2$ current gives us
the nonzero second order pole which is given by
$\frac{1}{2}\, d^{a b c} \, \hat{W}^{-1,c}(w)$.
Moreover, the first order pole gives
$ i\, f^{a b c} \, \hat{W}^{0,c}(w)$.
Note that there is no spin-$1$ singlet current.

The OPE between the charged spin-$2$ current 
$\hat{W}^{0,a}(z)$ and itself
can be calculated.
It turns out that
the fourth order pole is given by
$\frac{1}{4}\, \de^{a b}$, the third order pole is
$\frac{1}{4} \, i \, f^{a b c}\, \hat{W}^{-1,c}(w)$,
the second order pole is
$(\frac{1}{8} \, i \, f^{a b c}\, \pa \, \hat{W}^{-1,c}+
\frac{1}{2} \, \de^{a b} \, \hat{V}^0+ d^{a b c} \, \hat{W}^{0,c}-
\frac{1}{8} \, ( \hat{W}^{-1,a}\,  \hat{W}^{-1,b} +
\hat{W}^{-1,b}\,  \hat{W}^{-1,a}) )
(w)$. Note that
there are nonlinear terms in this pole.
Furthermore, the first order pole is given by
$(\frac{1}{24} \, i \, f^{a b c}\, \pa^2 \, \hat{W}^{-1,c}+
\frac{1}{4} \, \de^{a b} \, \pa \, \hat{V}^0+
\frac{1}{2} \, d^{a b c} \, \pa \, \hat{W}^{0,c}-
\frac{1}{16} \, \pa \,
( \hat{W}^{-1,a}\,  \hat{W}^{-1,b} +  \hat{W}^{-1,b}\,
\hat{W}^{-1,a}) )
(w)$. This implies that there is no new (quasi) primary operator
in this OPE.
Other OPEs can be determined without any difficulty.

Therefore, although we have decoupled the neutral spin-$1$ current in
the above analysis, the operators in Appendix
(\ref{secondred})
do not produce any new (quasi) primary operators due to the property
of the fermions. The charged spin-$1$ current which generates
the affine $SU(M)$ algebra produces the $W$ algebra between the
nonsinglet currents.



\end{document}